\documentclass[a4paper,english,14pt]{extarticle}
\usepackage[english]{babel} 
\usepackage[lmargin={3.5cm},rmargin={2.5cm},tmargin={2.5cm},bmargin={2cm}]{geometry}
\usepackage{graphicx}
\usepackage{fancyhdr}
\usepackage{helvet}
\usepackage{extramarks}
\usepackage{amsmath}
\usepackage{braket}
\usepackage{amsthm}
\usepackage{amssymb}
\usepackage{hyperref} 
\usepackage[right]{eurosym}
\usepackage[latin1]{inputenc}
\usepackage{fancybox} 
\usepackage[dvipsnames]{xcolor} 
\usepackage{color}
\usepackage{braket}
\usepackage{colortbl} 
\definecolor{myorange}{RGB}{247,150,0} 
\usepackage{float}
\usepackage{framed}
\usepackage{url}
\usepackage{verbatim} 
\usepackage[titles]{tocloft} 

\usepackage{natbib}

\usepackage{booktabs}
\usepackage{tabularx} 

\usepackage{multirow} 
\usepackage{rotating} 
\usepackage{rotfloat}

\usepackage[bottom]{footmisc} 
\usepackage{tikz} 
\usepackage{tcolorbox} 
\usepackage{pgfkeys}
\usepackage{multicol} 
\usepackage{colortbl} 
\usepackage{algorithm}
\usepackage{algpseudocode}
\usepackage{algorithmicx} 
\usepackage{xfrac} 
\usepackage{mathtools}
\usepackage{breqn}
\usepackage{pdfpages}
\DeclarePairedDelimiter{\norm}{\lVert}{\rVert}
\newtheoremstyle{mystyle}
{10pt}
{10pt}
{\itshape}
{}
{\bfseries}
{:}
{0.5em}
{\thmname{#1}\thmnumber{ #2}:\thmnote{ #3}}

\theoremstyle{mystyle}

\numberwithin{equation}{section}

\newcommand{\myparagraph}[1]{\paragraph{}\mbox{}\\}

\begin{document}
\begin{titlepage}


\begin{center}
		\Huge{Quantum Algorithms in Cybernetics}
		\vspace*{0.5cm}\\
		\Large{Doctoral Dissertation}\\
		\vspace*{0.5cm}
		\normalsize{
			Area: 5. Engineering
			\\
			Professional area: 5.2. Electrical engineering, electronics and automation
			\\
			Scientific specialty: Application of the principles and methods of cybernetics in various fields of science\\
		}
		\vspace*{0.5cm}
		\Large{by}\\
		\vspace*{0.5cm}
		\huge{Petar Nikolaev Nikolov}\\
		\vspace*{0.5cm}
		\Large{
			At the Faculty of German Engineering and Business Administration
			\\
			Technical University of Sofia
		}
		
	\end{center}
	\vspace*{0.375cm}
\large{
\begin{center}
\begin{tabular}[ht]{l l}
  Advisor: &Assoc. prof. Dr. Eng. Vassil Galabov \\
  Date of Submission: &28.02.2020\\
\end{tabular}
\end{center}
}

  \vspace*{0.75cm}
	
	\tiny{ TU-Sofia - Technical University of Sofia} \hfill
	\small{\textbf{www.tu-sofia.bg} }	
\end{titlepage}

\thispagestyle{empty}\cleardoublepage

\mbox{}\thispagestyle{empty}
\cleardoublepage

\include{Declaration}

\thispagestyle{empty}\cleardoublepage

{\Large \textbf{Abstract}} 

\bigskip

Quantum information theory and the quantum computing as the biggest part of this scientific area, is one of the fast growing emerging technologies nowadays. Quantum computers use the quantum mechanical effects like superposition, entanglement and coherence to process information. The main difference with the classical computer is that they process the information probabilistically, while the classical signal processing is deterministic. Quantum computers can handle two types of data - classical and quantum, although the classical data need to be prepared and structured before being processed. The quantum information finds patterns in the data by presenting those as certain quantum mechanical states and executes basic quantum routines.
\par
In general a quantum algorithm can be applied on different hardware platforms if it could be described as a quantum circuit. The circuit model for quantum computation applies different quantum logic gates over a quantum register which maps the initial register state to another desired state as a result. The quantum computation ends with a measurement operation, which is also a termination for the computational process.
\par 
The development of a new quantum algorithm requires the use of both existing and newly constructed quantum gates. And it can be generalized into two main tasks - constructing new quantum gates and procedure for solving the problem.
Having only two possible states, the binary homogeneous (the probability distribution does not change over time) Markov process, makes it possible the processing being executed on a quantum computer. The process's states can be represented by qubits, where the main challenge being the creation and tuning of the quantum logic gates connecting these qubits.
\par
A new method for simulation of a binary homogeneous Markov process using a quantum computer was proposed. This new method allows using the distinguished properties of the quantum mechanical systems - superposition, entanglement and probability calculations. Implementation of an algorithm based on this method requires the creation of a new quantum logic gate, which creates entangled state between two qubits. This is a two-qubit logic gate and it must perform a predefined rotation over the X-axis for the qubit that acts as a target, where the rotation accurately represents the transient probabilities for a given Markov process. This gate fires only when the control qubit is in state $\ket{1}$. It is necessary to develop an algorithm, which uses the distributions for the transient probabilities of the process in a simple and intuitive way and then transforms those into X-axis offsets. 
The creation of a quantum $C \sqrt[n]{X}$ gate using only the existing basic quantum logic gates at the available cloud platforms is possible, although the hardware devices are still too noisy, which results in a significant measurement error increase. The IBM's Yorktown "bow-tie" back-end performs quite better than the 5-qubit T-shaped and the 14-qubit Melbourne quantum processors in terms of quantum fidelity.
The simulation of the binary homogeneous Markov process on a real quantum processor gives best results on the Vigo and Yorktown (both 5-qubit) back-ends with Hellinger fidelity of near 0.82. The choice of the right quantum circuit, based on the available hardware (topology, size, timing properties), would be the approach for maximizing the fidelity.

\mbox{}\thispagestyle{empty}
\cleardoublepage
\rmfamily \pagestyle{fancy} \setcounter{secnumdepth}{4}

\setlength\cftparskip{-5pt}
\setlength\cftbeforesecskip{-1pt}
\setlength\cftaftertoctitleskip{-10pt}
\tableofcontents
\cleardoublepage
\phantomsection
\addcontentsline{toc}{section}{Acronyms} 

\markboth{ACRONYMS}{Acronyms}
\section*{Acronyms}
\begin{table}[H]
    \begin{tabular}{p{4cm} p{10cm}}
     AI & Artificial Intelligence \\
     IBM &  International Business Machines\\
     CNOT & Controlled-NOT \\
     PCA & Principal component analysis\\
     QPE & Quantum phase estimation\\
     HHL & Harrow, Hassidim, Lloyd. Quantum algorithm for solving linear systems of equations\\
     QSVM & Quantum support vector machine\\
     VQE & Vatiational-quantum-eigensolver \\
     QAOA & Quantum approximate optimization algorithm\\
    \end{tabular}%
\end{table}%

\cleardoublepage
\phantomsection
\addcontentsline{toc}{section}{List of Figures} 
\listoffigures
\cleardoublepage
\phantomsection
\addcontentsline{toc}{section}{List of Tables} 
\listoftables
\cleardoublepage

\interfootnotelinepenalty=10000 

\markboth{OVERVIEW}{Review of the Problem}
\section{Review of the Problem}
\label{sec:Review of the Problem}
The first proposition for a quantum computer was made by Richard Feynman back in 1981. He pointed out that simulating quantum-mechanical systems on classical computers would be inefficient, so new type of computers must be build - "built of quantum mechanical elements which obey quantum mechanical laws" \citep{Feynman1982}.
\par The quantum computing is one of the fastest growing research areas in recent years. The quantum computers are based on quantum mechanical physics laws and they use unitary operations for handling quantum information usually described as quantum logic gates. The quantum logic gates can be single-qubit gates and multiple-qubit gates. The single-qubit gates execute controlled quantum transitions from one quantum superposition to another. The multiple-qubit gates create so-called entangled states, where they create dependencies between more than one quantum systems (qubits) states, so that the state of one system can not be described without knowing the state of the others. Nowadays well-known fact is that classical computers can not fully simulate the dynamics of a quantum system effectively. Every classical simulation (a simulation executed on a classical computer) is exponentially slower than the actual process's dynamics.
\par 
The idea of the constantly growing scale of the problems which solution requires and is only possible using quantum computers is very scary. The available classical computer resources (hardware mostly) are not enough for solving some of the most interesting problems in science today. Here the quantum computers give hope - some of these problems would be solved efficiently. However, this brings new problem - the algorithms. Quantum algorithms as essential part of quantum computing, still not very advanced scientific field, could be classified as algorithms based on Quantum Fourier Transformation (based on the Shor's factorization algorithm) and based on Grover's search algorithm. The Quantum Fourier Transform based algorithms are exponentially faster than their classical equivalents, where the quantum amplification based algorithms are "only" quadratically more efficient.
\par 
The first applied quantum algorithm has been created by Peter Shor in 1994 \citep{Shor1997}. With his algorithm though quantum interference he succeeded to factorize prime number requiring only polynomial number of steps, while the classical equivalent would require exponential number of steps. 
\subsection{Solved and Unsolved Problems} 
\label{subsec:Solved and Unsolved Problems}

Classical computers work with bits, where each bit has a deterministic state - 0 or 1 at any moment of time. The fundamental quantum information-carrying elements are qubits (physically it can be single atom, electron, photo on superconducting circuit). \footnote{Classification and analysis of the technological clusters in quantum information theory has been made in author's work (\hyperref[publications:fdiba2019]{Author's publication iv}).}
\par The huge complexity (compared with the classical systems) of quantum-mechanical systems comes when there needs to be given a full description of highly entangled quantum states. One of the hardest challenges is to distinguish which problems are quantum hard and which classically hard \citep{quantumzoo,Montanaro2016}.
\par Why would quantum computers outperform the classical ones for high complexity problems? \citep{Preskill2018}
\begin{itemize}
\item Quantum algorithms for classical problems - these are problems known to be hard to solve using classical computers, but quantum algorithms could perform significantly more efficient. The best known example is factorization problem for large integers \citep{Shor1997, Chuang1995}.
\item Complexity theory arguments -these arguments are based on the fact that quantum states have super-classical properties. If a quantum register is measured, this is sampling from a correlated probability distribution, that can't be sampled efficiently by classical methods \citep{Lund2017, Harrow2017}
\item Classical computers can not simulate quantum computers efficiently \citep{Feynman1982}.
\end{itemize}
Richard Feynman also said: "if you want to make a simulation of Nature you better make it quantum mechanical, and by golly it's a wonderful problem because it doesn't look so easy." \citep{Feynman1982}. Here we could start to think why really quantum computing is hard, because simulating the Nature isn't really something possible nowadays on any classical computer. Building quantum computers \citep{Chuang1995simple} brings the contradiction problem for the qubits - on one side we want them to be fully isolated from the environment so there is no decoherence problem \citep{Zeh1970, bacon2003decoherence, Lidar2003, Zurek2007, Zwirn2016, Zurek1862}, but on the other side we want to fully control from the outside these qubits in the quantum systems and also to build highly entangled states.
\par The principle of quantum error correction could help scaling up the quantum computers \citep{Steane1996,Gottesman2010}. The main idea of this principle is that in order to protect a quantum system it should be encoded in a very high entangled state \citep{Preskill2018}.

\subsubsection{Factorization}
\label{subsubsec:shor}
The problem of prime factorization is a problem in number theory. It is the decomposition of large integer number to a product of small prime numbers. Peter Shor introduces his algorithm for quantum factorization in 1994 \citep{Shor1994}. Using quantum computers can speed-up dramatically the the task for prime factorization. The fastest classical algorithm which solves this problem is nearly exponential, while the quantum one is giving a solution in polynomial time \citep{Chuang1995}. 
\par This problem is a key point in modern cryptography because of its asymmetry for the difficulty of factorization and how easy it is to verify the result. Such an application widely used these days is the RSA (Rivest-Shamir-Adleman) algorithm. It depends on the fact that prime factorization of large numbers requires long time. The logic behind this algorithm is that there is so-called public and private keys. The public key is a product of two large numbers and the private key consists of these two large numbers. The public key is used to encrypt the information, while the decryption is made by using the private key. At the time when the algorithm was first introduced, the practicality of the quantum computers was questionable because of the quantum decoherence \citep{Chuang1995, Zurek1991}.
\par The quantum algorithm for prime factorization uses one of the fundamental properties of quantum systems - the coherence. The coherence describes the correlation between several wave packets. 
\par What is factoring?
\par The main part of the factoring algorithm is the period finding. If there is a periodic function \textit{f}, where \textit{f} maps some numbers $\{0,1,...,M-1\}$ to some set S, such that $\forall x$ , $f(x)= f(x+r)$. The task is to find the period \textit{r}. The number of the repetitions of the period is $M/r$. If this function is considered on a single period, then \textit{f} is 1-1: values are never repeated. This is the first condition to the period finding: that \textit{f} is 1-1 for each period. The second condition is that r divides M.  In order to solve the factoring problem, the ${M \over r} >> r$ must be true ($M \gg r^2$). 
\par How to find the periodic function and how hard is to solve the periodic finding?
\par Let's assume $M \approx 10^{1000}$ (1000 digits number) and $r \approx \sqrt{M}$ (500 digits number). The classical solution here is to pick random inputs until the pattern is found. Here another question arises - how the pattern looks like? There must be different numbers, for example x and y, where $f(x)=f(y)$. But still this is not enough information, since these numbers doesn't contain any other information other that the function has the same output for different input. At first look the count of the random inputs for trial as input for the \textit{f} function must be at least \textit{r}. Here the birthday paradox \citep{Naccache2011} says that actually the $\sqrt{r}$ are enough inputs until there is a collision. Still $\sqrt{r} \approx 250$ digits number, which makes it impossible to find the collision classically. 
\par What does the quantum algorithm do?
\par There is a quantum circuit which takes $\ket{x}$ and $\ket{0}$ as input and outputs $\ket{x}$ and $\ket{f(x)}$. It does this in superposition - an uniform superposition is set on the $\ket{x}s$ and the output of the circuit is ${1 \over \sqrt{M}} \sum_{x=0}^{M-1}{\ket{x}\ket{f(x)}}$. The next step to do is to measure the second register of the output: $\ket{f(x)}$, so the result will be some random values which form some period (for example: arithmetic progression). This superposition is the input for the next step, which is Fourier sampling, and the multiples of the period can be found as non-zero amplitudes in the superposition.
\par The result of Fourier sampling is: $\sqrt{r \over M} \sum_{j=0}^{M-1}{\ket{jr}}$.
\par Figuring the $M \over r$ from the result is the greatest common divisor of all random outputs. This task could be solved even using the Eucledian algorithm \citep{stark1970introduction}. And when \textit{M} and $M \over r$ are known it is easy to find the \textit{r}.
\par The algorithm could be divided in two parts \citep{Beckman1996} 
\begin{enumerate}
\item The first part is the order finding problem, which could be implemented classically.
\item The second part is a quantum algorithm which solves the order finding problem:
\renewcommand{\labelenumii}{\roman{enumii}}
\begin{enumerate}
\item Initialize a quantum register of length \textit{k} into $1\ket{0^k}$ state. Then create an equal superposition of all \textit{k} qubits using the  Walsh-Hadamard transformation. After this transformation there will be equal probability for each one of the $2^k$ possible states for the register.
\item Construct the $f(x)$ function as quantum function and apply it to the register. The result will be a superposition of \textit{k+n} qubits.
\par ${1 \over \sqrt{2^k}}\sum_{x=0}^{2^k-1}{\ket{x,f(x)}}$
\item Apply the inverse Quantum Fourier transform to the input register $\ket{x}$:
\par $U_{QFT}(\ket{x}) = {1 \over \sqrt{2^k0}}\sum_{y=0}^{2^k-1}{\omega^{xy}\ket{y}}$
\par this sum must be ordered according to the $f(x)=z$, $x \in \{0,1,...,2^k-1\}$
\begin{itemize}
\item $\omega = e^{2\pi i \over 2^k}$
\item \textit{r} is the period of \textit{f}
\item $x_0$ is the smallest
\end{itemize}
\item Perform a quantum measurement both on the input register and the output register. \textit{y} is the outcome from the input register and \textit{z} is the outcome from the output register. The probability of measuring the $\ket{y,z}$ state will be:
\par $\mathbb{P}(\ket{y,z}) = {1 \over {2^{2k}}}\left({\sin({{\pi m r y} \over {2^k}}) \over {\sin({{\pi r y} \over {2^k}}})}\right)$
\item Perform continued fraction expansion to find the appropriate period \textit{r}.
\end{enumerate}
\item Check if the found solution for the period is a prime factor. If so, here is the end.
\item Otherwise, obtain more candidates for \textit{r}.
\item If no candidates satisfy the conditions to be the period of the function, then return to step \textit{i}.
\end{enumerate}
Up to today, there are many available experimental realizations of this algorithm \citep{Lanyon2007}, including the use of nuclear magnetic resonance for the N = 15 with prime factors 3 and 5 \citep{Vandersypen2001}, N = 15 with Josephson phase qubit quantum processor \citep{Lucero2012}, N = 21 using recycling qubit instead of n-qubit control register which recycles n times \citep{MartnLpez2012}, N = 15 using photonic qubits \citep{Lu2007, Politi2009}, using qutrits for factoring \citep{Bocharov2017}, N=15,21,35 using the IBM Q \citep{Amico2019}, N = 15 at room temperature \citep{johansson2017realization}, scalable version of the algorithm \citep{Monz2016}.
\par This is one of the largest research areas in the quantum information theory and quantum computing with applied research in cryptography and number theory, including accuracy, implementation research and bottlenecks investigation \citep{Markov2013,Most2010, Shimoni2005, Gerjuoy2005, Azuma2017, Lawson2015, Most2010, GarcaMata2008, GarcaMata2007, Ukena2004, Wei2005,eker2017quantum}.
\subsubsection{Quantum search}
\label{subsubsec:grover}
Grover algorithm is a search algorithm over an unordered set of $N=2^n$ items to find a unique element which satisfies some predefined conditions \citep{10.1145/237814.237866, Grover1997QUANTUMCC, Grover1997, Grover1998, grover1998fast, grover2000searching, Grover2001, Grover2002, grover2004partial}. This algorithm requires only O($\sqrt{N}$) operations to perform the search, which is quadratic speed-up over its classical opponent. It is one of the most famous quantum algorithms not only because of its speed-up, but also because it is a good introduction to quantum algorithms. It demonstrates how the properties of quantum systems and their fundamental differences with the classical computing systems can be used to lower the runtime. Grover's algorithm is based on the quantum superposition and like many other algorithms, the first step of it is to initialize the system in equal superposition. This way an equal amplitude ($1/\sqrt{2^n}$) is associated with every possible state of the system, so there is equal probability of $1/\sqrt{2^n}$ that the system will be in any of the $2^n$ possible states. The next step in Grover's algorithm is to use the \textit{amplitude amplification} to selectively shift the phase of one state of the quantum system - the one that satisfies the search conditions, at each iteration. The phase shift of $\pi$ is equivalent to multiplication of the amplitude for this state by -1. Changing the amplitude doesn't change the probability for the state (the probability disregards the sign of the amplitude) \citep{Strubell2011}.
\par How this algorithm works?
\par The search problem needs to be translated into quantum-mechanical problem. So, let the system have $N=2^n$ states \{$S_1,S_2,...,S_n$\}. The representation of these states are bit strings - $n$ bit strings. A unique state $S_v$ must satisfy the condition $C(S_v)=1$, and all other states S, $C(S)=0$. The problem is to identify this $S_v$ state \citep{10.1145/237814.237866}.
The algorithm itself consists of three steps:
\renewcommand{\labelenumi}{\roman{enumi}}
\begin{enumerate}
\item System initialization, so that there is the same amplitude for each possible state: $\left( {1 \over \sqrt{N}},{1 \over \sqrt{N}},...,{1 \over \sqrt{N}} \right)$. This distribution can be obtained in $O[logN]$ steps and can be achieved by applying a "fair coin flip" operation on all the qubits in the quantum system \citep{Simon1997}. This operation is represented by the following matrix:
\par $M = {1 \over \sqrt{2}} \begin{bmatrix}
1 & 1 \\
1 & -1
\end{bmatrix}$
\par Applied on a signel qubit, the state 0 is transformed into two states: $\left({1 \over \sqrt{2}},{1 \over \sqrt{2}}  \right)$. And when applied to every qubit in a quantum system, the transition matrix representation will be of size $2^N X 2^N$ and an identical amplitude will be induced for every one of the $2^N$ possible states. This is known as Walsh-Hadamard transformation \citep{walshhadamard1992}.
\item Repetition of the following unitary operations O($\sqrt{N}$) times:
\begin{enumerate}
\item If $C(S)=1$: rotate phase by $\pi$ radians. Else if $C(S)=0$ leave the system unaltered.
\item Apply diffusion transform D. D=WRW, where W is the Walsh-Hadamard transformation and R is the rotation matrix. 
\par $R_{ij} = 0$ if $i \neq j$
\par $R_{ii} = 1$ if $i =0$; $R_{ii} = -1$ if $i \neq 0$
\par $W_{ij} = 2^{-n/2}(-1)^{\overline{i}.\overline{j}}$, where ${\overline{i}.\overline{j}}$ denotes the dot product. 
\end{enumerate}
\item  Sample the resulting states. The final state is $S_v$ with probability $\geq 1/2$ if $C(S_v)=1$
\end{enumerate}
The main part of this algorithm is the step $ii$ where the desired state amplitude is increased.
\par \textbf{Working example:}
\par The quantum system is formed of 8 states: $2^3$. This means that 3 qubits are required to represent these 8 states. And the queried state is $\ket{010}$.
\par $\ket{\Psi} = \alpha_0\ket{000} + \alpha_1\ket{001} + \alpha_2\ket{010} + \alpha_3\ket{011} + \alpha_4\ket{100} + \alpha_5\ket{101} + \alpha_6\ket{110} + \alpha_7\ket{111}$
\renewcommand{\labelenumi}{\roman{enumi}}
\begin{enumerate}
\item Initialize the system in 0:
\par Apply the identity gate to every qubit in the system so the result is:  $1\ket{000}$.
\par Then apply Hadamard transform to every one of the three qubits:
\par $H^3\ket{000} = {1 \over 2\sqrt{2}}\ket{000} + {1 \over 2\sqrt{2}}\ket{001} + {1 \over 2\sqrt{2}}\ket{010} + {1 \over 2\sqrt{2}}\ket{011} + {1 \over 2\sqrt{2}}\ket{100} + {1 \over 2\sqrt{2}}\ket{101} + {1 \over 2\sqrt{2}}\ket{110} + {1 \over 2\sqrt{2}}\ket{111} = \ket{\psi}$
\item The optimal number of Grover operations is ${\pi \over 4}\sqrt{2^3} = {\pi \over 2}\sqrt{2} \approx 2.22$.
\par Each operation consists of:
\begin{enumerate}
\item Call the quantum oracle $O$.
\item Inversion about the average (diffusion transform) - $2\ket{\psi}\bra{\psi} -I$.
\end{enumerate}
\par $\braket{\psi|\psi}=1$.
\par $\braket{\psi|010} = {1 \over 2\sqrt{2}}$ ($\ket{010}$ is one of the basis vectors).
\par $\left[2\ket{\psi}\bra{\psi} -I \right]\ket{x} = 2\ket{\psi}\braket{\psi|\psi} - \ket{\psi} - {2 \over \sqrt{2}}\ket{\psi}\braket{\psi|010} + {1 \over \sqrt{2}}\ket{010} = {1 \over 2}\ket{\psi} + {1 \over \sqrt{2}}\ket{010}$.
\par Substitution of $\ket{psi}$:
\par ${1 \over 2}{1 \over 2\sqrt{2}}\ket{000} + {1 \over 2\sqrt{2}}\ket{001} + {1 \over 2\sqrt{2}}\ket{010} + {1 \over 2\sqrt{2}}\ket{011} + {1 \over 2\sqrt{2}}\ket{100} + {1 \over 2\sqrt{2}}\ket{101} + {1 \over 2\sqrt{2}}\ket{110} + {1 \over 2\sqrt{2}}\ket{111} + {1 \over \sqrt{2}}\ket{010} = \\ = {1 \over 4\sqrt{2}}\ket{000} + {1 \over 4\sqrt{2}}\ket{001}  + {1 \over 4\sqrt{2}}\ket{011} + {1 \over 4\sqrt{2}}\ket{100} + {1 \over 4\sqrt{2}}\ket{101} + {1 \over 4\sqrt{2}}\ket{110} + {1 \over 4\sqrt{2}}\ket{111} + {5 \over 4\sqrt{2}}\ket{010}$
\par This completes the first iteration. Then the result of the second iteration is:
\par $-{1 \over 8\sqrt{2}}\ket{000} - {1 \over 8\sqrt{2}}\ket{001} - {1 \over 8\sqrt{2}}\ket{011} - {1 \over 8\sqrt{2}}\ket{100} - {1 \over 8\sqrt{2}}\ket{101} - {1 \over 8\sqrt{2}}\ket{110} - {1 \over 8\sqrt{2}}\ket{111} + {11 \over 8\sqrt{2}}\ket{010}$.
\item Now when this quantum-mechanical system is observed the probability for the state $\ket{010}$ will be:
\par $\left|{11 \over 8\sqrt{2}}\right|^2 \approx 95\%$.
\end{enumerate} 
\subsubsection{HHL}
\label{subsubsec:lin_sys_eq}
Linear algebra is a branch of mathematics which concerns operations with linear equations, linear functions, matrices. It is an essentially big part of many machine learning algorithms. Solving systems of linear equations is one of the most common problems not only in the linear algebra, but in all fields of science and engineering.
\par Approximating the solution for N linear equations in N unknown parameters takes time of order N with classical methods. A quantum algorithm \citep{Harrow_2009} in some cases can approximate the value of a function of the full solution to these N equations scaling logarithmically in N, and with some additional conditions and precision it could take polynomial time. This algorithm can achieve exponential speed-up in some special cases with additional restrictions applied.
\par How this algorithm works?
\par The problem: a Hermitian matrix $A=N \times N$ and a unit vector $\overrightarrow{b}$ are given.
\par Find $\overrightarrow{x}$, such that $A\overrightarrow{x}=\overrightarrow{b}$.
\par The algorithm itself consists of the following steps:
\renewcommand{\labelenumi}{\roman{enumi}}
\begin{enumerate}
\item Represent $\overrightarrow{b}$ as a quantum state $\ket{b} = \sum_{i=1}^N{b_i\ket{i}}$.
\item Use Hamiltonian simulation techniques to apply $e^{iAt}$ to $\ket{b}$ for a superposition of different times \citep{Berry_2006, Childs_2009}.
\item Use phase estimation to decompose $\ket{b}$ \citep{PhysRevA.54.4564, Cleve_1998, Bu_ek_1999}. The result is:
\par $\sum_{j=1}^N{\beta_j \ket{u_j} \ket{\lambda_j}}$ 
\item Use a non-unitary operation to do a linear mapping  $\ket{\lambda_j} \rightarrow C\lambda_j^{-1}\ket{\lambda_j}$. The result is:
\par $\sum_{j=1}^N{\beta_j \lambda_j^{-1} \ket{u_j}} = A^{-1} \ket{b} = \ket{x}$
\end{enumerate}
The real advantage in this algorithm is that it doesn't need to write all $A, \overrightarrow{b}, \overrightarrow{x}$ in the quantum registers, and requires only a register with length of $O(logN)$.
\par There are many potential applications of this algorithm since the matrix inversion is a widely spread problem across various science fields. Detailed explanation of the algorithm is given in part II of \citep{Harrow_2009}. This algorithm also had become an important subroutine for many quantum machine learning algorithms \citep{Buhrman2001, val, val-original,Klappenecker_2003,leyton2008quantum, grover2002creating}, but it has some limitations \citep{shao2018reconsider}, which must be taken into account when applying in practice.
\par Some of the problems that arise when applying this algorithm are:
\begin{itemize}
\item The preparation of the input state $\ket{b}$ - Lov Grover and Terry Rudolph have given an efficient process for generating certain probability distributions using quantum computers \citep{grover2002creating} ( forming a discrete approximation of any efficiently integrable probability density function). The quantum representation of a probability distribution is a superposition of the form:
\par $ \ket{\psi(\{p_i\})} = \sum_i{\sqrt{p_i}\ket{i}}$
\par where \textit{i} are orthonormal.
\par The quantum sates in this form can also be useful for Quantum Fourier Transform, since these can be efficiently fourier transformed \citep{Shor1997}.
\item The choice of \textit{t} parameter in the Hamiltonian simulation of $e^{iAt}$. Here the \textit{t} must satisfy:
\par $\left|\lambda t \right| < \pi$, $e^{2\pi i} = 1$. And theoretically $t=\pi/|\lambda_{max}|$, $|\lambda_{max}|$ is the maximal singular value of A.
\par For $|\lambda_{max}|$ exist several upper bounds and the choice of an upper bound will affect the complexity of the algorithm:
\begin{itemize}
\item $\sqrt{Tr(AA^{\dagger}}$
\item $||A||_1 = max_j \sum_i{|a_{ij}|}$
\item $||A||_2 = \sqrt{\sum_{i,j}{|a_{ij}|^2}}$
\item $M||A||_{max} = M_{max_{i,j}}|A_{ij}|$
\end{itemize}
\end{itemize}
\par Some of the potential application of the algorithm are:
\begin{itemize}
\item Linear regression \citep{PhysRevA.94.022342, PhysRevLett.109.050505, PhysRevA.96.012315}
\par $F^\dagger F x = F^\dagger b$
\par F is a data matrix and b is a given vector.
\item Supervised classification \citep{lloyd2013quantum} - this application is based on the distance comparison of a vector to the means of two clusters. The authors show a novel technique for a preparation of the desired quantum state.
\item Support vector machine - the main technique for the Hamiltonian simulation of a matrix is well described in \citep{Lloyd2014}. 
\item Hamiltonian simulations - a new method for a low rank non-sparse Hermitian matrix is proposed in \citep{PhysRevA.97.012327}.
\end{itemize}
\subsubsection{Quantum PCA}
\label{subsubsec:quantum_pca}
The principal component analysis (PCA) is a bedrock to dimensionality reduction technique for probability and statistics, commonly used in data science and machine learning applications, where there is big dataset with statistical distribution and the low-dimensional patterns must be uncovered.
\par Let's have data in form of vectors $\overrightarrow{v_j}$ in d-dimensional vector space, where covariance matrix of this data is:
\par $C=\sum_j{\overrightarrow{v_j}\overrightarrow{v_j^T}}$
\par where $\overrightarrow{v_j^T}$ is transposed vector.
\par This covariance matrix summarizes the correlations between the different components of the data.
\par The simplest form of PCA is the diagonalization of the covariance matrix:
\par $C=\sum_k{e_k\overrightarrow{c_k}\overrightarrow{c_k}^{\dagger}}$
\par $\overrightarrow{c_k}$ are eigenvectors of C.
\par $e_k$ are eigenvalues of C.
\par The eigenvectors $\overrightarrow{c_k}$ form an orthonormal set.
\par If the reminder are small or zero (few large values for $e_k$), the corresponding eigenvectors are called the principal components of C. Where each principal  component represents a correlation in the data. The classical algorithms to perform PCA have computational complexity of order $O(d^2)$.
\par 
\par In the quantum way this problem is translated to revealing properties of an unknown quantum state \citep{Lloyd2014}. A quantum coherence can be created among multiple copies of a randomly generated quantum state to perform a quantum principal component analysis, which reveals the eigenvectors corresponding to the large eigenvalues of this quantum state. This requires $O(d)$ operations in qRAM divided over $O(logd)$ steps, which could be executed in parallel.
\par The quantum tomography \citep{Nielsen2001QuantumIA, PhysRevA.77.032322} is a widely used tool where a given multiple copies of an unknown quantum state in d-dimensional Hilbert space are being measured with various techniques in order to extract useful information showing some features of the state \citep{Gross_2010, PhysRevLett.106.100401, Shabani_2011}. Multiple copies of the state can play active role in its own measurement and implement the unitary operator $e^{-i\rho t}$: energy operator or Hamiltonian, which generates transformations on other states.
\renewcommand{\labelenumi}{\roman{enumi}}
\begin{enumerate}
\item Exponentiate density matrix  - this exponentiates non-sparse matrices in $O(logn)$, which is exponential speed-up over its classical opponents \citep{Lloyd2014}.
\par Using Suzuki-Trotter expansion \citep{10.2307/2899535, 10.1145/780542.780546, Berry_2006, Wiebe_2010} the $e^{-iXt}$ is constructed for non-sparse positive matrix X.
\par $\sum_k{e_k}=1$ for X
\par $X=A^\dagger A$
\par $A = \sum_i{|\overrightarrow{a_i}|\ket{a_i}\bra{e_i}}$
\par Density matrix plays role in exploring its own features by decomposition of semi-definite Hermitian matrix to the eigenvectors of the largest eigenvalues \citep{10.5555/1162264, Murphy2012}.
\item Application to quantum phase estimation algorithm to find the eigenvalues and eigenvectors of the unknown density matrix. The quantum phase estimation algorithm takes any initial state 
$\ket{psi}\ket{0}$ to 
$\sum_i{\psi_i\ket{x_i}\ket{\tilde{r}_i}}$, where the eigenvectors are 
$\ket{psi_i}$ and the approximated (estimated) eigenvalues are 
$\tilde{r}_i$. There exists an improved method for phase estimation \citep{Harrow_2009} which requires 
$O(1/\epsilon^3)$ copies of the random quantum state 
$\rho$.
\end{enumerate}
\textbf{Advantages and future applications of the quantum self-tomography}
\begin{itemize}
\item Reveals eigenvectors and eigenvalues in time $O(Rlogd)$ compared to the compressive tomography ($O(Rdlogd)$) \citep{Gross_2010}.
\item The density matrix exponentiation is time-optimal \citep{Lloyd2014}.
\item Quantum self-tomography is comparable to group representation methods \citep{PhysRevA.64.052311}, but not only the spectrum is approximated - also as a result, the eigenvectors are found.
\item Quantum PCA applications in state discrimination and assignment, where the task is to assign a new set of states to already known other sets. The decomposition to eigenvectors and eigenvalues gives the possibility for assignment and also the magnitude of the measured eigenvalue is the confidence of the set assignment measurement: larger the magnitude is - the higher the confidence is \citep{Lloyd2014}.
\item Speed-up of some machine learning problems in clustering and pattern recognition \citep{lloyd2013quantum, Rebentrost2014}.
\end{itemize}
\subsubsection{Quantum Boltzmann Machines}
\label{subsubsec:quantum_bm}
Boltzmann machines (BMs) are recurrent neural networks which can also be represented as bidirectionally connected networks of stochastic processing units ( Markov Random Field - set of random variables, each having the Markov property) \citep{ACKLEY1985147, Brmaud1999, daphnekoller2009}. The usage of Boltzmann machines in practice is usually simplified by imposing restrictions on the topology of the network \citep{Fischer2012,6302931}. The learning process in Restricted Boltzmann machines could be summarized as adjusting the parameters of the Boltzmann machine so that the probability distribution of the network fits the input data. The constructive parts of the BMs are two layers - visible and hidden. The neurons in the visible layer correspond to the observed object ( for example: if the object is an image - then each neuron represents a pixel on this image). The hidden neurons represent the model of the object - dependencies and patterns between the components of the object (the neurons in the visible layers) \citep{Fischer2012, Hinton:2007}.
\par Restricted Boltzmann machines are reviewed well in \citep{10.1561/2200000006} and they have received attention in the scientific area after being proposed as building part of the deep belief networks (DBNs) \citep{Hinton2007, Hinton2006}.
\par The probability of a given visible and hidden layers configuration in Boltzmann machines is given by the Gibbs distribution \citep{wiebe2014quantum}:
\par $\mathbb{P}(v,h) = e^{-E(v,h)}/Z$,
\par where $v$ and $h$ are visible and hidden layers respectively and Z is the normalizing factor (partition function). The energy of a given configuration of $v$ and $h$, $E(v,h)$, is:
\par $E(v,h) = - \sum_i{v_i b_i} - \sum_j{h_j d_j} - \sum_{i,j}{\omega_{i,j}^{vh} v_i h_j} - \sum_{i,j}{\omega_{i,j}^v v_i v_j} - \sum_{i,j}{\omega_{i,j}^h h_i h_j}$
\par The $b$ and $d$ vectors are biases (energy penalty for unit value 1) and $\omega_{ij}^{vh}$, $\omega_{i,j}^v$, $\omega_{i,j}^h $ are weights. $n_v$ and $n_h$ are the numbers of visible and hidden units.
\par The learning for the Boltzmann machine is adjusting the strengths of the interactions within the graph to maximize the likelihood of the given observations to be produced by the method. The training process uses gradient decent to optimize the maximum-likelihood objective:
\par $O_{ML} := {1 \over N_{train}} \sum_{v \in X_{train}}{log(\sum_{h=1}^{n_h}{P(v,h)})} - {\lambda \over 2}\omega^T \omega$
\par $\lambda$ is the L2-norm. The derivative has the following look:
\par ${\partial O_{ML} \over \partial \omega_{i,j}} = \left\langle v_i h_j \right\rangle_{data} - \left\langle v_i h_j \right\rangle_{model} - \lambda \omega_{i,j}$
\par The gradient decent computation is exponentially hard (in $n_v$ and $n_h$) problem and the best classical approach is approximation through contrastive divergence \citep{MAL-006, 6789337, 10.1145/1273496.1273596, Tieleman08trainingrestricted, pmlr-v5-salakhutdinov09a}, which unfortunately does not provide gradient to any true objective function \citep{pmlr-v9-sutskever10a, Tieleman2009UsingFW, 6795523, Fischer2011}
\par Efficient alternatives to this classical method are provided in \citep{wiebe2014quantum} where two new quantum algorithms are proposed: Gradient Estimation via Quantum Sampling (GEQS) and Gradient Estimation via Quantum Aplitude Estimation (GEQAE).
\par \textbf{The quantum problem}
\par In Quantum Boltzmann machines the problem could be summarized to learning a set of Hamiltonian parameters $\omega_i$ for a fixed set of $H_i$, the input state $\rho$ is well approximated by $\sigma=e^{-\sum{\omega_i H_i}}/Tr(e^{-\sum{\omega_i H_i}})$ \citep{Biamonte_2017, Kieferov__2017, Amin_2018}. The quality of the approximation is measured by the quantum relative entropy:
\par $S(\rho || \sigma) = Tr(\rho log \rho - \rho log \sigma)$
\par where the upper bound is the distance between the two states. Minimizing the distance, minimizes the error in the approximated state $\sigma$ \citep{Biamonte_2017}.
\par Since it is hard to learn experimentally (calculating the quantum relative entropy), it is more practical (and easier) to estimate the gradient of the relative entropy:
\par $\sigma_{\omega_i} S(\rho || \sigma) = Tr(\sigma H_i)Tr(\rho H_i)$
\par The Stochastic Hamiltonians have all off-diagonal matrix elements to be real and non-negative (non-positive) to which no classical analogue is known \citep{Biamonte2008}.
\par The algorithms for quantum state generation, gradient calculation, gradient estimation and training via quantum amplitude estimation are well described in the Appendix in \citep{wiebe2014quantum}.
\vspace{3cm}

\subsubsection{Input and output problem}
\label{subsubsec:unsolved_input_output}
Loading classical data into a quantum computer is a bottleneck for some algorithms. Most quantum machine learning algorithms require exponential time procedures to load data into quantum states \citep{Aaronson2015}. One solution to this problem is using quantum Random Access Memory (qRAM), but it is a costly solution for big datasets \citep{Arunachalam2015}.
\par Similar problem is noticeable when a readout for a quantum system is required. Also known as the 'output problem'. It is a common problem for all linear algebra-based quantum machine learning algorithms, since it is exponentially hard to estimate the classical quantities for the solution vector of the qPCA algorithm.
\par The quantum information is very different from its classical counterpart, because it exists in a superposition and it's hard to measure it - every observation made on a quantum register leads to a collapse of this superposition.
\par In the Machine learning field the computer algorithms are being developed in such a way, that they learn from the history of observations. In general Machine learning algorithms can be classified as supervised and unsupervised learning algorithms depending on the input dataset. When the input dataset is labeled - it is supervised learning.
\par One of the main fundamental differences between classical and quantum computing is the representation of the information - in classical computing the smallest unit of information is bit, and it can be either in state 0 or in state 1. The quantum equivalent of the bit is the qubit (quantum bit) - it can be simultaneously in two orthonormal states, which means the qubit is in superposition, and using Dirac notation, the qubit representation is given as:
\par $ \ket{psi} = \alpha \ket{0} + \beta \ket{1}$
\par The collapse of the superposition means that, when measured the qubit is in either $\ket{0}$ or $\ket{1}$ state with probability $\left| \alpha \right|^2$ or $\left| \beta \right|^2$.
\par The knowledge about a particular quantum system is represented by the density matrix, which gives a complete description of what can be observed about the quantum system.
\par $\rho = \ket{\psi} \bra{\psi}$
\par $\bra{\psi}$ is the conjugate transpose of $\ket{\psi}$.
\par $\rho$ is a pure state when the latter being considered as a column vector in the Hilbert space.
\par A mixed state is an ensemble of pure states:
\par $\{(\alpha_1, \ket{\psi_1}),...,(\alpha_n,\ket{\psi_n})\}$
\par $\sum_{i=1}^n{\alpha_i} = 1$
\par $\alpha_i$ is the probability associated with the pure state $\ket{\psi_i}$.
\par $\rho = \sum_{i=1}^n{\alpha_i \ket{\psi_i} \bra{\psi_i}}$
\par The fundamental limits on operations with a quantum state are:
\begin{itemize}
\item No-cloning theorem \citep{Wootters1982} - no unknown quantum state can be cloned perfectly, unless it is known to belong to a set of pairwise orthogonal states \citep{Ameur2006}.
\item It is not possible to extract more than n bits of classical information from n qubits (Holevo's theorem \citep{holevo1973bounds}). For n qubits, all possible amplitudes are $2^n$, so only a small amount of the quantum information can be extracted and classically represented.
\end{itemize}
\par As mentioned above, the machine learning algorithms learn from history of observations (a training dataset). In classical dataset the observations are implicitly considered to be classical. In quantum machine learning a training dataset is also required, but this dataset obeys on the laws of quantum mechanics, and the entire learning process needs to be redesigned.
\par At the time of writing this dissertation, there exists several types of learning strategies for quantum datasets:
\begin{itemize}
\item Quantum tomography - quantum estimation technique through measurements on some copies of the quantum states.
\item One-time classifier - the copies of the quantum states are used only at the time of demand.
\item Hybrid - a combination between the other two.
\end{itemize}
\subsubsection{Benchmark problem}
\label{subsubsec:unsolved_benchmark}
The benchmark problem is a general problem not only for the quantum algorithms, but also a huge research area in classical computer science. In the quantum world the benchmark problem is connected not only with need for probing performance of quantum computers against their classical counterparts for identical (similar) problems, but also for a comparison between various quantum hardware backends. In \citep{blumekohout2019volumetric} the authors propose a large number of benchmarks which define a family of rectangular quantum circuits, allowing the study of the time/space performance trade-offs. 
\par The quantum computing is one of the fastest growing hardware areas and only for a few years the quantum processors scaled from a two coupled qubits to 49 (Intel), 50 (IBM) and 72 (Google) at 2018. In order to use efficiently quantum processors there need to be a tool for measuring the performance - a benchmark.
\par What is a quantum benchmark?
\par The quantum benchmark is a set of quantum circuits and instructions, analysis procedure and interpretation rules \citep{blumekohout2019volumetric}. There exists few families of quantum benchmarks, each of them measuring different metrics:
\begin{itemize}
\item Quantum volume - this is a benchmark proposed by IBM \citep{Cross_2019}, which measures the size of the accessible state space for a quantum processor. In the ideal world a quantum processor with \textit{n} qubits would have $2^n$ computational states, but in the practice parts of the state space is not accessible. This problem could be caused by various reasons: poor control for the qubits, noise in the quantum system. This benchmark gives the answer to the question "What's the largest number of qubits on which the processor can reliably produce a random state?" and it requires a "square" quantum circuit to be run on the hardware ( number of qubits = number of time steps), which is a considerable caveat, since not many existing quantum algorithms use "square" circuits.
\item Randomized benchmark - this is a method for measurement of the accuracy of the implementation of a coherent quantum transformation \citep{Emerson_2005, Emerson2007, Knill2008, Magesan2011, Magesan2012}. It estimates the gate fidelity and it is quite relevant for large-scale Hilbert spaces. It determines the noise in the quantum system by variation over different experimental arrangements and error-correction strategies.
\item Long-sequence gate set tomography - this technique provides an accurate complete tomographic description for every gate and is not only applicable on single qubit gates, but also on high-fidelity two-qubit gates \citep{BlumeKohout2017, blumekohout2013robust, greenbaum2015introduction, kim2015microwave,Dehollain2016}.
\item Volumetric benchmarks - this is a framework for large family of benchmarks, that had been inspired by the IBM's quantum volume benchmark \citep{blumekohout2019volumetric}.  It measures the quantum processor's ability to run an ensemble of rectangular quantum circuits.  
\end{itemize}

\subsection{Goals} 
\label{subsec:Goals}

The fundamental properties of the quantum computers give the quantum information theory a great advantage handling hard problems. The main goal of this dissertation is to create new quantum algorithm with as easy as possible implementation on real hardware platforms, which could be used for analysis of stochastic processes described by binary homogeneous Markov model. Essential result from this work is that this algorithm allows the stochastic process to be simulated, discretely over time, through its representation as a quantum mechanical system. The simulation of binary homogeneous Markov process with quantum computer returns  as a superposition of the quantum register the full set of all possible paths for the process. This way the required informational operations for the calculation of full set of paths are being reduced exponentially - $2^n$ possible paths could be calculated with $n$ operations. The computational complexity of the algorithm is linear to the number of discrete time steps of the process: $O(n)$, where $n$ is the number of the discrete time steps in the Markov process.

\subsection{Tasks} 
\label{subsec:Tasks}

The following tasks must be achieved in order to develop the method and algorithm for quantum simulation of binary homogeneous Markov model:
\begin{itemize}
\item An algorithm for estimation of the of qubit's rotation angle over the X axis, depending on the desired probability distribution for the states amplitudes.
\item Implementation of a quantum logic gate, which rotates the qubit over the X axis.
\item Implementation of a controlled quantum logic gate, which acts on two qubits and rotates the target over the X axis depending on the control qubit state.
\item Algorithm for quantum simulation of a binary homogeneous Markov process with application on real hardware platforms.
\item Development of quantum circuit representing the time evolution of a stochastic problem defined by binary homogeneous Markov model.
\item Development of an example: experimental quantum circuit, which could be executed on various hardware platforms including high-performance quantum simulator and real quantum processors in IBM's cloud platform "Quantum Experience".
\item Application of error correction and error mitigation techniques for the experimental circuit. Analysis of the results and comparative analysis for the performance of the different hardware back-ends.
\end{itemize}

\markboth{ANALYSIS}{Analysis of the Review}
\section{Analysis of the Review}
\label{subsec:Analysis of the Review}

\subsection{Motivation}
Markov chains are fundamental part in algorithms for problems in various scientific research areas: chemistry, physics, biology, economics, finance and many more. Using quantum computers for solving real scientific problems would cause a huge impact on modern science and technology, including the quantum computing.
\par This chapter describes the necessary fundamentals for understanding in-depth the quantum computing and quantum information theory behind the quantum simulation of stochastic processes and the necessary software frameworks and tools for quantum computation. Also attention is paid to the mathematical foundations of the probability theory, and processes described by Markov chains (binary homogeneous Markov process), more specifically. 

\begin{figure}[!h]
  \centerline{\includegraphics[width=1\textwidth]{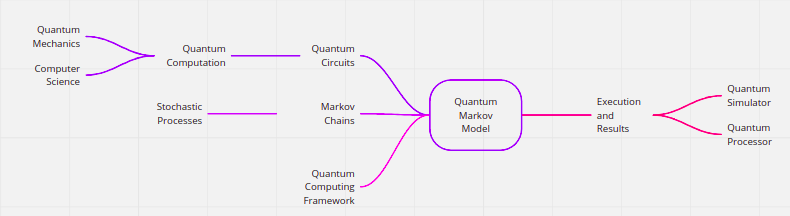}}
  \caption{Structure of the dissertation.}
  \label{fig:Structure}
\end{figure}

The workflow is defined by the structure of this dissertation, shown on figure \ref{fig:Structure}, where the structure could be divided down to 3 parts:
\begin{itemize}
	\item Survey (Review) of the fundamental ideas and modern techniques of quantum computing and quantum information. Some mathematics and physics fundamentals behind quantum computing theory - how quantum mechanics together with computer science theory lead to quantum computation. Fundamentals of probability theory and stochastic processes (Markov chains). Review of the required software framework for model development and experimental verification.
	\item Definition of the goals, challenges and problems this work is based on. The connection between quantum computing and quantum information theory with existing software frameworks and tools, required for solving the defined problems and challenges, and achieving the goals.
	\item The methodology and conception for development of new quantum algorithm, applied objects (software), running experiments and analysis of the results obtained. 
\end{itemize}

\subsection{Quantum Computing}
Quantum computers are based on the laws of quantum mechanics and offer a fundamentally different (from the classical computers) computational mechanism. The main fundamental differences between classical and quantum computers are three:
\begin{itemize}
	\item Superposition - this is one of the fundamental parts of quantum mechanics. In a given moment of time the quantum system is simultaneously in all of its possible states. In the context of quantum computing, this means that if a quantum register is in a superposition, its state is a linear combination of all possible states between 0 and 1 for the qubits in the register. In classical computers, a register can only be in one state at a given moment of time. The superposition collapses when a measurement is being performed on the quantum register. The collapse means that at the moment of measurement, the register is in one deterministic state. The quantum computations are performed by setting probabilities for all the possible states for a register.
	\item Probabilistic - the quantum system is probabilistic and not deterministic. Every possible state for this system can be observed when a measurement is performed over the quantum register. The quantum computation is performed by increasing the probability for the desired state of the register.
	\item Entanglement - this is a specific property for the quantum systems. A given quantum system is entangled if it can not be decomposed into more fundamental parts. And one part of this system can not be described without the knowledge for states of the others. 
\end{itemize}

\par 
Understanding mathematics used to describe quantum mechanical systems (states of quantum registers and quantum logic gates) requires knowledge of the fundamental concepts and notations. The Dirac notation (also known as bra-ket notation) is a way of describing vectors, where column-vector is called ket vector and row vector - bra vector.

\begin{equation}
v = \begin{bmatrix}
		   v_{0} \\
           v_{1} \\
           v_{2} \\
           \vdots \\
           v_{n}
         \end{bmatrix}
\label{eq:1}
\end{equation}

\begin{equation}
v = \overline{v^T} = \left[ \overline{v_0}, \overline{v_1}, \overline{v_2}, ..., \overline{v_n} \right]
\label{eq:2}
\end{equation}

Where $\overline{v}$ is the conjugate transpose of $v$. The Dirac notation e a convenient way of vector description in the Hilbert space. The Hilbert space is a vector space with a vector product and a norm defined by this product. The vector product of two vectors in the complex Hilbert space is the scalar product of the vectors $v$ and $u^T$ ( the complex conjugate of $u$):

\begin{equation}
\braket{v|u} = \overline{u^T}v = \left[ \overline{u_0}, \overline{u_1}, \overline{u_2}, ..., \overline{u_n} \right]\begin{bmatrix}
		   v_{0} \\
           v_{1} \\
           v_{2} \\
           \vdots \\
           v_{n}
         \end{bmatrix} =  \overline{u_0}v_0 +   \overline{u_1}v_1 +   \overline{u_2}v_2 + ... +   \overline{u_n}v_n
\label{eq:3}
\end{equation}

The vector product must satisfy the following conditions:
\begin{itemize}
	\item $\braket{v|v} >= 0$ where $\braket{v|v} = 0$ if and only if $\ket{v} = 0$.
	\item $\braket{u|v} = \overline{\braket{v|u}}$ for all $\ket{u}$ and $\ket{v}$ in the vector space.
	\item $\braket{u| \alpha_0 v + \alpha_1 w} = \alpha_0 \braket{u|v} + \alpha_1 \braket{u|w}$
\end{itemize}

The norm of $\ket{v}$ in the Hilbert space is defined as square root of the vector product of $\ket{v}$ with itself, which geometrically represented would be the distance from the origin of the coordinate system to the point $\ket{v}$ (also known as Euclidean distance).

\begin{equation}
\norm{ \ket{v}} = \sqrt{\braket{v|v}}
\label{eq:4}
\end{equation}

The tensor product of two vectors (also known as Kronecker product) describes the linear transformation between two vector spaces and is used for combining multiple vector spaces into a bigger one. The tensor product of two vector space describing two quantum mechanical systems gives the linear combination between all the vectors in the two spaces.

\begin{equation}
\ket{v} \bra{u}= \begin{bmatrix}
		   v_{0} \\
           v_{1} \\
           v_{2} \\
           \vdots \\
           v_{n}
         \end{bmatrix} \left[ \overline{u_0}, \overline{u_1}, \overline{u_2}, ..., \overline{u_n} \right] = 
         \begin{bmatrix}
		   v_{0} \overline{u_0} & v_{0} \overline{u_1} & ... & v_{0} \overline{u_n} \\
           v_{1} \overline{u_0} &  v_{1} \overline{u_1} & ... &  v_{1} \overline{u_n}\\
           v_{2}  \overline{u_0} & v_{2}  \overline{u_1} & ... & v_{2}  \overline{u_n}\\
           \vdots \\
           v_{n} \overline{u_0} & v_{n} \overline{u_1} & ... & v_{n} \overline{u_n}
         \end{bmatrix}
\label{eq:5}
\end{equation}

\par 
Like classical computers and information systems, the quantum computers are based on the quantum bit (qubit). The main properties of one qubit and a quantum register are presented bellow. Like in the theory of classical computing, where the bit represents a physical object, the qubits are also a representation of real physical objects, but for the purpose of this dissertation these will be treated as abstract mathematical objects, since there is no single quantum hardware realization - there are multiple hardware realizations with different physical properties. This representation of qubits gives the freedom to use the main theory for quantum computation and quantum information which does not depend on the hardware.
\par
One qubit pure state could be described by the wave function of its basis states:

\begin{equation}
\ket{\Psi(t)} = e^{-i\zeta_{0}(t)}c_0(t)\ket{0} + e^{-i\zeta_{1}(t)}c_1(t)\ket{1}
\label{eq:6}
\end{equation}

\par
The $\ket{0}$ and $\ket{1}$ are the two possible states for every qubit, and these are interpretation of the 0 and 1 states of classical bits. The difference here is that every qubit exists in the two states simultaneously - the superposition. This superposition is a linear combination of the two states:

\begin{equation}
\ket{\Psi} = \alpha \ket{0} + \beta \ket{1}
\label{eq:7}
\end{equation}

\par
Here $\alpha$ and $\beta$ are complex numbers. And the qubit's state is a vector in the two-dimensional complex vector space. When a measurement is performed over the qubit its superposition collapses and it goes into one of the two possible states - $\ket{0}$ or $\ket{1}$. Where the probability for the qubit to be in state $\ket{0}$ is $|\alpha^2|$ and in state $\ket{1}$ - $|\beta^2|$. The two probabilities must sum up to 1: $|\alpha^2| + |\beta^2| = 1$. Even if it doesn't sound intuitive, the superposition means that the qubit is in a state, which is neither  $\ket{0}$ nor $\ket{1}$, but somewhere in between. When a collapse occurs the state can be deterministic defined as one of the two possibilities.
The \ref{eq:6} could be transformed into one of the two equations bellow:

\begin{equation}
\ket{\Psi} = e^{i\gamma}\left(\cos{\left(\theta \over {2}\right)}\ket{0} + e^{i\phi}\sin{\left(\theta \over {2}\right)}\ket{1}\right)
\label{eq:8}
\end{equation}

And $e^{i\gamma}$ could be removed, so the \ref{eq:8} is simplified to:

\begin{equation}
\ket{\Psi} = \cos{\theta over 2}\ket{0} + e^{i\phi}\sin{\theta over 2}\ket{1}
\label{eq:9}
\end{equation}

\par 
The quantum registers are sequences of qubits, where the length of the register indicates how much information it can store. The superposition of a quantum register is an analogue of the superposition of a qubit, and when it is in a superposition, all the qubits in the register are in superposition. All possible bit configurations in the register's superposition are the tensor product of all consisting qubits. The vector space of a $n$-size quantum register is a linear combination of $n$ basis vectors, where every vector is with size $2^n$:

\begin{equation}
\ket{\Psi_n} = \sum_{i=0}^{2^n-1} {\alpha_i\ket{i}}
\label{eq:10}
\end{equation}

The quantum registers could be mathematically expressed as an extension to the qubits, where the probabilities for every possible state of the register satisfy the following condition:

\begin{equation}
\sum_{i=0}^{2^n-1} {\left|\alpha_i\right|} = 1
\label{eq:11}
\end{equation}

\par
A quantum computation (in the meaning of the computation analogue to the classical one) is achieved by the evolution of a quantum register (a qubit can be expressed as a register with length of 1) over time. And the evolution is achieved by the application of quantum logic gates over the registers. When applied over a register, the quantum logic gate, maps the superposition of this register from one to another. The mathematical representation of the quantum logic gates is transformation matrices and when applied to a quantum register the result is a tensor product of a transformation matrix with the matrix representation of a quantum register. All linear operators which represent quantum logic gates must satisfy the condition to be unitary. The unitary operations over a single qubit could be graphically represented by rotations over the $x,y,z$ axis on the Bloch sphere, where all linear combinations of $\alpha\ket{0} + \beta\ket{1}$ in $\mathbb{C}^2$ represent all the possible points on the surface of the Bloch sphere $(\theta, \Psi)$. 
\par Here is a list of all the elementary and newly generated quantum logic gates used in this dissertation:

\begin{itemize}
	\item Hadamard gate - a single qubit quantum logic gate which inducts a superposition with equal probabilities for the two possible qubit states ($\ket{0}$ and $\ket{1}$). It has the following matrix and graphical representation in quantum circuits:
\begin{equation}
H = {1 \over {\sqrt{2}}}  \begin{bmatrix}
		   1 & 1 \\
           1 & -1 \end{bmatrix} 
\label{eq:12}
\end{equation}
	\item Pauli-X gate - a single qubit quantum logic gate which acts on a qubit by a rotation over the $x$ axis of the Bloch sphere with $\pi$ radians. It does the following transformation of the qubit's states: $\ket{0} \rightarrow \ket{1}$ and $\ket{1} \rightarrow \ket{0}$:
\begin{equation}
X = \begin{bmatrix}
		   0 & 0 \\
           1 & 0 \end{bmatrix} 
\label{eq:13}
\end{equation}
	\item CNOT gate - a quantum logic gate which acts on two qubits, where one of the qubits plays the role of a control and the other one is target. When the control qubit is in state $\ket{1}$ the target state flips:
\begin{equation}
CNOT = \begin{bmatrix}
		   1 & 0 & 0 & 0 \\
           0 & 1 & 0 & 0 \\
           0 & 0 & 0 & 1 \\
           0 & 0 & 1 & 0 \end{bmatrix} 
\label{eq:14}
\end{equation}
	\item Phase-shift gate ($R_\phi$) - a single qubit quantum logic gate which doesn't change the basis state $\ket{0}$ and shifts the phase: $\ket{1} \rightarrow e^{i\phi}\ket{1}$. This way the probabilities for the states $\ket{0}$ and $\ket{1}$ when the qubit is measured with the main basis don't change, but the phase of the quantum state changes:
\begin{equation}
R_\phi = \begin{bmatrix}
		   1 & 0 \\
           0 & e^{i\phi} \end{bmatrix} 
\label{eq:15}
\end{equation}	
\end{itemize}

\begin{figure}[H]
  \centerline{\includegraphics[width=1\textwidth]{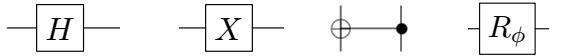}}
  \caption{Graphical representation of the symbols for Hadamard, Pauli-X, CNOT and Phase-shift quantum logic gates in quantum circuits.}
  \label{fig:Gates}
\end{figure}

\subsection{Markov Chains}
The term Markov chain is derived from the works of the Russian mathematician  Markov (1856 - 1922) where he studies a series of trials in which the outcome of each event depends only on the previous event. This is the simplest summary of a series of independent experiments. Today Markov processes find application in various fields such as biology, physics, computer and engineering sciences, economics and are useful in analysing practical problems.
Markov chains are mathematical models for time evolutionary stochastic processes. These processes are a sequence of random events and have multiple applications which makes them the most important example of stochastic processes. The main property of such a process is the fact that the process retains no memory \citep{0095301,HAAR659, Bochner1949} - no information about past states of the process is available. This means that the state in the next moment of time depends only on the current state. In this dissertation only discrete over time processes with finite discrete state space have being reviewed.
\par A discrete over time Markov process could be defined by the following definitions and assumptions:
\begin{itemize}
	\item $I$ is a finite countable set.
	\item Every $i \in I$ is called a state of the process, and $I$ is a state space.
	\item $\lambda = (\lambda : i \in I )$ is a measure of $I$ if $0 \leq \lambda \leq \infty$. 
	\item If $\sum{\lambda} = 1$, where $i \in I$, then $\lambda$ is a distribution.
	\item A random variable $X$ with values in $I$ is a function.
	\item $X:\Omega \rightarrow I$
\end{itemize}
\begin{equation}
\lambda_i =  \mathbb{P}(X=i) = \mathbb{P}(\omega : X(\omega) = i)
\label{eq:16}
\end{equation}	
Then $\lambda$ defines the distribution of $X$ and $X$ is a random state with values $i$ with probability $\lambda$. 
Markov process is a discrete over time stochastic process only if for every $i_0, i_1, i_2, ..., i_N \in I$ the following statement is satisfied:
\begin{equation}
\mathbb{P}(X_0=i_0, X_1=i_1, X_2=i_2,...,X_N=i_N) =  \lambda p_{i_0 i_1} p_{i_1 i_2}...p_{i_{N-1} i_N}
\label{eq:17}
\end{equation}	
A matrix $P = (p_{ij}:i,j \in I)$ is stochastic if every row in the matrix is a distribution.
There are three main Theorems which are used to define if a random process is Markov \citep{0095301,Stroock14,HAAR659}:
\begin{enumerate}
\item A discrete-time random process $(X_n)_0 \leq n \leq N $ is Markov ($\lambda, P$) if and only if $ \forall i_0,i_1,...,i_n \in I$:
\par $\mathbb{P}(X_0=i_0, X_1=i_1, X_2=i_2,...,X_N=i_N) =  \lambda p_{i_0 i_1} p_{i_1 i_2}...p_{i_{N-1} i_N}$ 
\item Let $(X_n)_n \geq 0$ be Markov ($\lambda,P$). Then, conditional on $X_m = i, (X_{m+n})_n \geq 0$ is Markov ($\delta, P$) and is independent of the random variables $X_0, X_1, ..., X_m$.
\item Let $(X_n)_n \geq 0$ be Markov ($\lambda, \mathbb{P}$). Then for all $n,m \geq 0$:
\begin{enumerate}
\item $\mathbb{P}(X_n=j)={(\lambda P^n)}_j $
\item $\mathbb{P}_i(X_n=j)=\mathbb{P}(X_{n+m}=j|X_m=i)={p_{ij}}^{(n)} $
\end{enumerate}
\end{enumerate}
In this dissertation a Binary Markov Process is notated as a process which has only two possible states at any given moment of time. The term Homogeneous over time \ref{ASENS1951} means it is a process with stationary transition probabilities - these transition probabilities do not change with time. When these two properties (limitations) are applied to the reviewed stochastic process, the result is a Binary Homogeneous Markov process which is going to be simulated with quantum computer.
\par If the Markov chain is defined as follows:
\par $\mathbb{P}(X_{n+1} = s_j| X_0 = s_0, X_1 = s_1, ..., X_n=s_i) = \\ \mathbb{P}(X_{n+1}=s_j|X_n=s_i) $
\par Then the Markov chain is homogeneous if:
\par $\mathbb{P}(X_{n+1}=j|X_n=i) = \mathbb{P}(X_{1}=j|X_0=i) , \forall n,i,j$
\par 
The probability $\mathbb{P}(X_{n+1}=j|X_n=i)$ is a probability for a transition from state $i$ to state $j$, and the term homogeneous means that the transition probabilities do not depend on the moment of time $n$.
\par $p_{ij} = \mathbb{P}(X_{n+1}=j|X_n=i)$
\par 
Classification of the Markov chain states \citep{pCIN75a,pKAR81a,pSNE94a,Ambrose1940}:
\begin{itemize}
\item \textbf{accessible} - the state $j$ is accessible from $i$ if, $\exists n \geq 1: {p_{ij}}^{(n)}>0$. If it is not accessible, then if the chain starts from $i$ state it will never go into the $j$ state.
\item \textbf{communicate} - if state $j$ is accessible from $i$ and $i$ is accessible from $j$ than these two states ($i$ and $j$) communicate: $i \\leftrightarrow j$. Where it must be true that every state is accessible by itself and communicate with itself - ${p_{ii}}^{(0)} = \mathbb{P}(X_0=i|X_0=i) = 1$. The $\leftrightarrow$ has the following properties:
\begin{itemize}
\item \textbf{reflexivity} - $i \leftrightarrow j$.
\item \textbf{symmetry} - $i \leftrightarrow j$ $\iff$ $j \leftrightarrow i$.
\item \textbf{transitivity} - $i \leftrightarrow k$, $k \leftrightarrow j$ $\Rightarrow $ $i \leftrightarrow j$.
\end{itemize}
\item \textbf{absorbing} - state $j$ is absorbing if no other state can be accessed by it: $p_{jj} = 1$.
\item \textbf{irreducible} - if every two states in the chain are communicating.
\item \textbf{transient} - state $i$ is transient if $\sum_{n=1}^{\infty}{p_{ii}^{(0)}} = \infty$.
\item \textbf{recurrent} - $\sum_{n=1}^{\infty}{p_{ii}^{(0)}} < \infty$
\end{itemize}
\par 
Using the communicate class of the Markov chains it is possible to decompose a Markov process into small pieces, analyse those separately and then together the whole process is understandable relatively easier. 
\par
Important properties of a Markov chain are the hitting times and absorption probabilities. These quantities can be calculated by the linear equations associated with the transition matrix P \citep{0095301}:
\begin{enumerate}
\item The hitting probabilities vector $\alpha(h)^A = (\alpha(h)_i^A : i \in I)  $is the minimal non-negative solution to the system of linear equations:
\par $
\begin{cases}
k_i^A = 0$  for $i \in A \\
k_i^A = 1 + \sum_{j \notin A}{p_{ij} k_j^A}$ for $i \notin A
\end{cases}
$
\par The minimality in this case means that $x_i \geq h_i$ for all $i$ if $x$ is another solution with $x_i \geq 0$ for $\forall i$.
\item The vector of mean hitting times $k^A = (k^A:i \in I)$ is the minimal non-negative solution to the system of linear equations:
\par $
\begin{cases}
k_i^A = 0$  for $i \in A \\
k_i^A = 1 + \sum_{j \notin A}{p_{ij} k_j^A}$ for $i \notin A
\end{cases}
$
\end{enumerate}
\par Strong Markov property
The Strong Markov property is based on the concept for the Markov property of a stochastic process, e.g. the probabilistic behaviour of the chain of events in the next moment of time depends only on the current state \ref{Ray1956}. Except that in addition the time T is a random quantity and has special properties:
\par $T$ is \textit{stopping time}, which takes values \{1,2,...\} such that $T=n$ can be determined by the chain values $X_0, X_1, ..., X_n$.
\par If $T$ is a \textit{stopping time}, for $m \geq 1$:
\par $\mathbb{P}(X_{T+m}=j|X_k=x_k,0 \leq k < T; X_T=i)= \mathbb{P}(X_{T+m}=j| X_T=i) $

\par \textbf{Applications}
\par Markov chains applications can be found in many scientific fields including biological modelling, queuing, Markov decision processes, Markov chain Monte Carlo and many more. Using Markov chains appropriate models for systems with high complexity could be built. An example of biological interpretation for a stochastic process described by Markov chains is a well known problem for a virus mutation \citep{0095301}. 
\par Imagine a virus can exists in N different strains and each time the virus either stay at the same strain or with a probability $\alpha$ mutates to another one chosen at random. What would be the probability that $n^{th}$ generation strain is the same as the starting one ($0^{th}$)?  
\par This process can be modelled as a Markov chain with $NxN$ transition matrix $P$, where:
\par $ p_{ii} = 1-\alpha , p_{ij} = {\alpha / (N-1)}$ for $i \neq j$.
\par The solution of this problem is found in the computation of $p_{11}^{(n)}$. At any moment of time a transition from initial state with probability $\alpha$ is made, and a transition to the initial state with probability $\alpha / (N-1)$.
\par If $\beta$ is put to be ${\alpha / (N-1)}$, the transition matrix has the form: 
\par $ P  =
\begin{pmatrix}
1-\alpha  & \alpha \\
\beta & 1-\beta

\end{pmatrix}$
\par Because of the recurrence relation for $p_{11}^{(n)}$: 
\par $p_{11}^{(n+1)} = (1-\alpha -\beta)p_{11}^{(n)} + \beta $, $p_{11}^(0)=1$.
\par The desired probability is:
\par ${1 \over N}+\left(1-{1 \over N}\right)\left(1-{\alpha N \over N-1}\right)^n$

\subsection{Software Tools and Frameworks for Quantum Computing}
Nowadays three are the most popular open-source quantum computing software frameworks: QISKIT, D-Wave Ocean and Forest SDK. QISKIT and Forest SDK are frameworks for gate-based (circuit) modelling in quantum computing and these are developed by IBM and Rigetti Computing respectively. For the purpose of this dissertation the IBM's QISKIT framework has been used. Also IBM has provided access to their real quantum processor available on the cloud platform.
\par 
What is QISKIT? QISKIT is a python framework that allows the creation and execution of algorithms for quantum computers. These algorithms are representation of quantum systems and contain all the information about how the systems should be created and manipulated.
\par The QISKIT framework consists of four main parts \citep{Qiskit}:
\begin{itemize}
\item Terra - this is the software foundation for the stack, containing tools for building quantum programs at level of circuits and pulses. It also provides possibility for optimization depending on the physical quantum processor and managing the execution of the quantum programs on cloud available backends. The software stack for Terra includes:
\begin{itemize}
\item User Inputs (Quantum Circuits and Pulse schedules).
\item Transpiler (Optimization passes).
\item Providers (Qiskit Aer. IBM Q, Others).
\item Visualization and Quantum Information Tools (Histograms, States, Entanglement).
\end{itemize}
\item Aer - it provides optimized C++ high-performance quantum simulator backends for execution of quantum circuits. Also it contains tools for performing realistic noisy simulations based on configurable noise models. Its software stack includes:
\begin{itemize}
\item Qiskit Terra.
\item Noise simulation (Noise models, Quantum errors).
\item Backends (QasmSimulator, StatevectorSimulator, NoisySimulator).
\item Jobs and Results (Counts, Memory, Statevector, Unitary, Snapshots).
\end{itemize}
\item Aqua - this is library of cross-domain quantum algorithms used to build near-term quantum applications. It is extensible with possibility for adding new custom quantum algorithms. The current version allows experiments on chemistry, AI, finance and optimization applications. The software stack includes:
\begin{itemize}
\item Qiskit Aqua Translations (Chemistry, Optimization, AI, Finance).
\item Quantum Algoritms (QPE, Grover, HHL, QSVM, VQE, QAOA).
\item Qiskit Terra
\item Providers (Qiskit Aer. IBM Q, Others).
\end{itemize}
\item Ignis - this is a framework containing experiments for understanding and mitigating noise in quantum circuits. The experiments are classified in three groups:
\begin{itemize}
\item Characterization - measure noise parameters.
\item Verification - verify gate and small circuits.
\item Mitigation - run calibration circuits.
\end{itemize}
Its software stack includes:
\begin{itemize}
\item Qiskit Ignis Experiments (Quantum circuits or pusle schedules).
\item Qiskit Terra
\item Providers (Qiskit Aer. IBM Q, Others).
\item Fitter/Filter (Fit to a model or plot results).
\end{itemize}
\end{itemize}
\par Using QISKIT the quantum processors can be remotely accessed. The main programming language is Python version at least 3.5. The main process in quantum software development could be divided in three steps:
\begin{itemize}
	\item Build: This step includes the design of the quantum circuit representing the problem's solution.
	\item Execute: Executing quantum circuits on various hardware back-ends (real quantum processors or quantum simulators).
	\item Analyze: Calculation and visualization (histograms) of generalized results and analysis.
\end{itemize}
The process above could be also described in more details:
\begin{itemize}
	\item Import the necessary libraries and modules in Python. The minimal required pack of libraries and modules used in this dissertation are the following:
	\begin{itemize}
	\item \textit{Numpy} - a fundamental packet in Python used for scientific programming. It consists of powerful N-size array type objects, complex mathematical functions, linear algebra solutions, random number generator, Fourier transform, integration tools for code in C++ and Fortran.
	\item \textit{QuantumCircuit} which is part of QISKIT framework and includes all available quantum operations. It could be also said that this is the "machine language" for the quantum systems description.
	\item \textit{Execute} is also modul in QISKIT and it executes the quantum circuits on varios back-ends. 
	\item \textit{Aer} module which enables the usage of simulation hardware. 
	\item \textit{plot\_histogram} is a method for creating histograms and results visualization.	
	\end{itemize}
\begin{figure}[H]
  \centerline{\includegraphics[width=1\textwidth]{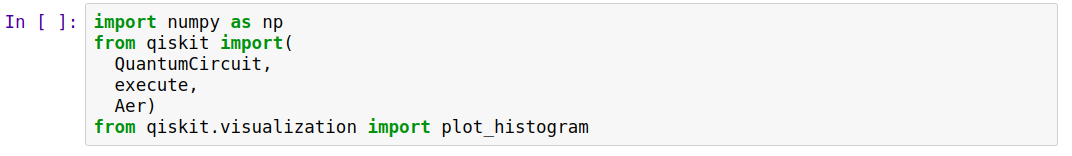}}
  \caption{Required libraries, packages and modules in Python for building a quantum program.}
  \label{fig:Includes}
\end{figure}
	\item Variables initialization.
	\item Adding quantum logic gates to design the desired quantum circuit. 
	\item Quantum circuit visualization and verification. This could be achieved by using the \textit{circuit.draw()} method. Using this method the circuit is visualized so that the qubit order is ascendant and the $0^{th}$ qubit is visualized on top of the circuit. The quantum circuits must be read from left to right and the gates located more to the right are being executed later in time by the quantum back-end. 
	\item Experiment simulation on the selected back-end.
\begin{figure}[H]
  \centerline{\includegraphics[width=1\textwidth]{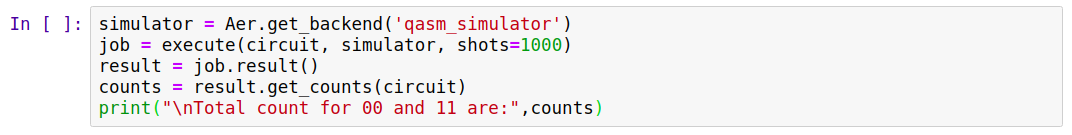}}
  \caption{Quantum experiment simulation on selected back-end and result extraction.}
  \label{fig:Simulation}
\end{figure}	
	\item Visualization of the results from the experiments. Beside the histogram plot, there are many other possibilities for result visualization in QISKIT, including interactive methods.
\end{itemize}
\subsection{Problem Description}
Quantum computers use the effects of quantum mechanics such as coherence and entanglement to process information differently from classical computers. Quantum computers can handle two types of data - classic and quantum. Classic data must be structured as input before being processed by quantum computers. Quantum information processing finds patterns (structures) in the data by presenting them as certain quantum mechanical states and then executing basic quantum subroutines. With only 2 possible states, the binary homogeneous Markov process, which transition probabilities does not change over time, enables these states to be represented by qubits, the main challenge being the creation and tuning of quantum logic gates.

\markboth{METHOD AND ALGORITHMS}{Method and Algorithms}
\section{Method and Algorithms}
\label{sec:Method and Algorithms}

\subsection{Method for quantum simulation of a binary homogeneous Markov process}
\label{subsec:Method}

The simulation of a binary homogeneous Markov process using quantum computer requires the simulated process to be described as a quantum-mechanical system. In every moment of time the process is represented by a single qubit, where its two possible states are the two qubit states - $\ket{0}$ and $\ket{1}$. The starting point for the simulation is a qubit in a superposition of the desired probability distribution for the stochastic process ( \ref{eq:7}, \ref{eq:11}).
\par 
This system updates its states by 2x1 transition matrix and for every discrete moment of time it goes from state $i$ to state $i+1$, where the system representation is a single qubit for every $i^{th}$ moment. The current state depends only on the state in the previous moment of time and the process retains no memory. This means no history must be remembered and the only required connections between states are those from current to the next state. The quantum representation of this connections is done by one of the fundamental properties of the quantum systems - entanglement. The quantum entangled states represent the connection between 2 or more qubits where the state of one qubit can not be described without the knowledge of the states of the others.
With this method for quantum simulation of a binary homogeneous Markov process, an entangled state is created between every two sequential qubits, where the each qubit plays the role of the control for the next one. This way a directed chain is constructed using entangled qubits. The state of each qubit defines the state of the next one, starting from qubit $q_0$ which acts as a control for the $q_1$, $q_1$ acts as a control for $q_2$ and so on. This way every qubit in the chain except the first and the last ($q_0$ and $q_N$) acts as a control qubit for the next one and a target qubit for the previous. If $N$ is the number of discrete time steps for a binary homogeneous Markov process, then an $N$-length quantum register is needed for the quantum simulation.
\par 
Markov process division on parts 
\par 

\begin{figure}[!h]
  \centerline{\includegraphics[width=1\textwidth]{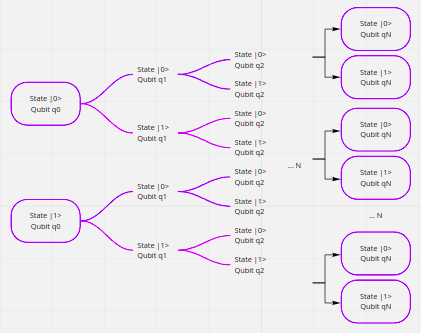}}
  \caption{Hierarchical Tree structure showing the full set of possible paths for a binary homogeneous Markov process which is described as a quantum-mechanical system.}
  \label{fig:Hierarchical_Markov}
\end{figure}

\begin{figure}[H]
  \centerline{\includegraphics[width=1\textwidth]{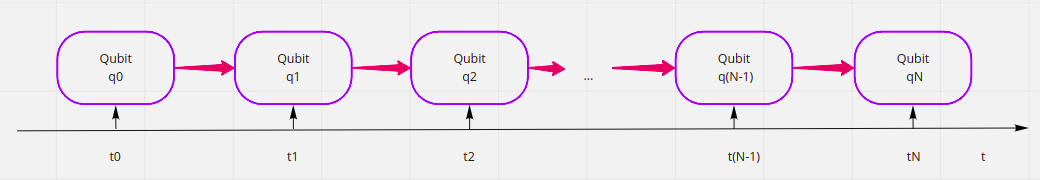}}
  \caption{A binary homogeneous Markov process visualization using chain of qubits. The red arrow shows the entanglement direction (the beginning of the arrow is the control qubit and the end of the arrow is the target qubit). In every discrete moment of time the process is represented by a single qubit, which is in an entangled state with the next one. For a quantum simulation of a process with N number of discrete time steps, a quantum register with N qubits is required.}
  \label{fig:Qubit_ladder}
\end{figure}

\subsection{Algorithm for determining the angle of rotation for a qubit over the X axis}
\label{subsec:algo_theta}
An implementation of a new quantum logic gate in QISKIT framework is required so that the rotation over the X axis could be achieved. This is the $\sqrt[n]{X}$ gate and it is a single qubit quantum logic gate with the following matrix representation:
\begin{equation}
\sqrt[n]{X} = {1 \over {2}} \begin{vmatrix}
1+e^{i \pi \over{n}} & 1-e^{i \pi \over{n}} \\
1-e^{i \pi \over{n}} & 1+e^{i \pi \over{n}}
\end{vmatrix}
\label{eq:18}
\end{equation}
The realization of this gate is achieved using some of the basis standard quantum logic gates in QISKIT:
\begin{itemize}
\item Hadamard gate (see equation \ref{eq:12}
\item U1 gate, which is the QISKIT equivalent of $R_z$ gate. This is a single qubit quantum logic gate, which does a rotation over the $Z$ axis where the angle is defined in radians. It has the following matrix representation:
\begin{equation}
U1 = R_z = \sqrt[n]{Z} = \begin{vmatrix}
1 & 0 \\
0 & e^{i \pi \over{n}}
\end{vmatrix}
\label{eq:19}
\end{equation}
\end{itemize}
\par Using the rule for enclosure of a Pauli-Z gate with Hadamard gates to create Pauli-X gate \citep{Aharonov2003ASP, Nielsen2001QuantumIA}:
\begin{equation}
H \sqrt[n]{Z} H = \sqrt[n]{X}
\label{eq:20}
\end{equation}
Where the probabilities for the two states of the qubit - $\ket{0}$ and $\ket{1}$ are respectively $\left|{{\alpha_0}^2}\right|$ and $\left|{{\alpha_1}^2}\right|$:
\begin{equation}
\left|{{\alpha_0}^2}\right| = \left|{1+e^{i\pi \over n}} \over 2\right| and \left|{{\alpha_1}^2}\right| = \left|{1-e^{i\pi \over n}} \over 2\right| 
\label{eq:21}
\end{equation}
or 
\begin{equation}
\left|{{\alpha_0}^2}\right| = \left|{1-e^{i\pi \over n}} \over 2\right| and \left|{{\alpha_1}^2}\right| = \left|{1+e^{i\pi \over n}} \over 2\right| 
\label{eq:22}
\end{equation}
The algorithm for determining the angle of rotation for a qubit over the X axis is actually reduced to solving the following equation, where \textbf{n} is the unknown variable:
\begin{equation}
\left|{{\alpha_0}^2}\right| = \left|{1+e^{i\pi \over n}}  2\right| 
\label{eq:23}
\end{equation}
Using the Euler's formula from complex analysis \citep{euler1748introductio} for connection between trigonometrical functions and the complex exponent the following solution of \ref{eq:23} is found:
\begin{equation}
n = {\pi \over acos(2\left|{{\alpha_0}^2}\right| -1)}
\label{eq:24}
\end{equation}
The denominator of the right-hand side of the equation is the so-called 'lambda' parameter in the QISKIT software framework, which determines the angle of rotation of the qubit along the X axis.
\subsection{Building a quantum logic gate C$\sqrt[n]{X}$}
\label{subsec:build_CnX}
The construction of a quantum logical gate that acts on two qubits, creating an etangled state between them, is accomplished by a unitary transformation of the states of the qubits. Thus, at two points in time, the evolution of the state change from point 1 to point 2 is described by a unitary matrix. This quantum logic gate will be called controlled- $\sqrt[n]{X}$ ($C\sqrt[n]{X}$). Knowing the matrix representation of the gate shown in \ref{eq:18}, it is possible to construct a controlled gate \citep{Barenco1995Elementary} with the following matrix representation:
\begin{equation}
C\sqrt[n]{X} = \begin{vmatrix}
1 & 0 & 0 & 0 \\
0 & 1 & 0 & 0 \\
0 & 0 & {1+e^{i \pi \over{n}}} \over 2 & {1-e^{i \pi \over{n}}} \over 2 \\
0 & 0 & {1-e^{i \pi \over{n}}} \over 2 & {1+e^{i \pi \over{n}}} \over 2
\end{vmatrix}
\label{eq:25}
\end{equation}
Thus, the state of the first qubit that plays the role of control does not change, but the state of the second qubit (target) changes depending on the state of the first.
The roots of the controlled Pauli-Z gate (two-qubit gate that performs rotation along the Z axis on the target qubit, depending on the state of the control qubit) can be constructed by phase shifting by half the desired angle over the Z axis with opposite direction, after which the two qubits are phase-shifted along the Z axis by half the desired angle, applying a CNOT gate between the two qubit turns. Thus, enclosing this operation for the second qubit (the target qubit) with Adamar gates, is obtained the $C\sqrt[n]{X}$.

\begin{figure}[H]
	\centerline{\includegraphics[width=0.5\textwidth]{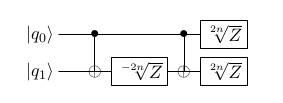}}
	\caption{Quantum circuit for construction of a two-qubit quantum logic gate that does a controlled phase-shift over the Z axis for the target qubit. }
\label{fig:CnZ}
\end{figure}

\begin{figure}[H]
	\centerline{\includegraphics[width=0.7\textwidth]{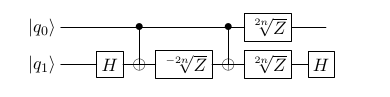}}
	\caption{Quantum circuit for construction of a two-qubit quantum logic gate that does a controlled phase-shift over the X axis for the target qubit.  }
\label{fig:CnX}
\end{figure}
As seen from figures \ref{fig:CnZ} and \ref{fig:CnX} the construction of $C\sqrt[n]{X}$ is achieved only by using three basic quantum logic gates which are supposed to be available on any modern platform for gate-based quantum computing.
\vspace{4cm}
\subsection{Algorithm for finding the complete set of possible paths for a stochastic process described by a binary homogeneous Markov model}
\label{subsec:algo_Markov}
This algorithm consists of four steps described bellow.
\renewcommand{\labelenumi}{\roman{enumi}}
\begin{enumerate}
	\item Calculate the rotation over the X axis for the $\sqrt[n]{X}$ and C$\sqrt[n]{X}$ gates so that these are representing the Markov process transition probabilities. 
    \item Initializing the system in the desired superposition of all qubits by applying $\sqrt[n]{X}$ to each qubit in the register.
For example, if the desired binary homogeneous Markov model is absorbent in the state $\ket{0}$ - all the qubits must be initialized in the state $\ket{0}$ with a probability of 1 (deterministic state) (Note: if $q_0$ is initialized in the state $\ket{0}$ with probability of 1, the process will not undergo any development because it will enter its absorbing state from the first discrete moment of time).
    \item Repeat the following unitary operation n-1 times: $C\sqrt[n]{X}$ for all consecutive pairs of qubits from $q_0$ to $q_{n-1}$.
    \item Measuring the resultant state of a quantum system. As a result of the measurement, the probability distribution for each possible path of the binary homogeneous Markov process is obtained. Each possible state of the quantum register corresponds to one possible path.
\end{enumerate}

Conclude 
Using the method presented in this chapter to simulate a binary homogeneous Markov process, the distinguishing properties of quantum computers - superposition, probability calculations and entangled states can be effectively used.
To implement an algorithm based on this method, a new quantum logic gate was implemented to create entangled states between two qubits. This gate performs a rotation along the X-axis for the qubit acting as a target with a predetermined angle $\Theta$, this angle (offset) accurately represents the transient probabilities for the given Markov process when the control qubit is in a state $\ket{1}$. \footnote{Representation of the transient probabilities for the given Markov in author's work (\hyperref[publications:amee2]{Author's publication iii}).}
An algorithm has been developed to determine the rotation of the qubit along the X axis, by which the necessary distributions for the transient probabilities of a binary homogeneous Markov process are introduced. \footnote{Algorithm for determining qubit rotation over the X axis in author's work (\hyperref[publications:fdiba2017]{Author's publication i and ii}).}
An algorithm has been developed to find the full set of possible paths for a stochastic process, described by a binary homogeneous Markov model. 
 
\markboth{APPLICATIONS}{Applications}
\section{Applications}
\label{sec:Applications}

\subsection{Implementation of a quantum logic gate which rotates the qubit with custom angle over the X axis}
The realization of a quantum logic gate which does a phase-shift over the X axis for a single qubit is somehow straightforward and easy task using QISKIT framework. It could be achieved by using the following standard quantum logic gate available in the framework:
\begin{itemize}
\item Hadamard gate - equation \ref{eq:12}
\item U1 gate, which is the equivalent of a $R_z$ gate - equation \ref{eq:19}
\end{itemize}
And using the rule for bracketing the Pauli-Z gate with Hadamard gates to create a Pauli-X gate ( equation \ref{eq:20}), the following quantum circuit is constructed:
\begin{figure}[H]
	\centerline{\includegraphics[width=0.5\textwidth]{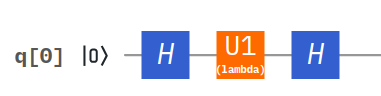}}
	\caption{Graphical representation of a quantum logic gate, which shifts the qubit over the X axis with custom angle in QISKIT framework. }
\label{fig:nX}
\end{figure}

\subsection{Implementation of a controlled quantum logic gate which rotates the qubit with custom angle over the X axis}
This is so-called controlled $C\sqrt[n]{X}$ gate and it acts on two qubits where the first one plays the role of control and the second one is the target. The purpose of this gate is to change the state of the target qubit only if the control is in state $\ket{1}$. The quantum circuit which represents this logic gate is shown on figure \ref{fig:CnX}.
The implementation in the QISKIT framework is similar to the one done for the $\sqrt[n]{X}$ gate, and a validation circuit should be designed, so that the gate's conditional fire is validated.
The verification and validation of the $C\sqrt[n]{X}$ gate could be achieved by using a modified variant of the quantum circuit shown on figure \ref{fig:CnX}. And the new circuit gets the following look:
\begin{figure}[H]
	\centerline{\includegraphics[width=0.9\textwidth]{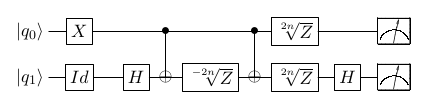}}
	\caption{Quantum circuit used for verification and validation of the $C\sqrt[n]{X}$ gate, where the gate will always fire.}
\label{fig:CnX_verification}
\end{figure}
The difference between \ref{fig:CnX} and \ref{fig:CnX_verification} is the initialization of the qubits in the beginning of the quantum operation, before the actual $C\sqrt[n]{X}$ gate application. As it is shown on the figure \ref{fig:CnX_verification}, the control qubit $q_0$ is in state $\ket{1}$ and the target qubit $q_1$ is in state $\ket{0}$. The fact that the control qubit will always be in state $\ket{1}$ means that the $C\sqrt[n]{X}$ gate will always fire and the target qubit $q_1$ will be rotated over its X axis with the desired angle. The most right part in the quantum circuit is the quantum measurement operator for the two qubits. Using this circuit the quantum operation will verify the $C\sqrt[n]{X}$ gate.
Also another quantum circuits could be use for more detailed experiments and results, regarding the validation and verification of this quantum logic gate. For example the control qubit $q_0$ could be superposed using the Hadamard gate when initialized, so that it will be with equal probabilities to be in state $\ket{0}$ and $\ket{1}$. An example quantum circuit is shown on figure 
\begin{figure}[H]
	\centerline{\includegraphics[width=0.9\textwidth]{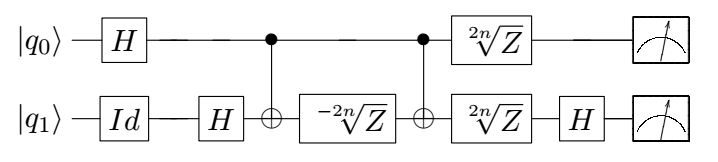}}
	\caption{Quantum circuit example for verification and validation of the $C\sqrt[n]{X}$ gate, where the gate will fire only 50\% of the time.}
\label{fig:CnX_H_init}
\end{figure}

\subsection{Development of a quantum circuit allowing modelling on a real quantum processor of any stochastic problem described by a binary homogeneous Markov model.}
\label{subsec:algo_dev}
Let's look at a system that has two possible states, and at any one time is in one of those two states. By presenting this system as quantum information system - by using 1 qubit, the two orthonormal states of the qubit will be analogous to the states of the system. The system is updated as states are defined only in certain values over time - i.e. it can be considered as a discrete signal. When updating its state from moment $i$ to moment $i + 1$, it is represented by a single qubit. The connection between the present and the next state of the system is realized by one of the basic quantum properties - the entangled states. Each qubit in the quantum circuit is in an entangled state with the next one by applying the $C\sqrt[n]{X}$ gate between these, so that the qubit $q_1$ is control qubit for the quantum logic gate and the qubit $q_{2}$ is the target, $q_2$ is control for $q_3$ and so on $q_{k-1}$ is control for $q_k$. Where $k$ is the length of the quantum register, which represents the number of the discrete time steps in the evolution of the process. 

\begin{figure}[H]
	\centerline{\includegraphics[width=0.8\textwidth]{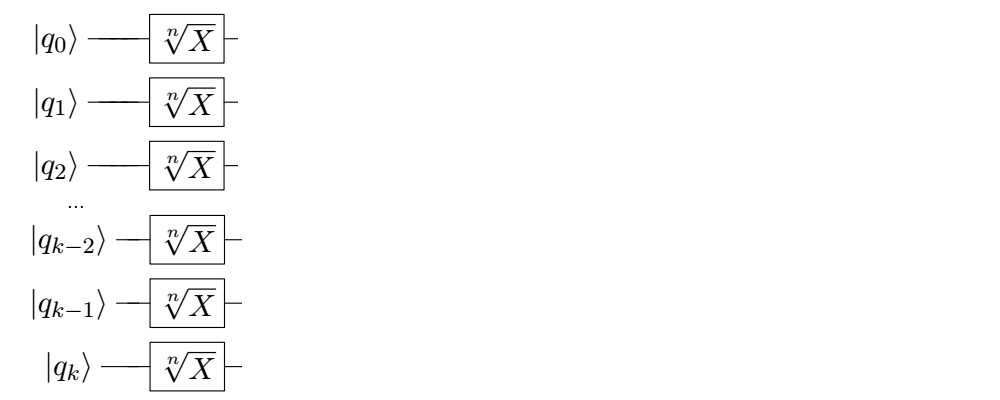}}
	\caption{Quantum circuit showing the initialization of the binary homogeneous Markov process.}
\label{fig:Markov_init}
\end{figure}

\begin{figure}[H]
	\centerline{\includegraphics[width=1\textwidth]{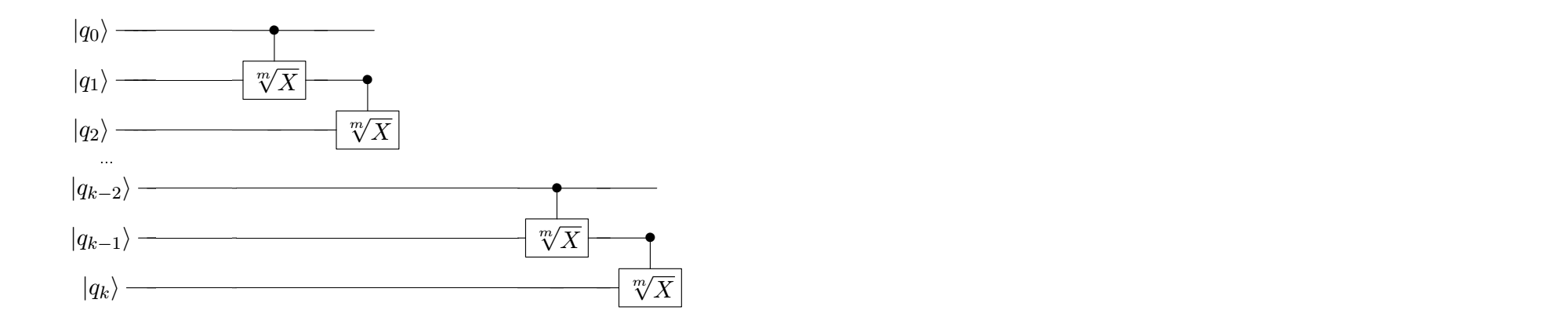}}
	\caption{Quantum circuit showing the repetitive unitary operation for the binary homogeneous Markov process.}
\label{fig:Markov_controlled}
\end{figure}

\begin{figure}[H]
	\centerline{\includegraphics[width=0.9\textwidth]{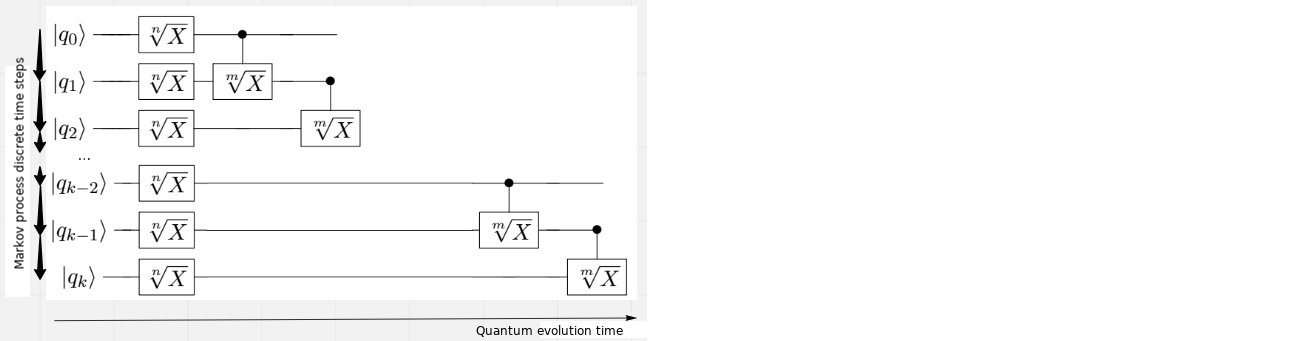}}
	\caption{Quantum circuit showing the schematic representation of an algorithm for quantum simulation of a binary homogeneous Markov process.}
\label{fig:Markov_alg}
\end{figure}

The quantum circuit model for application of this algorithm on a real hardware is shown on figure \ref{fig:Markov_alg} and this circuit can be simplified in two steps - first the quantum register is initialized in the desired quantum state using the $\sqrt[n]{X}$ quantum logic gate applied on every qubit in this register ( figure \ref{fig:Markov_init} and second - a unitary operation is applied throught the controlled quantum logic gate $C\sqrt[n]{X}$ between every two consecutive qubits like described in the previous paragraph (step $iii$ in the \ref{subsec:algo_Markov}).

\subsection{Development of an experimental quantum circuit, which to be implemented on both a high-performance quantum simulator and real quantum processors at IBM's cloud service through the QISIKT software framework.}
Since the algorithm described in the previous chapter (\ref{subsec:algo_dev}) allows for many variations of quantum circuits depending on the structure of the process for which a model is to be constructed, in this dissertation an example of a process will be described starting with an initial probability distribution for its two states - $\alpha$ for a state $\ket{0}$, and $\alpha -1$ for state $\ket{1}$. As state $\ket{0}$ it will be an absorbing state for the process.
For convenience, as well as allowing the algorithm to run on the maximum number of real quantum processors in the IBM cloud platform, there will be a limit on the number of steps for the stochastic process to be 3. Let k = 2, which means that $i_0, i_1, i_2$ and the distribution P form a Markov chain in the following order:
\begin{equation}
X_0 \rightarrow X_1 \rightarrow X_2
\label{eq:26}
\end{equation}
Following the Markov rule for a stochastic process, all the information about the $X_2$ state is in $X_1$, so if the task is to find the $X_2$ state, there is no need to know the $X_0$ state. Every $X_i$ for the process is represented in the quantum system with its corresponding qubit $q_i$ and the Markov property (the process retains no memory) is represented in a quantum way by applying the $C\sqrt[n]{X}$ gate as described in \ref{subsec:algo_dev}. 

\begin{figure}[H]
	\centerline{\includegraphics[width=0.6\textwidth]{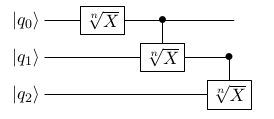}}
	\caption{Quantum circuit showing the quantum simulation of a binary homogeneous Markov process with absorbing state in $\ket{0}$.}
\label{fig:quantum_Markov_absorbing}
\end{figure}
N=3 defines the required length of the quantum register for a successful simulation ($k=2^N-1$). All possible states for the quantum register  are 8, so its superposition hypothetically could be constructed by 8 ket vectors:
\begin{equation}
\begin{split}
\ket{\Psi} = \alpha_0\ket{000} +\alpha_1\ket{001} + \alpha_2\ket{010} + \alpha_3\ket{011} + \\
 + \alpha_4\ket{100} + \alpha_5\ket{101} + \alpha_6\ket{110} + \alpha_7\ket{111}
\end{split}
\label{eq:27}
\end{equation}
The first step of this quantum simulation is the register to be initialized in state $1\ket{000}$.
The next step is the application of a $\sqrt[n]{X}$ quantum logic gate on the first qubit ($q_0$) - the first discrete time step for the binary homogeneous Markov process. This way the initial probability distribution is initialized as probability $\alpha$ for state $\ket{0}$ and $1-\alpha$ for state $\ket{1}$. Where state $\ket{0}$ is absorbing state for the process following the limitations described in \ref{subsec:algo_dev}.
\begin{equation}
\ket{\sqrt[n]{X^{q_0}}} = {{1+e^{i\pi \over n}} \over 2} \ket{000} + {{1-e^{i\pi \over n}} \over 2} \ket{001}
\label{eq:28}
\end{equation}
The final superposition of the quantum register is achieved with applying the $C\sqrt[n]{X}$ quantum logic gate as shown in figure \ref{fig:quantum_Markov_absorbing}. 
\subsection{Conclusion}
The constructed quantum logic gates are created by using the three standard basic quantum logic gates available in each modern quantum computing platform. The development of software in the Python programming language and the use of the Jupyter Notebook open source software development environment allow experimental research on various hardware devices - both on a real quantum processor and on high-performance quantum simulators.
The synthesized algorithm simulates a binary homogeneous Markov process by representing it through the superposition of the quantum register. When measuring the quantum register, the full set of possible paths for the process is obtained. The next chapter will introduce the algorithm in the QISKIT software framework on various hardware devices on the IBM cloud platform. \footnote{Binary homogeneous Markov process simulation with quantum computer in author's work (\hyperref[publications:amee2]{Author's publication iii}).}
 
\markboth{VERIFICATION}{Experimental Verification}
\section{Experimental Verification}
\label{sec:Experimental Verification}
\subsection{Implementation considerations}
\label{subsec:considerations}
The algorithm for binary homogeneous Markov process simulation seems easy to implement in modern quantum computing systems compared to other quantum-mechanical algorithms, and the following advantages can be highlighted:
\begin{itemize}
\item All the quantum transformations that are needed are:
	\begin{itemize}
	\item Hadamard gate.
	\item Z axis rotation gate.
	\item CNOT gate.
	\end{itemize}
These transformations are relatively simple, and are available as ready-to-use solutions in the QSIKIT software framework.
\item The implementation requires only a CNOT gate, as a two-qubit gate, with entanglement between the control qubit and the target qubit are between any two consecutive ones, which is possible in a large number of hardware configurations.
\item The controlled phase rotation at the $C\sqrt[n]{X}$ gate can be accomplished by using classical computer memory to preserve the probability distributions for the stochastic process. The quantum measuring gives as a result the complete set of possible paths for the binary homogeneous Markov process, given as the superposition of the measured quantum register.
\end{itemize}
When implementing quantum algorithms on real hardware platforms, one of the main considerations is the hardware limitations of quantum processors. In an ideal world, it would be great to have full control over the qubits so that operations can be performed allowing entangled states between any two (or more) qubits in the system. Unfortunately, in today's quantum processor hardware solutions, there are many limitations associated with this quantum property. A fundamental rule of gate-based superconducting quantum processors in the IBM platform is that for multiple-qubit gates, the control qubit(s) must be at a higher frequency of the resonant oscillator, than the target one. With exceptions to this rule, the direction of the gate must be reversed if degenerative effects on the target qubit are observed as a result of the higher frequency of the controller. The next paragraph will look at existing quantum processors in the IBM cloud platform that will be used to verify and validate the quantum gate $C\sqrt[n]{X}$, as well as the development of an experimental quantum circuit for the implementation of the algorithm for finding the full set of possible paths for a stochastic process described by a binary homogeneous Markov model.
\begin{figure}[H]
	\centerline{\includegraphics[width=1\textwidth]{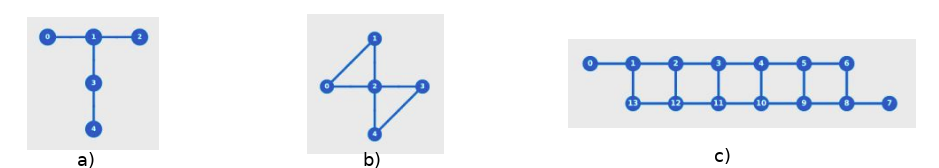}}
	\caption{Configuration maps of quantum processors in the IBM cloud platform that have been subjected to experimental verification and validation.	
a) T-shape layout of the qubits used in Ourence, Vigo, London, Burlington, Essex processors.
b) Diagram with bow-tie type used with the Yorktown processor.
c) Directional ladder geometry diagram of Melbourne processor.}
\label{fig:config_maps}
\end{figure}
The available quantum processors in the IBM cloud platform that have been experimented with in this thesis are the following 5-qubit (Ourence, Vigo, London, Burlington, Essex, Yorktown) and a 14-qubit (Melbourne) processors:
\begin{itemize}
\item Ourence, Vigo, London, Burlington, Essex - IBM's 5-qubit quantum systems, whose configuration maps have a graphical representation of figure \ref{fig:config_maps}a.
\item Yorktown - IBM's 5-qubit quantum system. Connectivity in the device is accomplished by two coplanar waveguide resonators with resonances of about 6.6 Ghz and 7 Ghz. Each of the qubits has its own waveguide for control and measurement. The configuration map has a graphical representation shown in figure \ref{fig:config_maps}b.
\item Melbourne - IBM quantum directional ladder system. The connectivity of the device is made up of 22 coplanar waveguide "bus" resonators, each of which is connected to 2 qubits, using 3 different configuration frequencies - 6.25 GHz, 6.45 GHz and 6.65 GHz. The configuration map has a graphical representation shown in figure \ref{fig:config_maps}c.
\end{itemize}
\subsection{Verification and validation of $C\sqrt[n]{X}$ quantum logic gate}
\label{subsec:results_CnX}
As described in \ref{sec:Applications} and the circuits shown in figures \ref{fig:nX} and \ref{fig:CnX_validation}, quantum circuits have been designed to verify and validate the $C\sqrt[n]{X}$ gate.
\begin{figure}[H]
	\centerline{\includegraphics[width=1\textwidth]{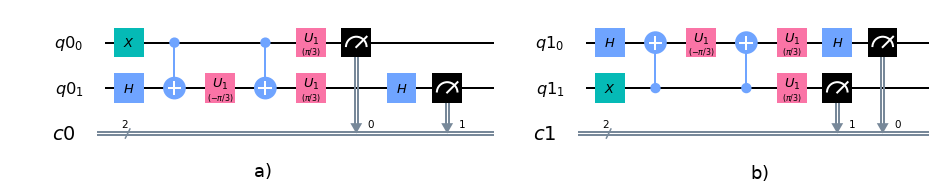}}
	\caption{Quantum circuits for gate verification and validation. \newline a) Circuit in which the controlled qubit is $q_1$. \newline b) Circuit with inverted qubit functions: $q_0$ is the objective and $q_1$ is the control.}
\label{fig:QISKIT_CnX}
\end{figure}
Both circuits, that are shown on figure \ref{fig:QISKIT_CnX} are applied to both a high-performance quantum simulator and the following real quantum processors in the IBM cloud platform:
\vspace{15mm}

\begin{itemize}
\item High-performance quantum simulator
\begin{figure}[H]
	\centerline{\includegraphics[width=1\textwidth]{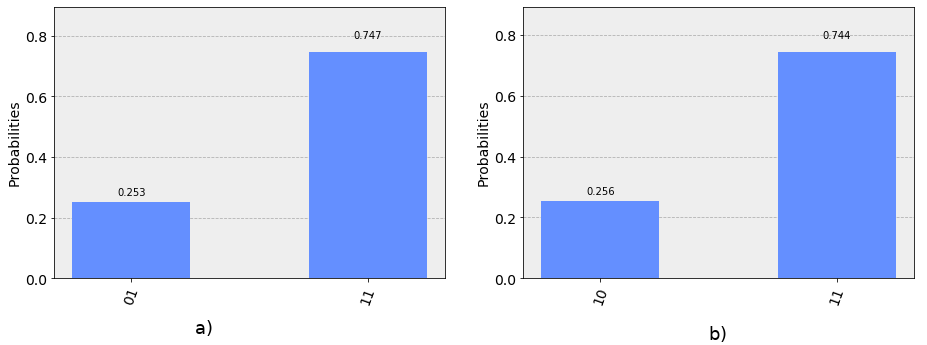}}
	\caption{Results from the implementation of $C\sqrt[n]{X}$ quantum gate validation and verification circuits on a high-performance quantum simulator.
\newline a) Results of circuit \ref{fig:QISKIT_CnX}a.
\newline b) Outcome of circuit \ref{fig:QISKIT_CnX}b}
\label{fig:QISKIT_CnX_simulator}
\end{figure}

\item Ourence (Avg. T1: 88.7, Avg. T2: 63.3)
\begin{figure}[H]
	\centerline{\includegraphics[width=1\textwidth]{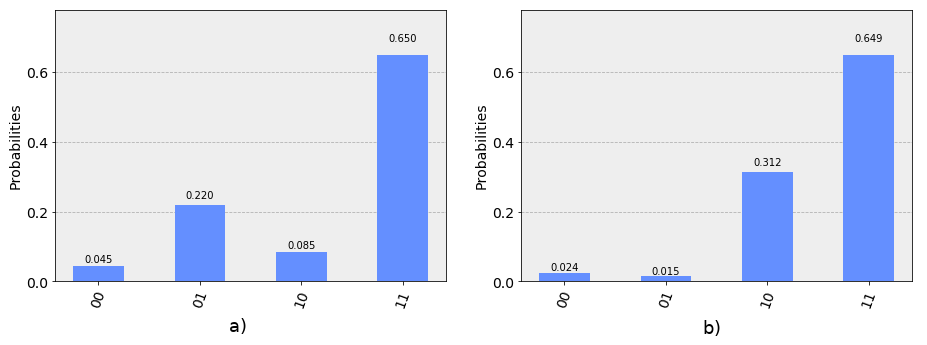}}
	\caption{Results from the implementation of $C\sqrt[n]{X}$ quantum gate validation and verification circuits on a real quantum processor Ourence.
\newline a) Results of circuit \ref{fig:QISKIT_CnX}a.
\newline b) Results of circuit \ref{fig:QISKIT_CnX}b
}
\label{fig:QISKIT_CnX_ourence}
\end{figure}
\vspace{15mm}

\item Vigo (Avg. T1: 99.1, Avg. T2: 72.5)
\begin{figure}[H]
	\centerline{\includegraphics[width=1\textwidth]{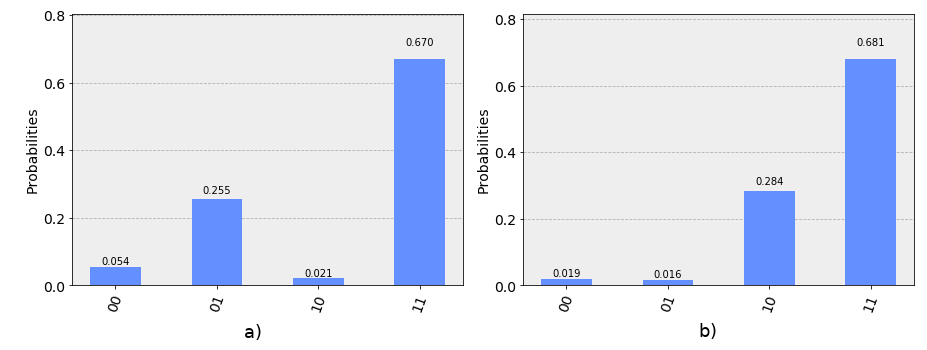}}
	\caption{ \newline Results from the implementation of $C\sqrt[n]{X}$ quantum gate validation and verification circuits on a real quantum processor Vigo.
\newline a) Results of circuit \ref{fig:QISKIT_CnX}a.
\newline b) Results of circuit \ref{fig:QISKIT_CnX}b}
\label{fig:QISKIT_CnX_vigo}
\end{figure}

\item London (Avg. T1: 66.1, Avg. T2: 71.7)
\begin{figure}[H]
	\centerline{\includegraphics[width=1\textwidth]{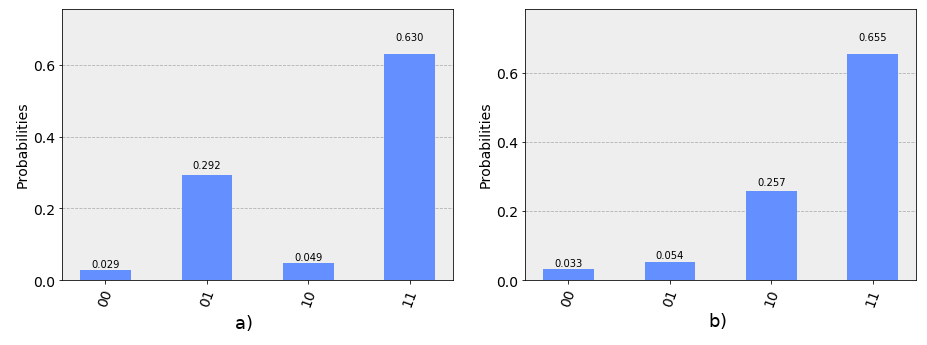}}
	\caption{Results from the implementation of $C\sqrt[n]{X}$ quantum gate validation and verification circuits on a real quantum processor London.
\newline a) Results of circuit \ref{fig:QISKIT_CnX}a.
\newline b) Results of circuit \ref{fig:QISKIT_CnX}b
}
\label{fig:QISKIT_CnX_london}
\end{figure}
\vspace{15mm}

\item Burlington (Avg. T1: 67.8, Avg. T2: 73.7)
\begin{figure}[H]
	\centerline{\includegraphics[width=1\textwidth]{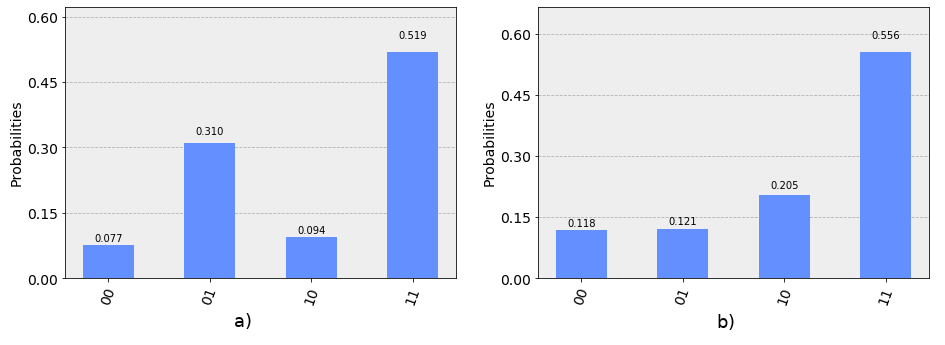}}
	\caption{Results from the implementation of $C\sqrt[n]{X}$ quantum gate validation and verification circuits on a real quantum processor Burlington.
\newline a) Results of circuit \ref{fig:QISKIT_CnX}a.
\newline b) Results of circuit \ref{fig:QISKIT_CnX}b}
\label{fig:QISKIT_CnX_burlington}
\end{figure}

\item Essex (Avg. T1: 118.5, Avg. T2: 81.6) \newline
\begin{figure}[H]
	\centerline{\includegraphics[width=1\textwidth]{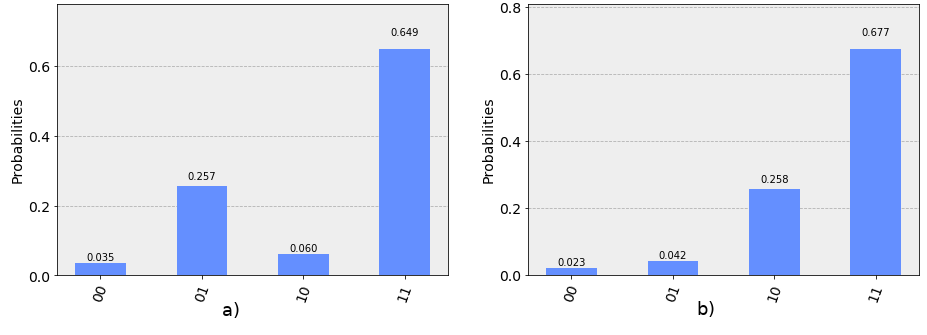}}
	\caption{Results from the implementation of $C\sqrt[n]{X}$ quantum gate validation and verification circuits on a real quantum processor Essex.
\newline a) Results of circuit \ref{fig:QISKIT_CnX}a.
\newline b) Results of circuit \ref{fig:QISKIT_CnX}b
}
\label{fig:QISKIT_CnX_essex}
\end{figure}
\vspace{15mm}

\item{Yorktown (Avg. T1: 56.7, Avg. T2: 47.6)
\begin{figure}[H]
	\centerline{\includegraphics[width=1\textwidth]{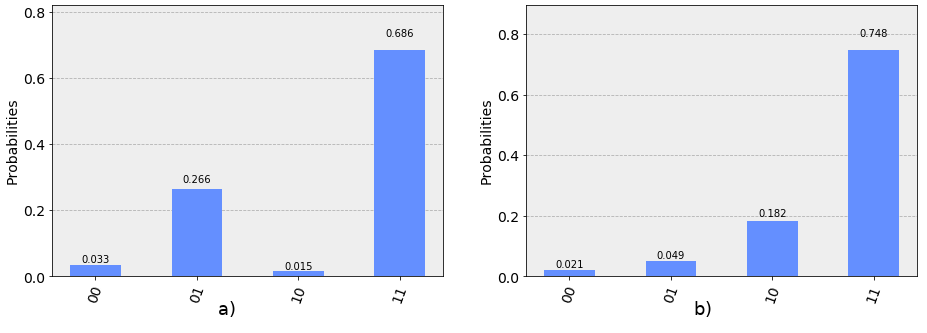}}
	\caption{Results from the implementation of $C\sqrt[n]{X}$ quantum gate validation and verification circuits on a real quantum processor Yorktown.
\newline a) Results of circuit \ref{fig:QISKIT_CnX}a.
\newline b) Results of circuit \ref{fig:QISKIT_CnX}b}
\label{fig:QISKIT_CnX_yorktown}
\end{figure}}

\item Melbourne (Avg. T1: 51.7, Avg. T2: 71.9)
\begin{figure}[H]
	\centerline{\includegraphics[width=1\textwidth]{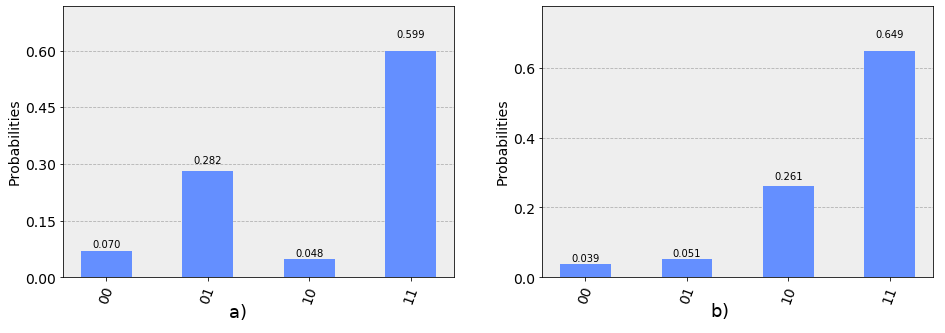}}
	\caption{Results from the implementation of $C\sqrt[n]{X}$ quantum gate validation and verification circuits on a real quantum processor Melbourne.
\newline a) Results of circuit \ref{fig:QISKIT_CnX}a.
\newline b) Results of circuit \ref{fig:QISKIT_CnX}b
\newline
\newline
\newline}
\label{fig:QISKIT_CnX_melbourne}
\end{figure}
\end{itemize}

\subsection{Results of quantum simulation of a binary homogeneous Markov process on various hardware back-ends}
\label{subsec:results_markov}

The quantum circuit used to simulate the binary homogeneous Markov process explained in \ref{sec:Applications} shown in figure \ref{fig:quantum_Markov_absorbing} is implemented in two variants - the CNOT gates are respectively \{CX\_01, CX\_12\} and \{CX\_21, CX\_10\}. The two circuits are as follows:

\begin{figure}[!h]
	\centerline{\includegraphics[width=1\textwidth]{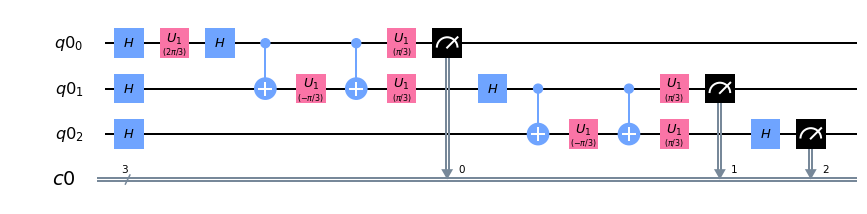}}
	\caption{Quantum circuit for simulation of a binary homogeneous Markov process with absorbin state in $\ket{0}$ on a high-performance quantum simulator and real quantum processors in IBM cloud platform with CNOT gates directions \{CX\_21, CX\_10\} respectively.}
\label{fig:QISKIT_Markov_1}
\end{figure}

\begin{figure}[!h]
	\centerline{\includegraphics[width=1\textwidth]{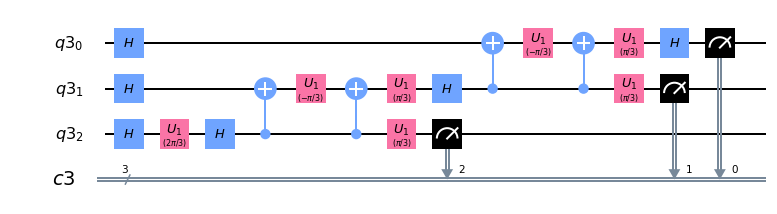}}
	\caption{Quantum circuit for simulation of a binary homogeneous Markov process with absorbing state in $\ket{0}$ on a high-performance quantum simulator and real quantum processors in IBM cloud platform with CNOT gates directions \{CX\_01, CX\_12\} respectively.}
\label{fig:QISKIT_Markov_2}
\end{figure}

The two quantum circuits \ref{fig:QISKIT_Markov_1} and \ref{fig:QISKIT_Markov_2} have been executed both on high-performance simulator and several real quantum processors in the cloud platform of IBM. The results are as follows:

\begin{itemize}
\item High-performance quantum simulator
\begin{figure}[H]
	\centerline{\includegraphics[width=1\textwidth]{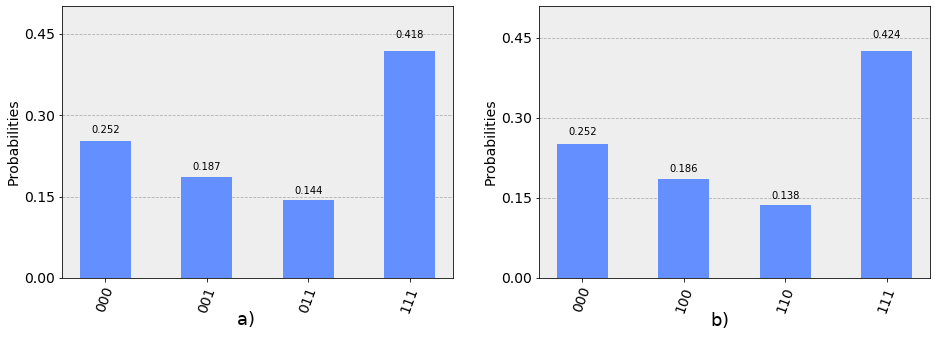}}
	\caption{Results of the execution of quantum circuits for the simulation of a homogeneous binary Markov process with an absorbing state in $\ket{0}$ on a high-performance quantum simulator.
\newline a) Results of circuit \ref{fig:QISKIT_Markov_1}.
\newline b) Outcome of circuit \ref{fig:QISKIT_Markov_2}.}
\label{fig:QISKIT_Markov_simulator}
\end{figure}

\item Ourence (Avg. T1: 88.7, Avg. T2: 63.3)
\begin{figure}[H]
	\centerline{\includegraphics[width=1\textwidth]{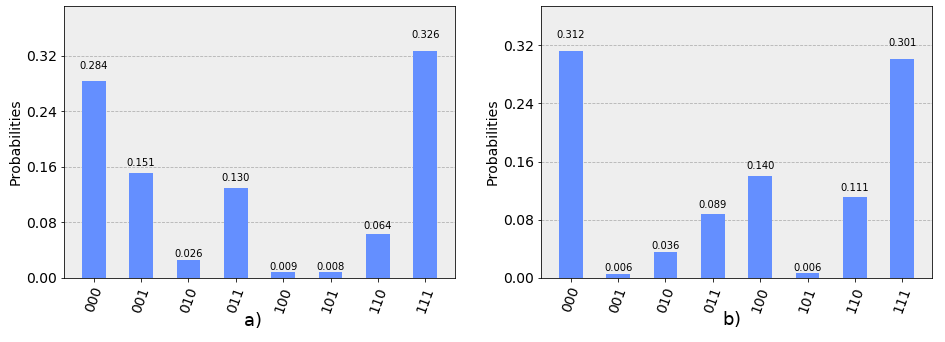}}
	\caption{Results of the execution of quantum circuits for the simulation of a homogeneous binary Markov process with an absorbing state in $\ket{0}$ on a real quantum processor Ourence.
\newline a) Results of circuit \ref{fig:QISKIT_Markov_1}.
\newline b) Outcome of circuit \ref{fig:QISKIT_Markov_2}.}
\label{fig:QISKIT_Markov_ourence}
\end{figure}

\item Vigo (Avg. T1: 99.1, Avg. T2: 72.5)
\begin{figure}[H]
	\centerline{\includegraphics[width=1\textwidth]{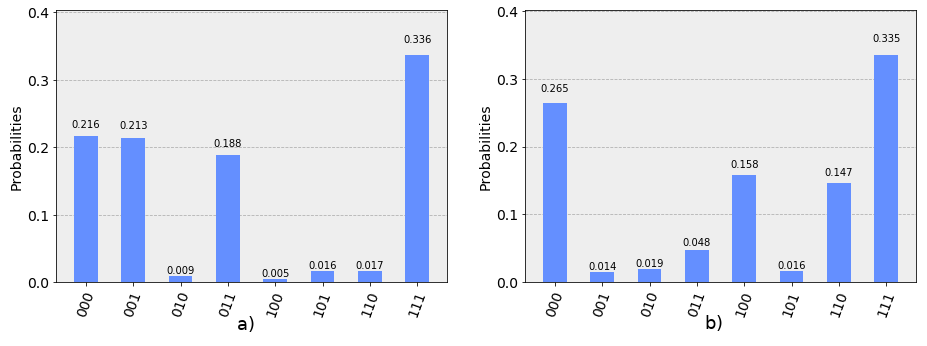}}
	\caption{Results of the execution of quantum circuits for the simulation of a homogeneous binary Markov process with an absorbing state in $\ket{0}$ on a real quantum processor Vigo.
\newline a) Results of circuit \ref{fig:QISKIT_Markov_1}.
\newline b) Outcome of circuit \ref{fig:QISKIT_Markov_2}.}
\label{fig:QISKIT_Markov_vigo}
\end{figure}

\item London (Avg. T1: 66.1, Avg. T2: 71.7)
\begin{figure}[H]
	\centerline{\includegraphics[width=1\textwidth]{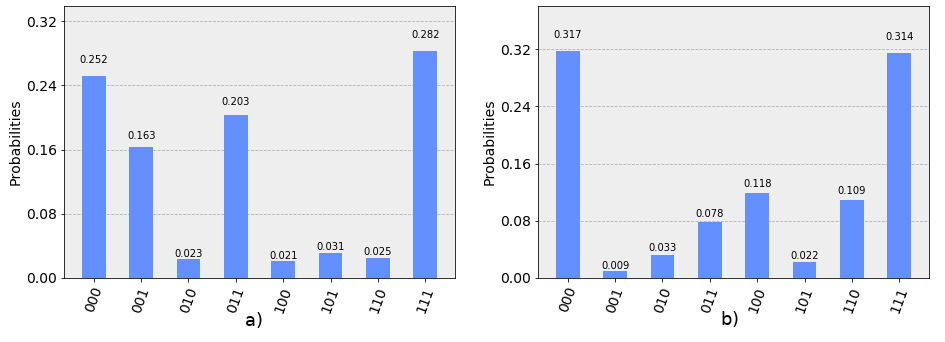}}
	\caption{Results of the execution of quantum circuits for the simulation of a homogeneous binary Markov process with an absorbing state in $\ket{0}$ on a real quantum processor London.
\newline a) Results of circuit \ref{fig:QISKIT_Markov_1}.
\newline b) Outcome of circuit \ref{fig:QISKIT_Markov_2}.
}
\label{fig:QISKIT_Markov_london}
\end{figure}
\vspace{15mm}
\item Burlington (Avg. T1: 67.8, Avg. T2: 73.7)
\begin{figure}[H]
	\centerline{\includegraphics[width=1\textwidth]{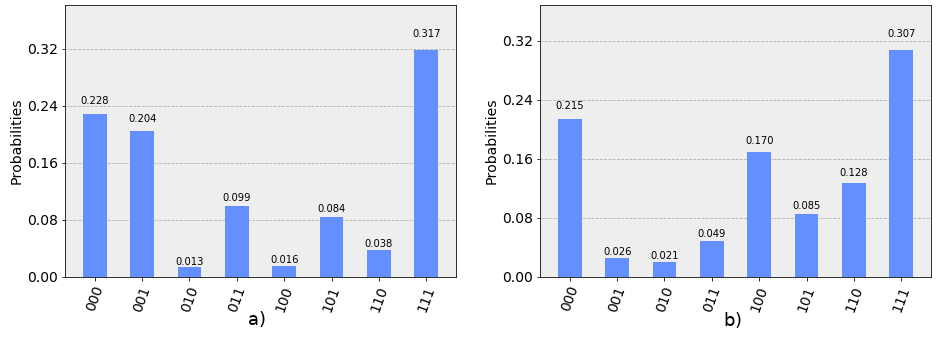}}
	\caption{Results of the execution of quantum circuits for the simulation of a homogeneous binary Markov process with an absorbing state in $\ket{0}$ on a real quantum processor Burlington.
\newline a) Results of circuit \ref{fig:QISKIT_Markov_1}.
\newline b) Outcome of circuit \ref{fig:QISKIT_Markov_2}.}
\label{fig:QISKIT_Markov_burlington}
\end{figure}

\item Essex (Avg. T1: 118.5, Avg. T2: 81.6)
\begin{figure}[H]
	\centerline{\includegraphics[width=1\textwidth]{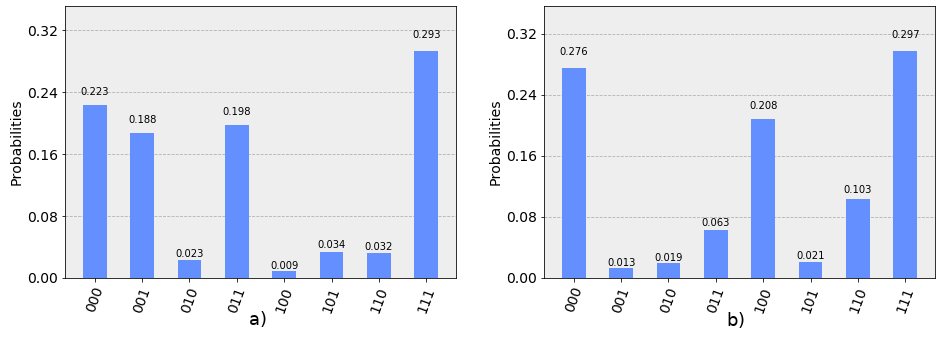}}
	\caption{Results of the execution of quantum circuits for the simulation of a homogeneous binary Markov process with an absorbing state in $\ket{0}$ on a real quantum processor Essex.
\newline a) Results of circuit \ref{fig:QISKIT_Markov_1}.
\newline b) Outcome of circuit \ref{fig:QISKIT_Markov_2}.}
\label{fig:QISKIT_Markov_essex}
\end{figure}
\vspace{15mm}

\item Yorktown (Avg. T1: 56.7, Avg. T2: 47.6)
\begin{figure}[H]
	\centerline{\includegraphics[width=1\textwidth]{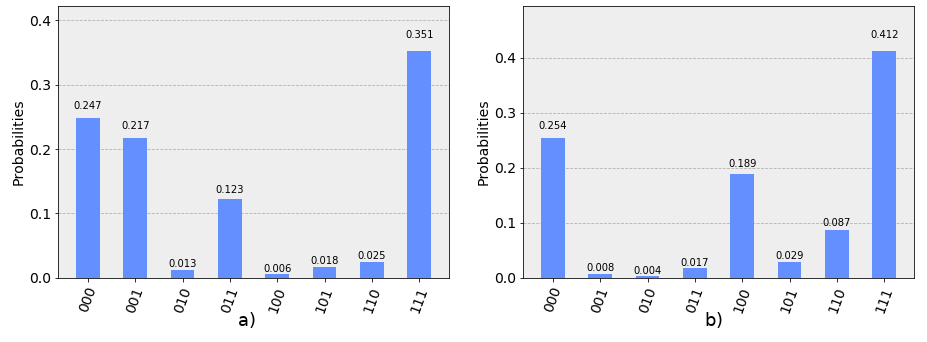}}
	\caption{Results of the execution of quantum circuits for the simulation of a homogeneous binary Markov process with an absorbing state in $\ket{0}$ on a real quantum processor Yorktown.
\newline a) Results of circuit \ref{fig:QISKIT_Markov_1}.
\newline b) Outcome of circuit \ref{fig:QISKIT_Markov_2}.}
\label{fig:QISKIT_Markov_yorktown}
\end{figure}

\item Melbourne (Avg. T1: 51.7, Avg. T2: 71.9)
\begin{figure}[H]
	\centerline{\includegraphics[width=1\textwidth]{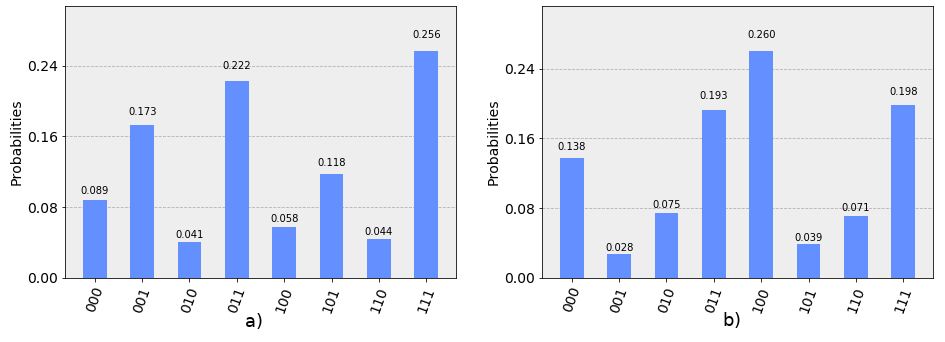}}
	\caption{Results of the execution of quantum circuits for the simulation of a homogeneous binary Markov process with an absorbing state in $\ket{0}$ on a real quantum processor Melbourne.
\newline a) Results of circuit \ref{fig:QISKIT_Markov_1}.
\newline b) Outcome of circuit \ref{fig:QISKIT_Markov_2}.}
\label{fig:QISKIT_Markov_melbourne}
\end{figure}
\vspace{15mm}
\end{itemize}

\subsection{Analysis of the results}
\label{subsec:analysis}
\par
Since quantum computers are still in their early stage of development, the qubits (observed environment) interact with the (unobserved) environment, which causes the system and its environment to become entangled driving decoherence of the system. This decoherence can be reversed in theory by applying a recovery transformation, which entangles the data with ancilla. The ancilla introduced can be discarded and replaced with another fresh ancilla for any future rounds in the error correction. Another obstacle on the road of mass usage of quantum computers is the limited lifetime of a qubit. Usually the best qubits are available for much less than a second due to this decoherence. Large-scale quantum algorithms need qubit lifetime in order of hours or even days. 
\par
The approaches that are used in IBM's platform for quantum computing could be classified in two classes - hardware improvement for better qubits (reduce noise coming from the environment, build tools for better qubit control, etc.) and error correction techniques. Despite the enormous improvements in the hardware in the past decade, there are a lot of limitations of this approach. On the other hand, the quantum error correction is a relatively new approach, which allows researchers and engineers to achieve better performance of the quantum computers. The techniques used here are redundant encoding of information, diagnose and correct errors, and in theory - unlimited qubit lifetime. 
\par \textbf{Redundant encoding} is a technique for encoding information from one qubit into many qubits, which allows the diagnostics and correction of quantum errors caused by decoherence. The redundancy was first introduced by Peter Shor in \citep{Shor1995}, where he reduces the effect of decoherence by assuming that the decoherence acts independently on each qubit in the quantum system. Those quantum codes introduced by Shor find applications in many areas of quantum research - better and scalable quantum computers (Extend qubit lifetime and improve precision of quantum logic gates), quantum Shannon theory (Reliable transmission of qubits through noisy quantum channels), topological quantum order (Explain robustness of ground state degeneracy to local pertubations), quantum gravity and more (Holographic quantum codes).
\par 
Quantum error correction could simply be explained this way:
\par $\ket{\Psi} \rightarrow Encoding \ket{\overline{\Psi} } \rightarrow Error \rightarrow Decoding \ket{\Psi} \rightarrow \ket{\Psi}$
\par and one of main tasks here is how to map this (one qubit) state $\ket{\Psi}$ to multi qubit. The actual error correction and diagnosis happens in the decoding state, where from multiple $\ket{\overline{\Psi}}$ states, the result is a single  $\ket{\Psi}$ state.
\begin{itemize}
\item \textbf{Encoding}
The task here is to encode k logical qubits into n physical qubits and it could basically be represented as a linear subspace of orthonormal basis. The k and n define the dimension of the subspace. 
\par $C = (\mathbb{C}^2)^{\otimes n}$
\par $dim(C) = 2^k$
\par Where the codespace is C, the codeword is any state $ \ket{\overline{\Psi} } \in C$, and the error is n-qubit linear operator, which might not be unitary.
\par 
\item \textbf{Error} The conditions to detect an error E with quantum code are:
\par 
$\begin{cases}
\bra{\overline{x}} E \ket{\overline{y}} = 0 \\
\bra{\overline{x}} E \ket{\overline{x}} = \bra{\overline{y}} E \ket{\overline{y}}
\end{cases}$
for all $ x \neq y$
\par * All diagonal matrix elements must be zeroes. And all diagonal matrix elements must be the same. 

\item \textbf{Decoding} The quantum code C can detect an error E if and only if:
\par $\Pi_C E \Pi_C = \eta \Pi_C$
\par The $\eta$ is a complex number.
\par The $\Pi_C$ is the codespace projector.

\end{itemize}
\par 

\par
The relations in quantum mechanical systems are part of the foundational features in quantum physics - non-linearity and Bell's inequality violations. As shown in \citep{Spehner2017}, the geometric discord and the discord of response with Hellinger distance are reliable enough and offer non-computational expensive measures. 
If a bipartite quantum system AB is being considered, formed by putting together two quantum systems - A and B, with a finite dimensional Hilbert space $\mathcal{H}_{AB}$:
\par $n_a = dim(\mathcal{H}_{A}) < \infty$ and $ n_b = dim(\mathcal{H}_{B}) < \infty$. 
\par A state of AB is then given by a density matrix $\rho$, which is a non-negative operator on $\mathcal{H}_{AB}$ and with unit trace tr $\rho$ = 1. 
\par If there is a finite sample space ${1,2,...,n}$ and the $\mathcal{E}_{clas} = {p \in \mathcal{R}_{+}^n ; \sum{p_k} = 1}$. The classical Hellinger distance $d_{clas}$ on $\mathcal{E}_{clas}$ is:
\par $d_{clas}(\rho,\sigma) = \left( \sum_{k=1}^{n}{ \left( \sqrt{\rho_k} - \sqrt{\sigma_k} \right)^2 } \right)^{1/2} $
\par
In general, even with the current hardware solutions for quantum processors, two-qubit gates are possible for realization between two adjacent qubits that are connected via a superconducting bus resonator. IBM hardware uses cross-resonance impact as the basis for the CNOT gate, a higher frequency control and a lower frequency target. The functions of the two qubits must be reversed if a degenerate effect on the qubit playing the target is observed, and if it is connected to a third qubit that has the same or higher frequency. Due to the possibility of these exceptions, the quantum circuit \ref{fig:quantum_Markov_absorbing} was implemented in two variants - \ref{fig:QISKIT_Markov_1} and \ref{fig:QISKIT_Markov_2}, respectively, with two-qubit gates for figure \ref{fig:QISKIT_Markov_1} - \{CX_01\}, \{CX_12\}, and for figure \ref{fig:QISKIT_Markov_1} - \{CX_21, CX_10\}.
\par 
In quantum information theory, fidelity is a measure of the "closeness" between two quantum-mechanical states. Criteria for fidelity and distances between probability distributions are a topic of study in both classical and quantum information sciences. Hellinger's fidelity is the main measure adopted to study the proximity of two quantum distributions.
\par 
Let's $P = \{pi\} i \in [n], Q = \{qi\} i \in [n]$ are two probability distributions, then the Hellinger distance will have the following look:
\begin{equation}
h(P,Q) = {1 \over \sqrt{2}} \left| \sqrt{P} - \sqrt{Q} \right|_2
\label{eq:29}
\end{equation}
Where $h(P,Q) \leq 1$ for all distributions, and the fidelity is:
\begin{equation}
F[P,Q] = 1- h(P,Q)
\label{eq:30}
\end{equation}

\par
Increasing the number of execution for each quantum circuit on the respective IBM quantum systems does not lead to significant error mitigation. The experiments performed are at the maximum value of the input parameter for the number of executions - namely 8192 executions.

\begin{table}[H]
\centering
\begin{tabular}{ |p{3.5cm}||p{5cm}|p{5cm}|  }
 \hline
 IBM Q Backend & Quantum Circuit on fig. \ref{fig:QISKIT_Markov_1} & Quantum Circuit on fig. \ref{fig:QISKIT_Markov_2} \\
 \hline
 Burlington & 0.71475	& 0.689795 \\
 Essex & 0.760052 &	0.744267 \\
 London & 0.753879 & 0.714005 \\
 Melbourne & 0.595152 & 0.546999 \\
 Ourense & 0.756185 & 0.721887 \\
 Vigo & 0.832508 & 0.773089 \\
 Yorktown & 0.818702 & 0.821362 \\
 \hline
\end{tabular}
\caption{Hellinger fidelity calculated from the distance between the high-performance quantum simulator and various hardware backends in IBM cloud platform for the quantum circuits shown on figures \ref{fig:QISKIT_Markov_1} and \ref{fig:QISKIT_Markov_2}.}

\end{table}

\begin{figure}[H]
	\centerline{\includegraphics[width=1\textwidth]{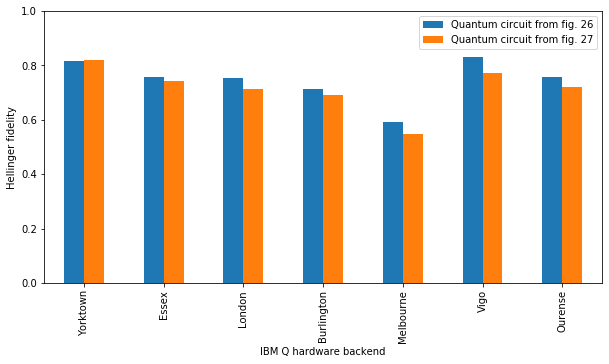}}
	\caption{Hellinger fidelity calculated from the distance between the high-performance quantum simulator and various hardware backends in IBM cloud platform for the quantum circuits shown on figures \ref{fig:QISKIT_Markov_1} and \ref{fig:QISKIT_Markov_2}.}
\label{fig:Hellinger_markov}
\end{figure}
\vspace{3cm}
The quantum gates construction and the simulation of a binary homogeneous Markov process on various hardware backends have been verified and validated. It has been shown that creating a quantum gate with the basic gates available on multiple platforms is possible, although the hardware devices are still too noisy and this results in a significant increase in measurement error. \\
In the $C\sqrt[n]{X}$ gate implementation, IBM's 5-qubit quantum T-shaped systems, along with the 14-qubit Melbourne quantum system, perform significantly worse than the 5-qubit Yorktown processor with a "bow-tie" configuration map. \\
In the simulation of a homogeneous binary Markov model, the best performing quantum processors are Vigo and Yorktown. Improving accuracy (~ 0.82) is achieved by using circuit from figure \ref{fig:QISKIT_Markov_1}, and this is fully expected since, as mentioned at the beginning of this chapter, this scheme is applicable when there are no exceptions to the rule for determining the direction of the gate and the gate direction should not be reversed - there are no degenerative effects on the target qubit as a result of the higher frequency of the control. \\
Choosing the right quantum circuit based on the available hardware backends, as well as choosing a hardware device with the number of qubits as close as possible to the simulation required, would be the right approach to achieve maximum fidelity.

\markboth{CONTRIBUTIONS}{Scientific and Scientifically Applied Contributions}
\section{Scientific and Scientifically Applied Contributions}
\label{subsec:Scientific and Scientifically Applied Contributions}

\par A method has been developed to find the complete set of possible paths of a stochastic process, described by a homogeneous binary Markov model with the apparatus of quantum computational operations.
\label{contribution:1}

\par An algorithm for determining the phase shift of the qubit along the X axis is developed, depending on the desired probability distribution for the amplitudes of its states, with application in the modelling of stochastic processes through the apparatus of quantum computational operations.
\label{contribution:2}

\par Based on the method developed to find the complete set of possible random process paths, described by a binary homogeneous Markov model, an algorithm applicable to real quantum processors has been synthesized (a test version of the algorithm is implemented on quantum processors in the IBM Quantum Experience platform).
\label{contribution:3}

\par Software has been developed in the Jupyter Notebook environment with Python 3.7 software kernel, which implements the phase shifting algorithm and its applicability in comparison to the solutions used so far.
\label{contribution:4}

\par A quantum circuit has been developed to allow modelling on a real quantum processor of any stochastic problem described by a binary homogeneous Markov model.
\label{contribution:5}

\par An analysis of the scientific sectors in quantum information theory and a classification, including 8 technological clusters with the definition of perspective and key problems have been made.
\label{contribution:6}
 
\appendix 
\section{Appendix A}
\label{sec:Appendix A}
Author's work

\renewcommand{\labelenumii}{\roman{enumii}}
\begin{enumerate}
\item Petar N. Nikolov,Vassil T. Galabov, "Experimental realization of Controlled Square Root of Z Gate Using IBM`s Cloud Quantum Experience Platform" in FDIBA Conference Proceedings, vol. 1, dec. 2017, pp. 11-13
\label{publications:fdiba2017}

\item Nikolov, P. (2019). Controlled nth root of X gate on a real quantum computer. PROCEEDINGS OF THE 45TH INTERNATIONAL CONFERENCE ON APPLICATION OF MATHEMATICS IN ENGINEERING AND ECONOMICS (AMEE'19). PROCEEDINGS OF THE 45TH INTERNATIONAL CONFERENCE ON APPLICATION OF MATHEMATICS IN ENGINEERING AND ECONOMICS (AMEE'19). 
\newline https://doi.org/10.1063/1.5133583  
\label{publications:amee1}

\item Nikolov, P., Galabov, V. (2019). Markov process simulation on a real quantum computer. PROCEEDINGS OF THE 45TH INTERNATIONAL CONFERENCE ON APPLICATION OF MATHEMATICS IN ENGINEERING AND ECONOMICS (AMEE'19). PROCEEDINGS OF THE 45TH INTERNATIONAL CONFERENCE ON APPLICATION OF MATHEMATICS IN ENGINEERING AND ECONOMICS (AMEE'19). 
\newline https://doi.org/10.1063/1.5133584 
\label{publications:amee2}

\item Petar Nikolov, Vassil Galabov, "The Journey of Quantum Information Technology," in FDIBA Conference Proceedings, vol. 3, Nov. 2019, pp. 11-25
\label{publications:fdiba2019}
\end{enumerate} 
\section{Appendix B}
\label{sec:Appendix B}
Here is attached a sample code in the form of a Jupyter Notebook used for execution of the quantum circuits shown on figure \ref{fig:QISKIT_Markov_1} and figure \ref{fig:QISKIT_Markov_2}. The results here are not the ones used in the dissertation, but as mentioned - this is a sample code, and should be used for experiment replication. 



\section*{Supplementary material}

```
[1]: %matplotlib inline
     import qiskit
     import matplotlib.pyplot as plt
     import numpy as np
     import math
     # Importing standard Qiskit libraries and configuring account
     from qiskit import QuantumCircuit, execute, Aer, IBMQ, BasicAer
     from qiskit.compiler import transpile, assemble
     from qiskit.visualization import *
     # Loading your IBM Q account(s)
     provider = IBMQ.load_account()
```

/home/hp/anaconda3/lib/python3.7/site-packages/qiskit/providers/models/backendconfiguration.py:337: UserWarning: `dt` and `dtm` now have units of seconds(s) rather than nanoseconds(ns).
  warnings.warn('`dt` and `dtm` now have units of seconds(s) rather '

```
[2]: qiskit.__qiskit_version__
```

```
[2]: {'qiskit-terra': '0.11.0',
      'qiskit-aer': '0.3.4',
      'qiskit-ignis': '0.2.0',
      'qiskit-ibmq-provider': '0.4.4',
      'qiskit-aqua': '0.6.1',
      'qiskit': '0.14.0'}
```

```
[3]: qiskit.tools.monitor.backend_overview()
```

```
ibmq_rome                   ibmq_essex                  ibmq_burlington
---------                   ----------                  ---------------
Num. Qubits:   5            Num. Qubits:   5            Num. Qubits:   5
Pending Jobs:  5            Pending Jobs:  4            Pending Jobs:  0
Least busy:    False        Least busy:    False        Least busy:    True
Operational:   True         Operational:   True         Operational:   True
Avg. T1:       59.0         Avg. T1:       85.2         Avg. T1:       79.9
Avg. T2:       88.9         Avg. T2:       123.9        Avg. T2:       81.5


ibmq_london                 ibmq_16_melbourne           ibmqx2
-----------                 -----------------           ------
Num. Qubits:   5            Num. Qubits:   15           Num. Qubits:   5
Pending Jobs:  1            Pending Jobs:  8            Pending Jobs:  5
Least busy:    False        Least busy:    False        Least busy:    False
Operational:   True         Operational:   True         Operational:   True
Avg. T1:       70.5         Avg. T1:       55.0         Avg. T1:       58.3
Avg. T2:       78.5         Avg. T2:       63.8         Avg. T2:       61.9
```

```
              ibmq_vigo                        ibmq_ourense                    ibmq_armonk
              ---------                        ------------                    -----------
Num. Qubits:     5                Num. Qubits:     5                Num. Qubits:     1
Pending Jobs:    6                Pending Jobs:    1                Pending Jobs:    0
Least busy:      False            Least busy:      False            Least busy:      False
Operational:     False            Operational:     False            Operational:     False
Avg. T1:         108.4            Avg. T1:         117.0            Avg. T1:         103.9
Avg. T2:         72.5             Avg. T2:         85.2             Avg. T2:         119.7
```

[49]:
```python
q = qiskit.QuantumRegister(3)
c = qiskit.ClassicalRegister(3)
qc = qiskit.QuantumCircuit(q, c)
qc1 = qiskit.QuantumCircuit(q, c)
shots = 8192
credits = 5
```

[5]:
```python
x = 0.25
phase = math.acos(2*x-1)
phase
#phase = 2.793427 # 3% |0> and 97% |1>
```

[5]: 2.0943951023931957

[50]:
```python
qc.h(q[0])
qc.u1(phase,q[0])
qc.h(q[0])

qc.h(q[1])
qc.cx(q[0],q[1])
qc.u1(-phase/2, q[1])
qc.cx(q[0],q[1])
qc.u1(phase/2, q[0])
qc.u1(phase/2, q[1])
qc.h(q[1])

qc.h(q[2])
qc.cx(q[1],q[2])
qc.u1(-phase/2, q[2])
qc.cx(q[1],q[2])
qc.u1(phase/2, q[1])
qc.u1(phase/2, q[2])
qc.h(q[2])

qc.measure(q,c)
```

[50]: `<qiskit.circuit.instructionset.InstructionSet at 0x7f604942b3c8>`

[51]: `qc.draw(line_length=180, output='mpl')`

[51]:

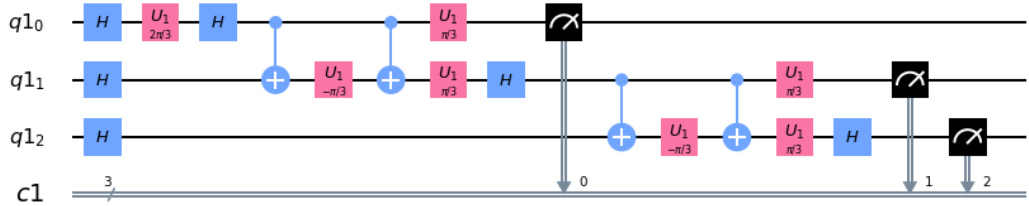

[52]:
```
qc1.h(q[2])
qc1.u1(phase,q[2])
qc1.h(q[2])

qc1.h(q[1])
qc1.cx(q[2],q[1])
qc1.u1(-phase/2, q[1])
qc1.cx(q[2],q[1])
qc1.u1(phase/2, q[2])
qc1.u1(phase/2, q[1])
qc1.h(q[1])

qc1.h(q[0])
qc1.cx(q[1],q[0])
qc1.u1(-phase/2, q[0])
qc1.cx(q[1],q[0])
qc1.u1(phase/2, q[1])
qc1.u1(phase/2, q[0])
qc1.h(q[0])

qc1.measure(q,c)
```

[52]: `<qiskit.circuit.instructionset.InstructionSet at 0x7f6049303390>`

[53]: `qc1.draw(line_length=180, output='mpl')`

[53]:

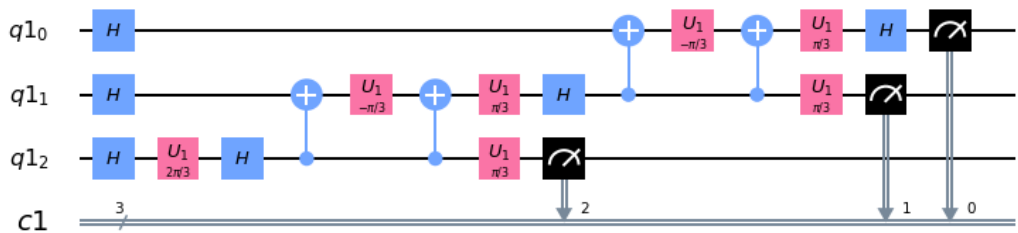

[8]: ```
golden_label_counts = {'111': 0.75*0.75*0.75*8192, '010': 0, '101': 0, '110': 0, '001': 0.75*0.25*8192, '100': 0, '000': 0.25*8192, '011': 0.75*0.75*0.25*8192}
```

[58]: ```
golden_label_counts_1 = {'111': 0.75*0.75*0.75*8192, '010': 0, '101': 0, '110': 0.75*0.75*0.25*8192, '001': 0, '100': 0.75*0.25*8192, '000': 0.25*8192, '011': 0}
```

[61]: ```
backend = BasicAer.get_backend('qasm_simulator')
job = qiskit.execute(qc, backend = backend, shots=shots, max_credits=credits)
qasm_simulator_result = job.result()
qasm_simulator_counts = qasm_simulator_result.get_counts()
qiskit.visualization.plot_histogram(qasm_simulator_result.get_counts())
```

[61]:

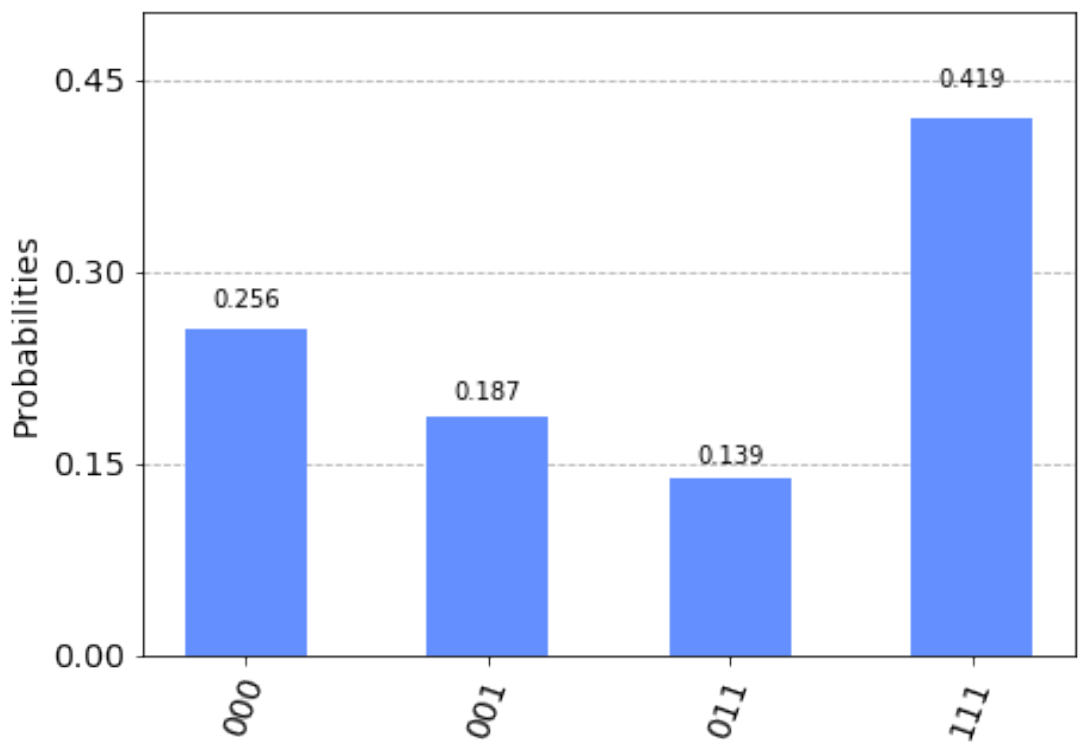

[11]: ```
qasm_simulator_hellinger_fidelity = qiskit.quantum_info.hellinger_fidelity(golden_label_counts, qasm_simulator_result.get_counts(qc))
qasm_simulator_hellinger_fidelity
```

[11]: 0.9940876906816659

[56]: ```
backend = BasicAer.get_backend('qasm_simulator')
job = qiskit.execute(qc1, backend = backend, shots=shots, max_credits=credits)
qasm_simulator_result_1 = job.result()
```

```python
qasm_simulator_counts_1 = qasm_simulator_result_1.get_counts()
qiskit.visualization.plot_histogram(qasm_simulator_counts_1)
```

[56]:

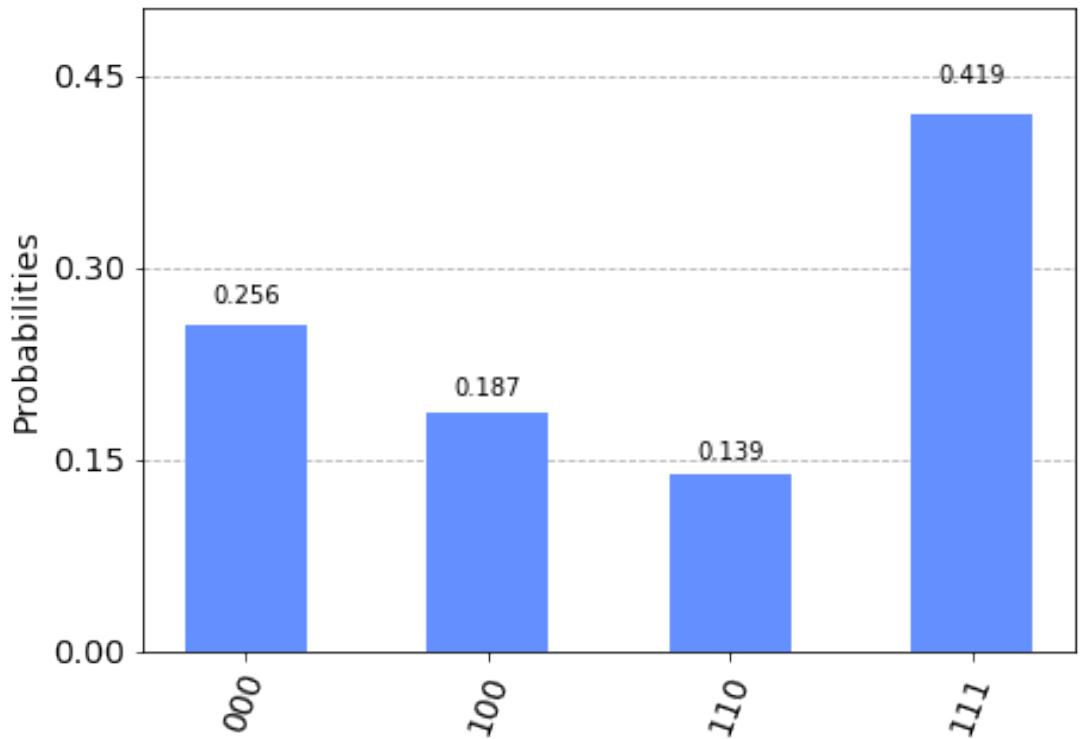

[59]:
```python
qasm_simulator_hellinger_fidelity_1 = qiskit.quantum_info.
 ↪hellinger_fidelity(golden_label_counts_1, qasm_simulator_counts_1)
qasm_simulator_hellinger_fidelity_1
```

[59]: 0.9951888822463194

[ ]:

[ ]:

[ ]:
```python
backend = provider.get_backend('ibmqx2')
job = qiskit.execute(qc, backend = backend, shots=shots, max_credits=credits)
ibmqx2_result = job.result()
ibmqx2_counts = ibmqx2_result.get_counts(qc)
```

[14]:
```python
print(ibmqx2_counts)
qiskit.visualization.plot_histogram(ibmqx2_counts)
```

```
{'001': 1350, '101': 191, '110': 285, '111': 2756, '010': 116, '000': 2294,
'011': 1159, '100': 41}
```

[14]:

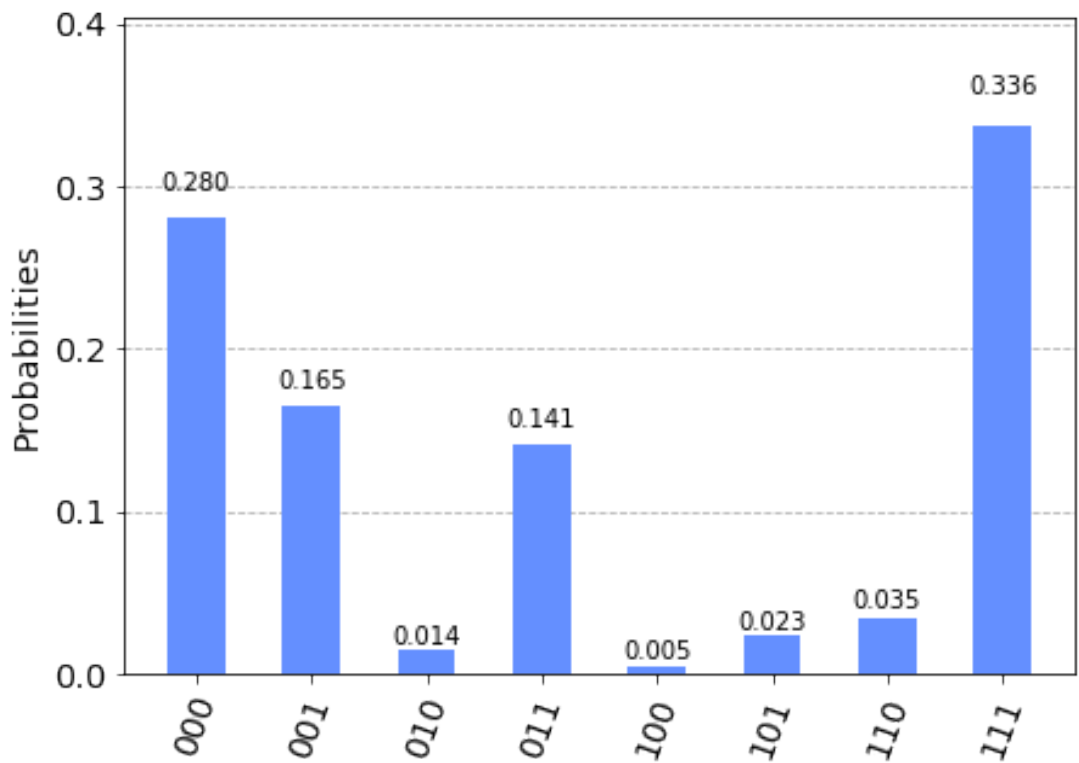

```
[62]: ibmqx2_hellinger_fidelity = qiskit.quantum_info.
       ↪hellinger_fidelity(qasm_simulator_counts, ibmqx2_counts)
      ibmqx2_hellinger_fidelity
```

[62]: 0.7962356868446028

```
[63]: backend = provider.get_backend('ibmqx2')
      job = qiskit.execute(qc1, backend = backend, shots=shots, max_credits=credits)
      ibmqx2_result_1 = job.result()
      ibmqx2_counts_1 = ibmqx2_result_1.get_counts(qc1)
```

```
[64]: print(ibmqx2_counts_1)
      qiskit.visualization.plot_histogram(ibmqx2_counts_1)
```

    {'001': 80, '101': 278, '110': 1032, '111': 2672, '010': 380, '000': 2120,
    '011': 201, '100': 1429}

[64]:

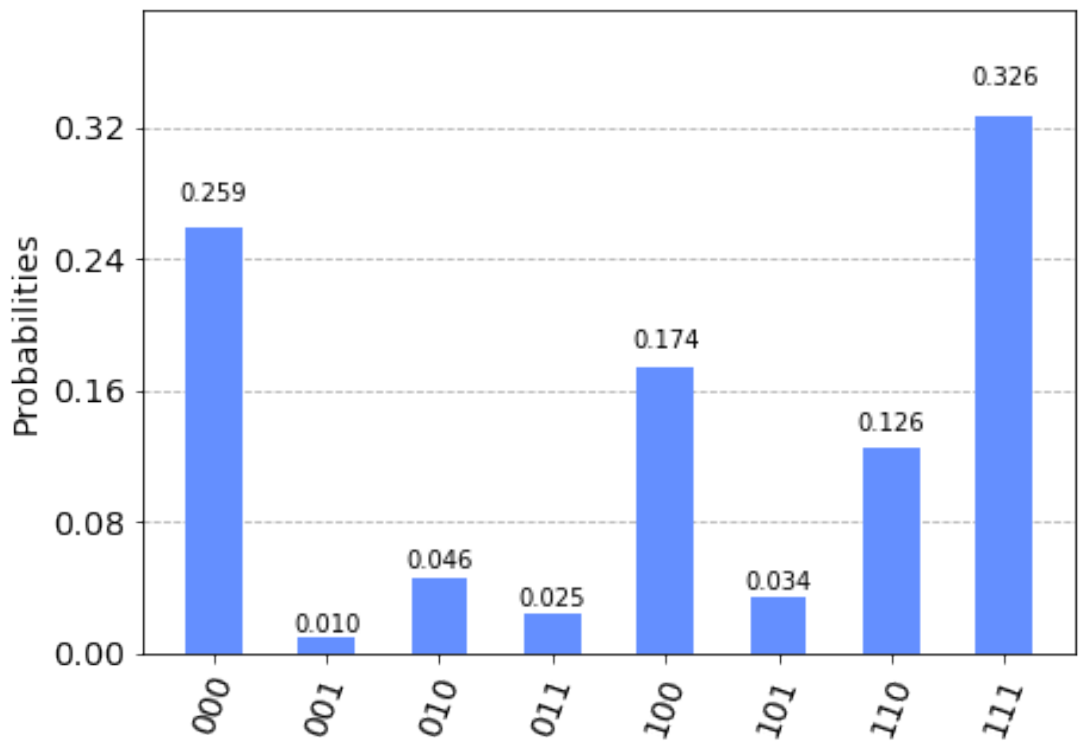

```
[65]: ibmqx2_hellinger_fidelity_1 = qiskit.quantum_info.
      ↪hellinger_fidelity(qasm_simulator_counts_1, ibmqx2_counts_1)
      ibmqx2_hellinger_fidelity_1
```

[65]: 0.7541032789630906

[ ]:

[ ]:

```
[16]: backend = provider.get_backend('ibmq_essex')
      job = qiskit.execute(qc, backend = backend, shots=shots, max_credits=credits)
      ibmq_essex_result = job.result()
      ibmq_essex_counts = ibmq_essex_result.get_counts(qc)
```

```
[17]: print(ibmq_essex_counts)
      qiskit.visualization.plot_histogram(ibmq_essex_counts)
```

    {'001': 1242, '101': 296, '110': 303, '111': 3020, '010': 215, '000': 1423,
    '011': 1623, '100': 70}

[17]:

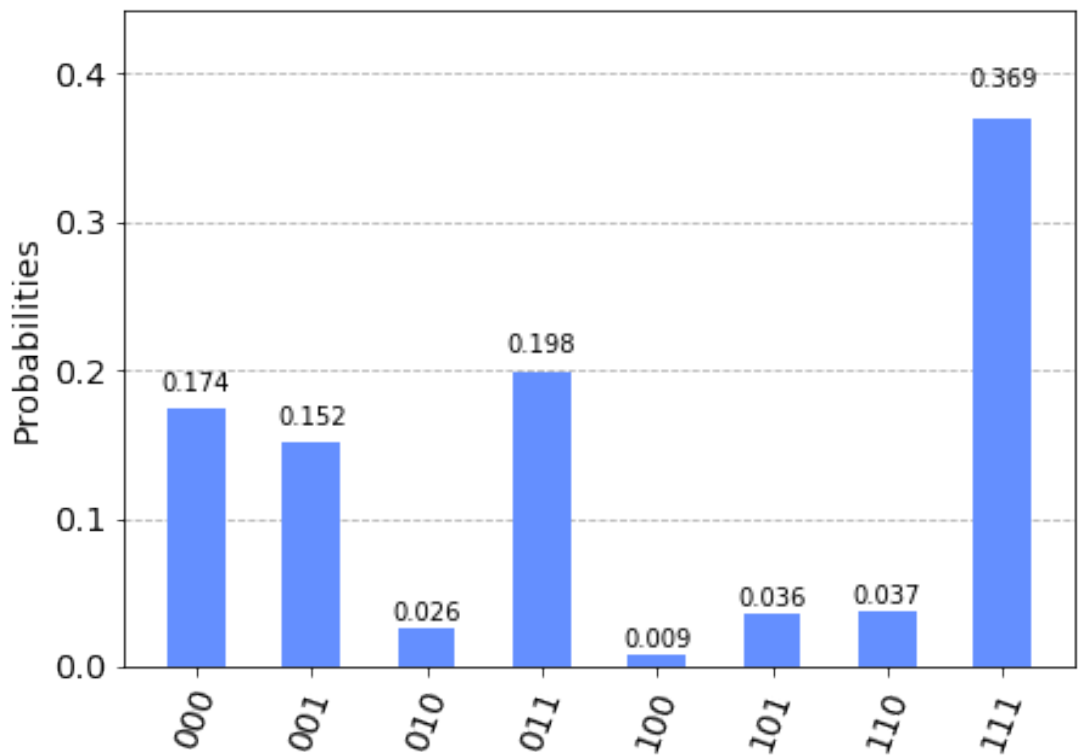

[18]: ```
ibmq_essex_hellinger_fidelity = qiskit.quantum_info.
 ↪hellinger_fidelity(qasm_simulator_result.get_counts(), ibmq_essex_result.
 ↪get_counts(qc))
ibmq_essex_hellinger_fidelity
```

[18]: 0.7525123193418292

[66]: ```
backend = provider.get_backend('ibmq_essex')
job = qiskit.execute(qc1, backend = backend, shots=shots, max_credits=credits)
ibmq_essex_result_1 = job.result()
ibmq_essex_counts_1 = ibmq_essex_result_1.get_counts(qc1)
```

[67]: ```
print(ibmq_essex_counts_1)
qiskit.visualization.plot_histogram(ibmq_essex_counts_1)
```

{'001': 100, '101': 120, '110': 861, '111': 2537, '010': 196, '000': 2659, '011': 577, '100': 1142}

[67]:

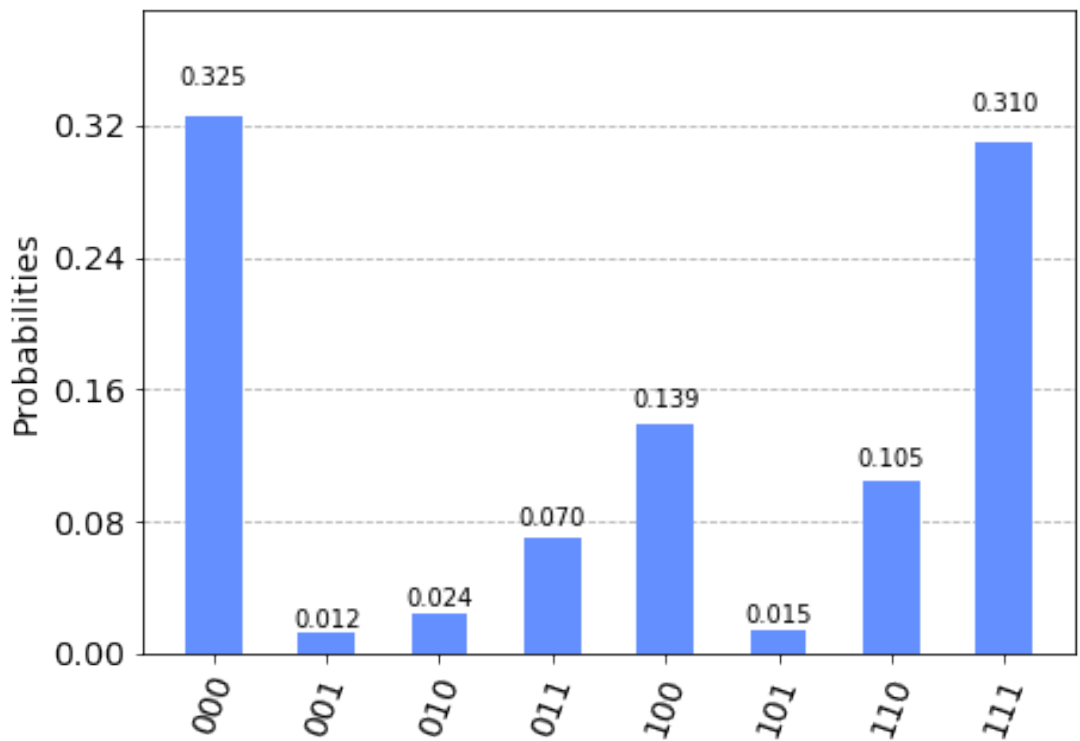

```
[68]: ibmq_essex_hellinger_fidelity_1 = qiskit.quantum_info.
      ↪hellinger_fidelity(qasm_simulator_counts_1, ibmq_essex_counts_1)
      ibmq_essex_hellinger_fidelity_1
```

[68]: 0.7360730741309539

[ ]:

[ ]:

```
[19]: backend = provider.get_backend('ibmq_burlington')
      job = qiskit.execute(qc, backend = backend, shots=shots, max_credits=credits)
      ibmq_burlington_result = job.result()
      ibmq_burlington_counts = ibmq_burlington_result.get_counts(qc)
```

```
[20]: print(ibmq_burlington_counts)
      qiskit.visualization.plot_histogram(ibmq_burlington_counts)
```

{'001': 2254, '101': 371, '110': 413, '111': 1875, '010': 274, '000': 1361, '011': 1543, '100': 101}

[20]:

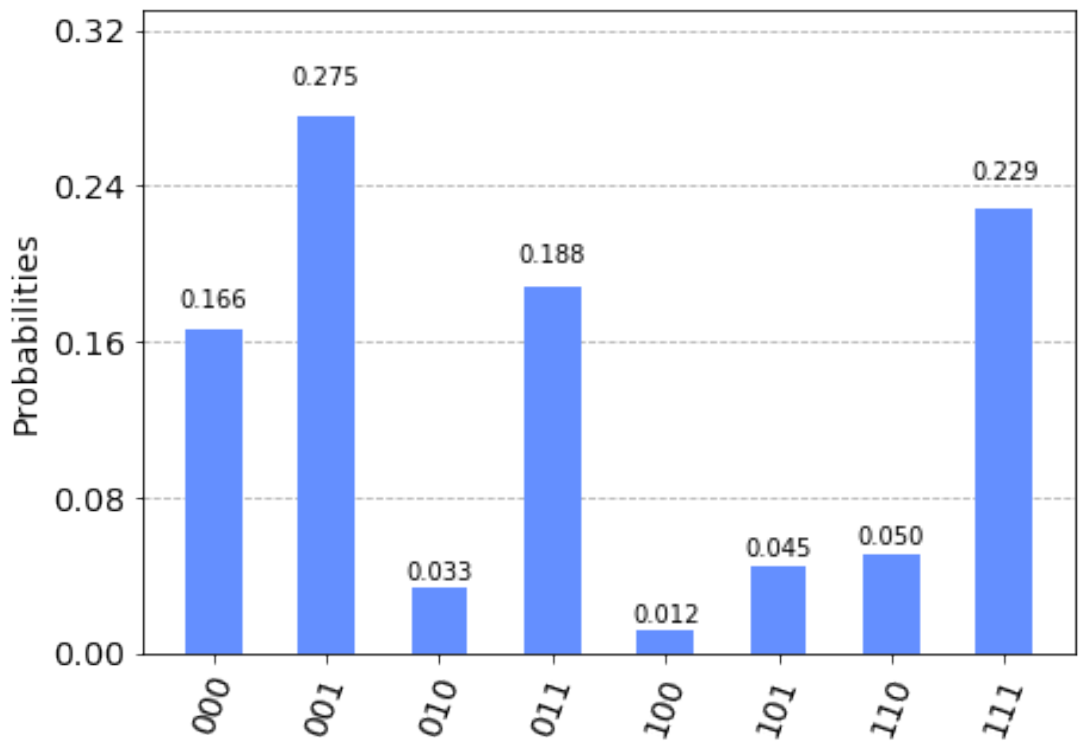

```
[21]: ibmq_burlington_hellinger_fidelity = qiskit.quantum_info.
      ↪hellinger_fidelity(qasm_simulator_result.get_counts(),
      ↪ibmq_burlington_result.get_counts(qc))
      ibmq_burlington_hellinger_fidelity
```

[21]: `0.6929537945927091`

```
[70]: backend = provider.get_backend('ibmq_burlington')
      job = qiskit.execute(qc1, backend = backend, shots=shots, max_credits=credits)
      ibmq_burlington_result_1 = job.result()
      ibmq_burlington_counts_1 = ibmq_burlington_result_1.get_counts(qc1)
```

```
[71]: print(ibmq_burlington_counts_1)
      qiskit.visualization.plot_histogram(ibmq_burlington_counts_1)
```

    {'001': 349, '101': 904, '110': 1212, '111': 2021, '010': 347, '000': 1582,
    '011': 542, '100': 1235}

[71]:

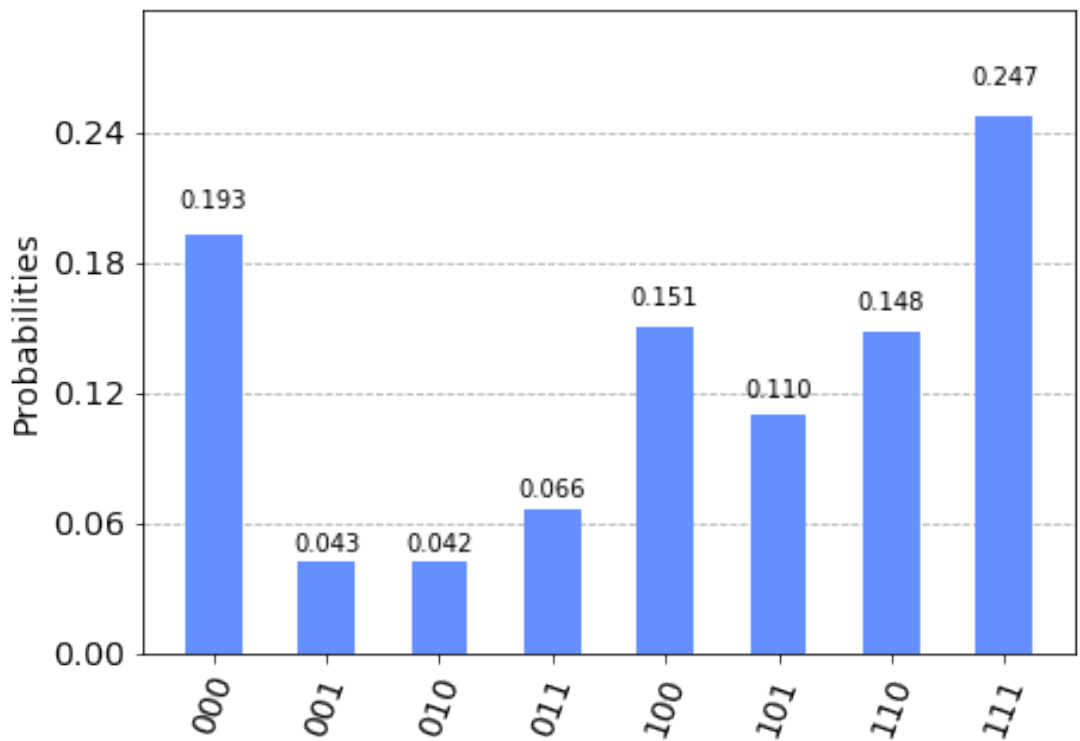

```
[72]: ibmq_burlington_hellinger_fidelity_1 = qiskit.quantum_info.
      ↪hellinger_fidelity(qasm_simulator_counts_1, ibmq_burlington_counts_1)
      ibmq_burlington_hellinger_fidelity_1
```

[72]: 0.6188179242105111

[ ]:

[ ]:

```
[22]: backend = provider.get_backend('ibmq_london')
      job = qiskit.execute(qc, backend = backend, shots=shots, max_credits=credits)
      ibmq_london_result = job.result()
      ibmq_london_counts = ibmq_london_result.get_counts(qc)
```

```
[23]: print(ibmq_london_counts)
      qiskit.visualization.plot_histogram(ibmq_london_counts)
```

{'001': 1078, '101': 151, '110': 255, '111': 2706, '010': 147, '000': 2347,
'011': 1444, '100': 64}

[23]:

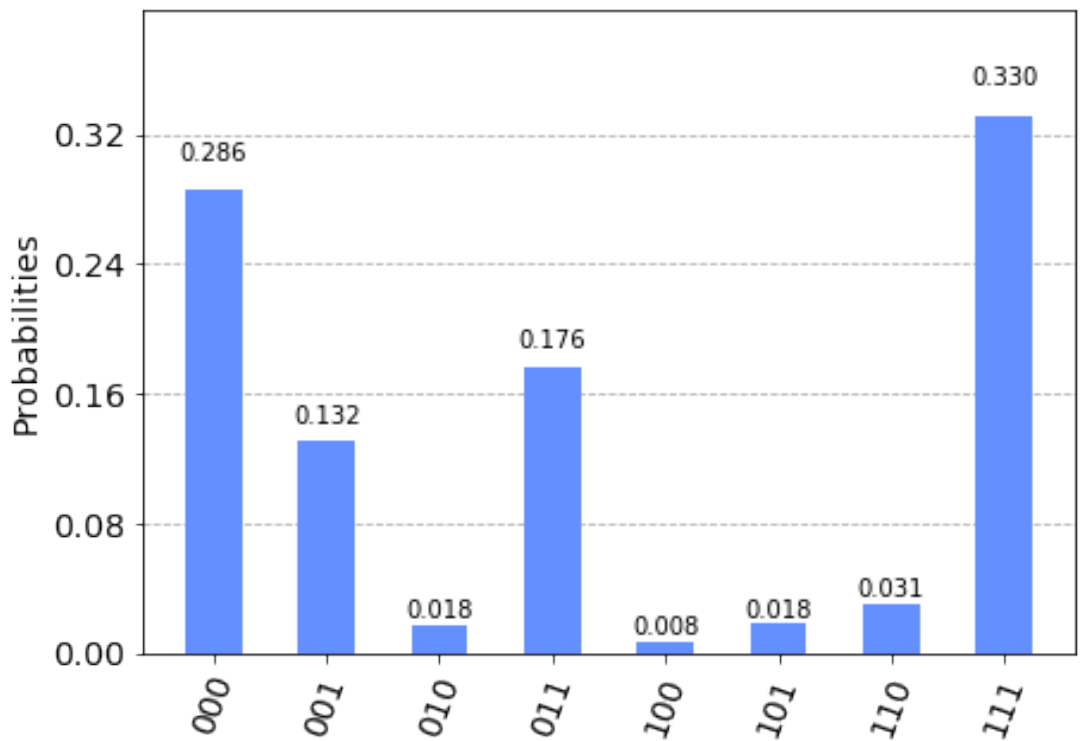

```
[24]: ibmq_london_hellinger_fidelity = qiskit.quantum_info.
       ↪hellinger_fidelity(qasm_simulator_result.get_counts(), ibmq_london_result.
       ↪get_counts(qc))
      ibmq_london_hellinger_fidelity
```

[24]: `0.7902814034323097`

```
[73]: backend = provider.get_backend('ibmq_london')
      job = qiskit.execute(qc1, backend = backend, shots=shots, max_credits=credits)
      ibmq_london_result_1 = job.result()
      ibmq_london_counts_1 = ibmq_london_result_1.get_counts(qc1)
```

```
[75]: print(ibmq_london_counts_1)
      qiskit.visualization.plot_histogram(ibmq_london_counts_1)
```

    {'001': 59, '101': 176, '110': 1156, '111': 2862, '010': 192, '000': 2396,
    '011': 407, '100': 944}

[75]:

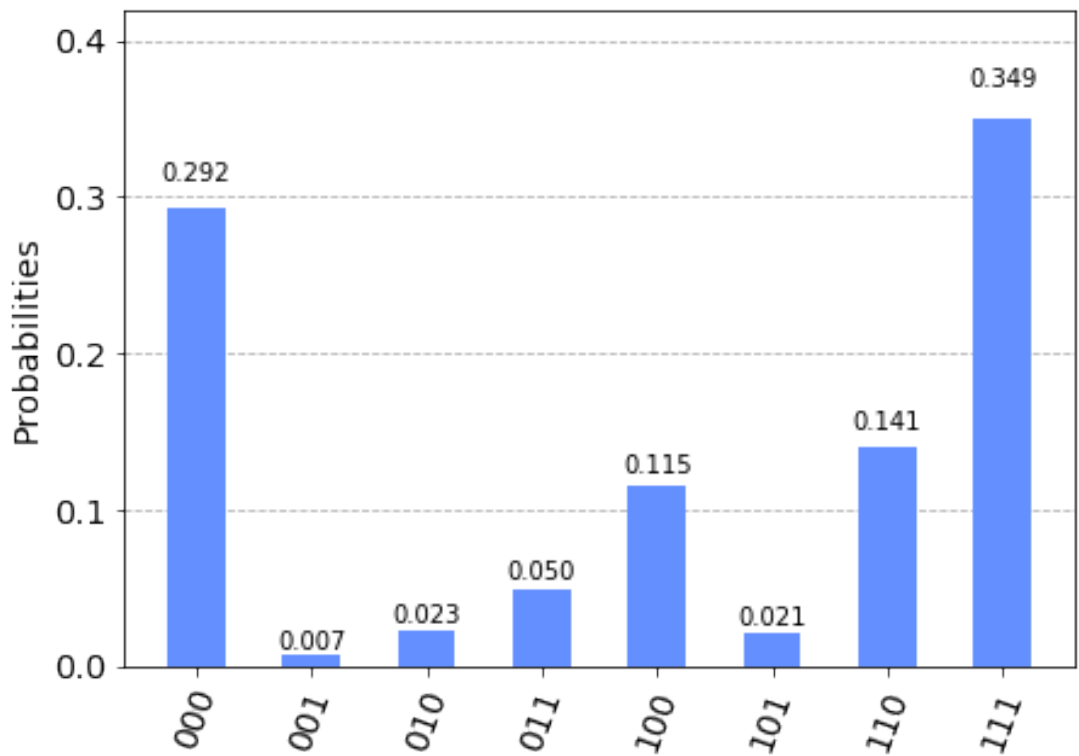

```
[83]: ibmq_london_hellinger_fidelity_1 = qiskit.quantum_info.
       ↪hellinger_fidelity(qasm_simulator_counts_1, ibmq_london_counts_1)
      ibmq_london_hellinger_fidelity_1
```

[83]: 0.7603983077670277

```
[ ]:
```

```
[ ]:
```

```
[ ]: backend = provider.get_backend('ibmq_ourense')
     job = qiskit.execute(qc, backend = backend, shots=shots, max_credits=credits)
     ibmq_ourense_result = job.result()
     ibmq_ourense_counts = ibmq_ourense_result.get_counts(qc)
```

```
[ ]: print(ibmq_ourense_counts)
     qiskit.tools.visualization.plot_histogram(ibmq_ourense_counts)
```

```
[ ]: ibmq_ourense_hellinger_fidelity = qiskit.quantum_info.
      ↪hellinger_fidelity(qasm_simulator_result.get_counts(), ibmq_ourense_result.
      ↪get_counts(qc))
     ibmq_ourense_hellinger_fidelity
```

```
[ ]: backend = provider.get_backend('ibmq_ourense')
     job = qiskit.execute(qc1, backend = backend, shots=shots, max_credits=credits)
```

```python
ibmq_ourense_result_1 = job.result()
ibmq_ourense_counts_1 = ibmq_ourense_result_1.get_counts(qc1)
```

```python
print(ibmq_ourense_counts_1)
qiskit.tools.visualization.plot_histogram(ibmq_ourense_counts_1)
```

```python
ibmq_ourense_hellinger_fidelity_1 = qiskit.quantum_info.
 ↪hellinger_fidelity(qasm_simulator_counts_1, ibmq_ourense_counts_1)
ibmq_ourense_hellinger_fidelity_1
```

```python
```

```python
```

```python
backend = provider.get_backend('ibmq_vigo')
job = qiskit.execute(qc, backend = backend, shots=shots, max_credits=credits)
ibmq_vigo_result = job.result()
ibmq_vigo_counts = ibmq_vigo_result.get_counts(qc)
```

```python
print(ibmq_vigo_counts)
qiskit.tools.visualization.plot_histogram(ibmq_vigo_counts)
```

```python
ibmq_vigo_hellinger_fidelity = qiskit.quantum_info.
 ↪hellinger_fidelity(qasm_simulator_result.get_counts(), ibmq_vigo_result.
 ↪get_counts(qc))
ibmq_vigo_hellinger_fidelity
```

```python
backend = provider.get_backend('ibmq_vigo')
job = qiskit.execute(qc1, backend = backend, shots=shots, max_credits=credits)
ibmq_vigo_result_1 = job.result()
ibmq_vigo_counts_1 = ibmq_vigo_result_1.get_counts(qc1)
```

```python
print(ibmq_vigo_counts_1)
qiskit.tools.visualization.plot_histogram(ibmq_vigo_counts_1)
```

```python
ibmq_vigo_hellinger_fidelity_1 = qiskit.quantum_info.
 ↪hellinger_fidelity(qasm_simulator_counts_1, ibmq_vigo_counts_1)
ibmq_vigo_hellinger_fidelity_1
```

```python
```

```python
```

```python
[29]: backend = provider.get_backend('ibmq_16_melbourne')
job = qiskit.execute(qc, backend = backend, shots=shots, max_credits=credits)
ibmq_16_melbourne_result = job.result()
ibmq_16_melbourne_counts = ibmq_16_melbourne_result.get_counts(qc)
```

```python
[31]: print(ibmq_16_melbourne_counts)
qiskit.visualization.plot_histogram(ibmq_16_melbourne_counts)
```

{'001': 1572, '101': 246, '110': 201, '111': 2736, '010': 179, '000': 2138, '011': 1075, '100': 45}

[31]:

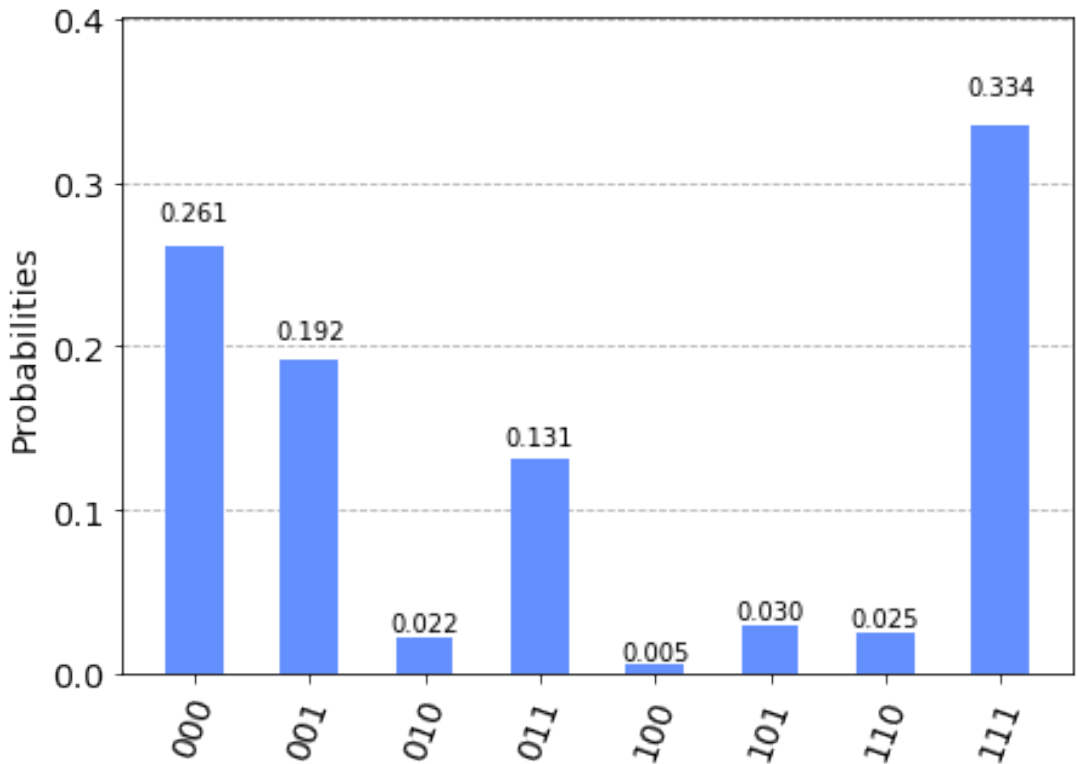

[32]:
```
ibmq_16_melbourne_hellinger_fidelity = qiskit.quantum_info.
 ↪hellinger_fidelity(qasm_simulator_result.get_counts(),
 ↪ibmq_16_melbourne_result.get_counts(qc))
ibmq_16_melbourne_hellinger_fidelity
```

[32]: 0.791719115005206

[76]:
```
backend = provider.get_backend('ibmq_16_melbourne')
job = qiskit.execute(qc1, backend = backend, shots=shots, max_credits=credits)
ibmq_16_melbourne_result_1 = job.result()
ibmq_16_melbourne_counts_1 = ibmq_16_melbourne_result_1.get_counts(qc1)
```

[77]:
```
print(ibmq_16_melbourne_counts_1)
qiskit.visualization.plot_histogram(ibmq_16_melbourne_counts_1)
```

    {'001': 56, '101': 289, '110': 1080, '111': 2570, '010': 242, '000': 2198,
    '011': 227, '100': 1530}

[77]:

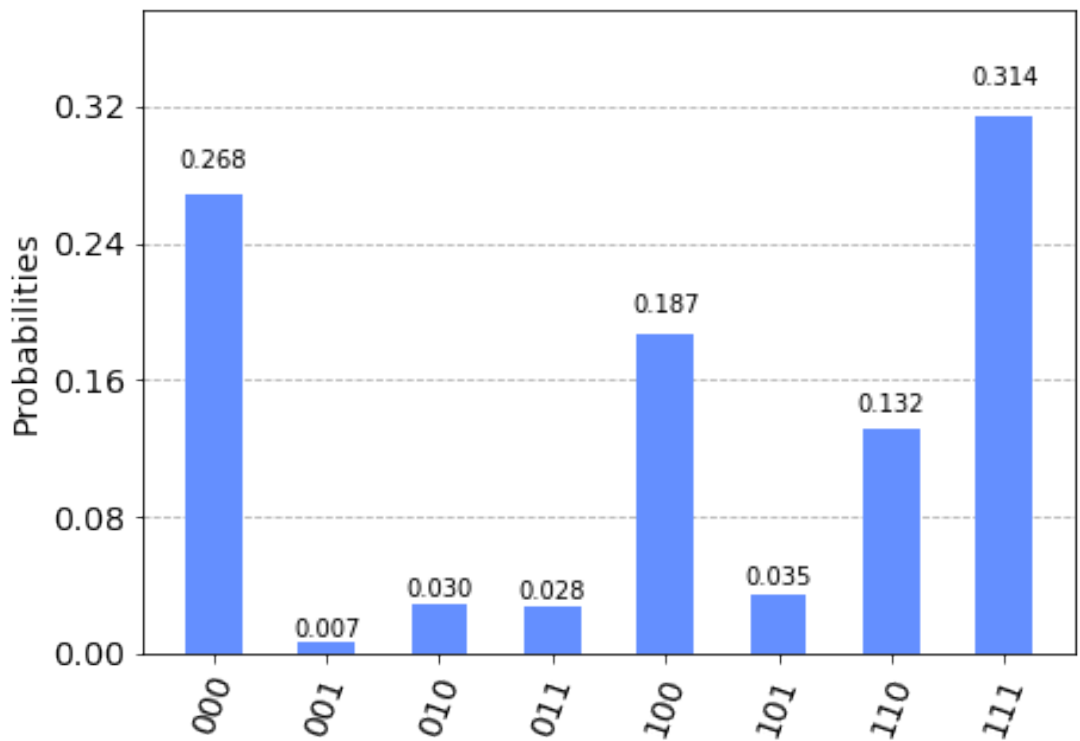

[78]: ```
ibmq_16_melbourne_hellinger_fidelity_1 = qiskit.quantum_info.
 ↪hellinger_fidelity(qasm_simulator_counts_1, ibmq_16_melbourne_counts_1)
ibmq_16_melbourne_hellinger_fidelity_1
```

[78]: 0.7685064431178723

[ ]:

[ ]:

[33]: ```
backend = provider.get_backend('ibmq_qasm_simulator')
job = qiskit.execute(qc, backend = backend, shots=shots, max_credits=credits)
ibmq_qasm_simulator_result = job.result()
ibmq_qasm_simulator_counts = ibmq_qasm_simulator_result.get_counts(qc)
```

[34]: ```
print(ibmq_qasm_simulator_counts)
qiskit.visualization.plot_histogram(ibmq_qasm_simulator_counts)
```

{'001': 1528, '000': 2094, '111': 3435, '011': 1135}

[34]:

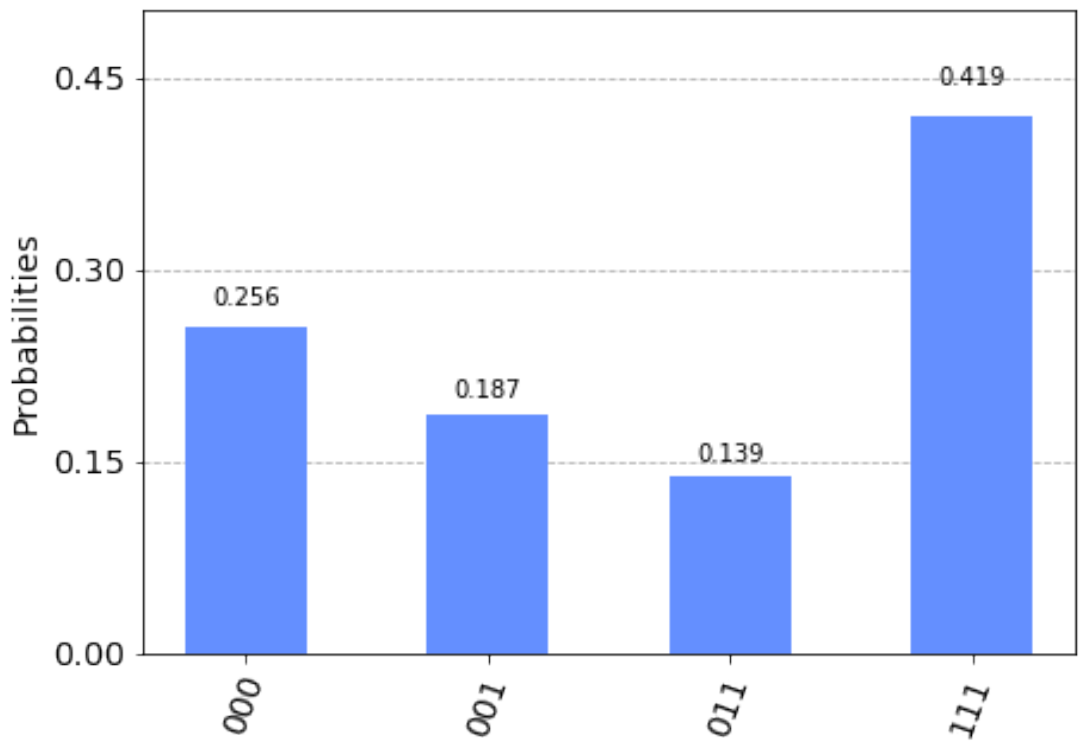

```
[35]: ibmq_qasm_simulator_hellinger_fidelity = qiskit.quantum_info.
       ↪hellinger_fidelity(qasm_simulator_result.get_counts(),
       ↪ibmq_qasm_simulator_result.get_counts(qc))
      ibmq_qasm_simulator_hellinger_fidelity
```

[35]: `0.9922734073102918`

```
[79]: backend = provider.get_backend('ibmq_qasm_simulator')
      job = qiskit.execute(qc1, backend = backend, shots=shots, max_credits=credits)
      ibmq_qasm_simulator_result_1 = job.result()
      ibmq_qasm_simulator_counts_1 = ibmq_qasm_simulator_result_1.get_counts(qc1)
```

```
[80]: print(ibmq_qasm_simulator_counts_1)
      qiskit.visualization.plot_histogram(ibmq_qasm_simulator_counts_1)
```

    {'110': 1110, '000': 2069, '100': 1545, '111': 3468}

[80]:

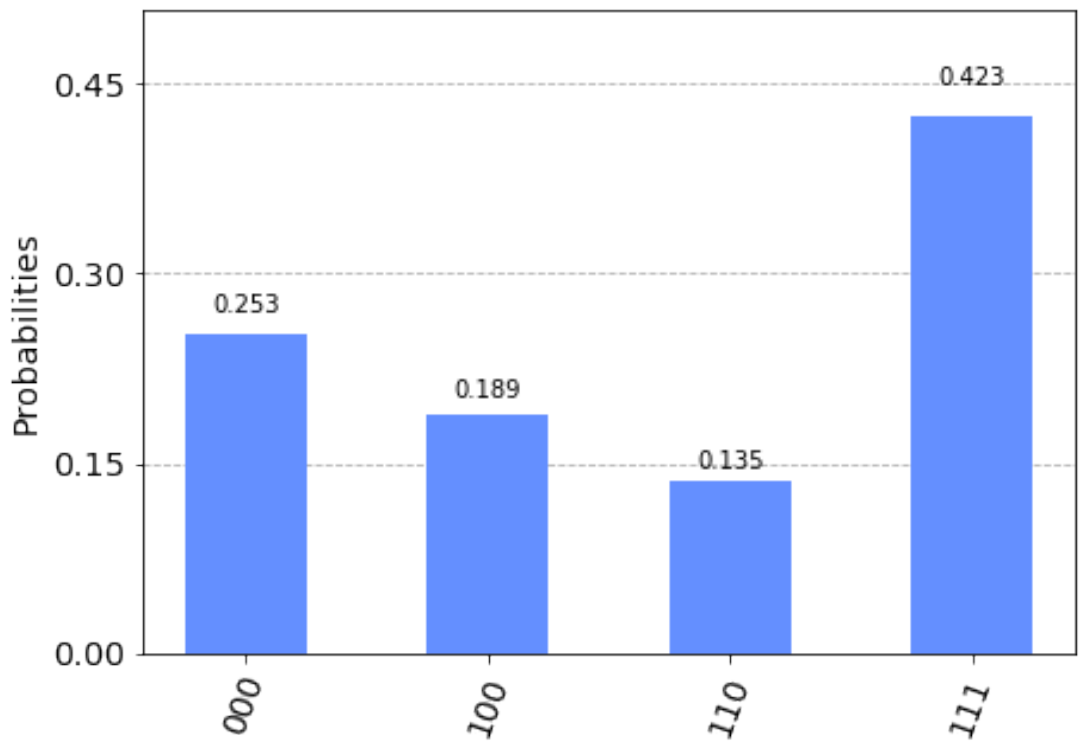

```
[81]: ibmq_qasm_simulator_hellinger_fidelity_1 = qiskit.quantum_info.
      ↪hellinger_fidelity(qasm_simulator_counts_1, ibmq_qasm_simulator_counts_1)
      ibmq_qasm_simulator_hellinger_fidelity_1
```

[81]: 0.9953684788547141

[ ]:

[ ]:

```
[38]: #hellinger_fidelity_dict = {'Yorktown':ibmqx2_hellinger_fidelity, 'Essex':
      ↪ibmq_essex_hellinger_fidelity, 'London':ibmq_london_hellinger_fidelity,
      ↪ 'Burlington':ibmq_burlington_hellinger_fidelity, 'Melbourne':
      ↪ibmq_16_melbourne_hellinger_fidelity, 'Vigo':ibmq_vigo_hellinger_fidelity,
      ↪'Ourense':ibmq_ourense_hellinger_fidelity}
```

```
[39]: hellinger_fidelity_dict = {'Yorktown':ibmqx2_hellinger_fidelity,
                                 'Essex': ibmq_essex_hellinger_fidelity,
                                 'London':ibmq_london_hellinger_fidelity,
                                 'Burlington':ibmq_burlington_hellinger_fidelity,
                                 'Melbourne': ibmq_16_melbourne_hellinger_fidelity,
                                 'Vigo':0, #ibmq_vigo_hellinger_fidelity Vigo is not
      ↪operational at the moment
                                 'Ourense':0} #ibmq_ourense_hellinger_fidelity
      ↪Ourense is not operational at the moment
```

```python
[84]: hellinger_fidelity_dict_1 = {'Yorktown':ibmqx2_hellinger_fidelity_1,
                                   'Essex': ibmq_essex_hellinger_fidelity_1,
                                   'London':ibmq_london_hellinger_fidelity_1,
                                   'Burlington':ibmq_burlington_hellinger_fidelity_1,
                                   'Melbourne': ibmq_16_melbourne_hellinger_fidelity_1,
                                   'Vigo':0, #ibmq_vigo_hellinger_fidelity_1 Vigo is
       ↪not operational at the moment
                                   'Ourense':0} #ibmq_ourense_hellinger_fidelity_1
       ↪Ourense is not operational at the moment
```

```python
[85]: hellinger_fidelity_dict
```

```python
[85]: {'Yorktown': 0.7962592386446301,
       'Essex': 0.7525123193418292,
       'London': 0.7902814034323097,
       'Burlington': 0.6929537945927091,
       'Melbourne': 0.791719115005206,
       'Vigo': 0,
       'Ourense': 0}
```

```python
[86]: hellinger_fidelity_dict
```

```python
[86]: {'Yorktown': 0.7962592386446301,
       'Essex': 0.7525123193418292,
       'London': 0.7902814034323097,
       'Burlington': 0.6929537945927091,
       'Melbourne': 0.791719115005206,
       'Vigo': 0,
       'Ourense': 0}
```

```python
[42]: import pandas as pd
```

```python
[87]: df = pd.DataFrame({'Quantum Circuit Figure 26': {x:y for x,y in
       ↪hellinger_fidelity_dict.items()},
                          'Quantum Circuit Figure 27': {x:y for x,y in
       ↪hellinger_fidelity_dict_1.items()}})
```

```python
[88]: df.index.name = 'IBM Q BackEnd'
```

```python
[89]: df
```

```
[89]:                Quantum Circuit Figure 26  Quantum Circuit Figure 27
      IBM Q BackEnd
      Burlington                      0.692954                   0.618818
      Essex                           0.752512                   0.736073
      London                          0.790281                   0.760398
      Melbourne                       0.791719                   0.768506
      Ourense                         0.000000                   0.000000
      Vigo                            0.000000                   0.000000
      Yorktown                        0.796259                   0.754103
```

```python
[46]: df.index
```

[46]: Index(['Burlington', 'Essex', 'London', 'Melbourne', 'Ourense', 'Vigo',
       'Yorktown'],
      dtype='object', name='IBM Q BackEnd')

[ ]:

[90]:
```
ax = df.plot(ylim=(0,1.0), kind='bar',figsize=(10,5))
ax.set_xlabel('IBM Q BackEnd')
ax.set_ylabel("Hellinger Fidelity")
```

[90]: Text(0, 0.5, 'Hellinger Fidelity')

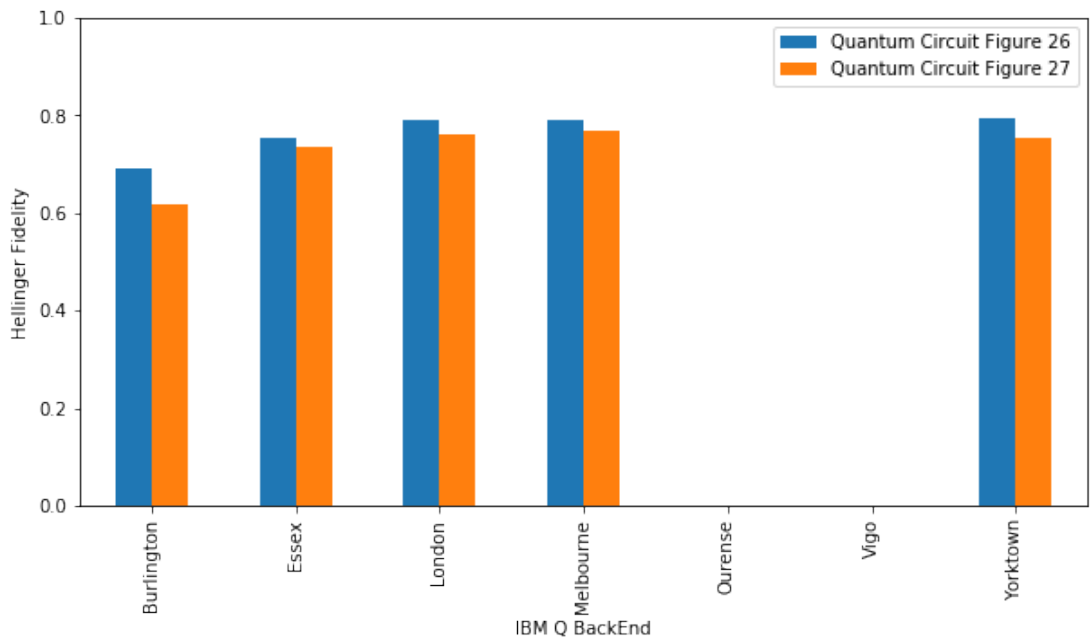

[92]: `df['Quantum Circuit Figure 26'].sort_values()`

[92]:
```
IBM Q BackEnd
Ourense       0.000000
Vigo          0.000000
Burlington    0.692954
Essex         0.752512
London        0.790281
Melbourne     0.791719
Yorktown      0.796259
Name: Quantum Circuit Figure 26, dtype: float64
```

[93]: `df['Quantum Circuit Figure 27'].sort_values()`

[93]:
```
IBM Q BackEnd
Ourense       0.000000
Vigo          0.000000
```

```
    Burlington    0.618818
    Essex         0.736073
    Yorktown      0.754103
    London        0.760398
    Melbourne     0.768506
    Name: Quantum Circuit Figure 27, dtype: float64
```

[ ]:

 

\renewcommand\bibname{References}
\bibliographystyle{apalike}
\bibliography{References.bib}

\end{document}